\documentclass[12pt]{article}
\usepackage[T2A]{fontenc}
\usepackage[utf8]{inputenc}
\usepackage[english]{babel}
\usepackage{bm,epsfig,amssymb,amsfonts,amsmath,cite,xcolor,adjustbox}
\usepackage[colorlinks=true,linkcolor=blue,unicode=true,breaklinks=true,linktocpage=true]{hyperref}
\usepackage{color}
\graphicspath{{Figs.dir/}}

\newcommand{\be}{\begin{equation}}
\newcommand{\ee}{\end{equation}}
\newcommand{\beq}{\begin{eqnarray}}
\newcommand{\eeq}{\end{eqnarray}}
\newcommand{\bea}{\begin{eqnarray}}
\newcommand{\eea}{\end{eqnarray}}
\newcommand{\ds}{\displaystyle}

\DeclareMathSymbol{\varGamma}{\mathord}{letters}{"00}
\newcommand{\Br}{\mbox{Br}}

\newcommand{\jp}{\ensuremath{J/\psi}}
\newcommand{\Dn}{\ensuremath{{D^0}}}
\newcommand{\Db}{\ensuremath{\bar{D}^0}}
\newcommand{\Kn}{\ensuremath{{K^0}}}
\newcommand{\Kb}{\ensuremath{\bar{K}^0}}
\newcommand{\Bn}{\ensuremath{{B^0}}}
\newcommand{\Bb}{\ensuremath{\bar{B}^0}}
\newcommand{\BBb}{\ensuremath{{\Bn\Bb}}}
\newcommand{\Bs}{\ensuremath{{B^0_s}}}
\newcommand{\Bbs}{\ensuremath{\bar{B}^0_s}}
\newcommand{\BBs}{\ensuremath{{\Bs\Bbs}}}
\newcommand{\Bd}{\ensuremath{{B^0_d}}}
\newcommand{\Bbd}{\ensuremath{\bar{B}^0_d}}
\newcommand{\BBd}{\ensuremath{\Bd\Bbd}}
\newcommand{\Ks}{\ensuremath{K^0_S}}
\newcommand{\jpks}{\ensuremath{\jp\Ks}}
\newcommand{\Vub}{\ensuremath{\left|V_{ub}\right|}}
\newcommand{\Vcb}{\ensuremath{\left|V_{cb}\right|}}

\newcommand{\Vud}{\ensuremath{\left|V_{ud}\right|}}
\newcommand{\Vcd}{\ensuremath{\left|V_{cd}\right|}}

\renewcommand{\vec}{\bm}

\newcommand{\ac}{(ATLAS Collab.)~}
\newcommand{\bc}{(Belle Collab.)~}
\newcommand{\bbc}{(Belle II Collab.)~}
\newcommand{\bac}{(BaBaR Collab.)~}
\newcommand{\cc}{(CMS Collab.)~}
\newcommand{\lc}{(LHCb Collab.)~}
\newcommand{\pc}{(PDG Collab.)~}
\newcommand{\kc}{(KLOE Collab.)~}
\newcommand{\kkc}{(KLOE-2 Collab.)~}
\newcommand{\sbc}{(SuperB Collab.)~}
\newcommand{\bsc}{(BESIII Collab.)~}
\newcommand{\arc}{(ARGUS Collab.)~}
\newcommand{\clc}{(CLEO Collab.)~}
\newcommand{\muc}{(Muon g-2 Collab.)~}
\newcommand{\hfc}{(HFLAV Collab.)~}
\newcommand{\oc}{(OPAL Collab.)~}
\newcommand{\alc}{(ALEPH Collab.)~}

\def\bbletal{et~al.}
\def\bblin{in}

\title{\bf Superfactory of bottomed hadrons Belle II}
\author{
S. I. Eidelman$^{\rm a,b,c}$, 
A. V. Nefediev$^{\rm a}$, 
P. N. Pakhlov$^{\rm a}$, 
V. I. Zhukova$^{\rm a}$ \\[5mm]
${\rm ^a}$ {\small\it P.N. Lebedev Physical Institute of RAS,}\\
{\small\it 119991, Leninsky pr. 53, Moscow, Russia}\\
${\rm ^b}$ {\small\it Budker Institute of Nuclear Physics of Siberian Branch of RAS,}\\
{\small\it 630090, Acad. Lavrentieva Pr. 11, Novosibirsk, Russia}\\
${\rm ^c}$ {\small\it Novosibirsk State University,}\\
{\small\it 630090, Pirogova str. 2, Novosibirsk, Russia}
}
\date{}

\begin{document}

\maketitle

\begin{abstract}
In 2018 the Belle II experiment, aimed at detailed studies of $B$-mesons, started operation at the electron-positron collider Super\-KEKB at KEK (Japan).  This was preceded by a long and quite successful work of the $B$-factories of previous generations, including the Belle experiment for which Belle II is a successor. This experiment is unique and has no analogues or competitors in the world. The spectrum of problems it is aimed at is quite broad: from studies of hadronic states containing heavy quarks to precision measurements and searches for New Physics beyond the Standard Model. This review describes specific features of the Belle II experiment, its ambitious goals and specific tasks, expected results of its work and the hopes related to its successful accomplishment.
\end{abstract}

\tableofcontents

\section{Introduction}\label{chapter:introduction}

The Theory of Everything --- a fundamental theory explaining all physical phenomena from first principles is probably just an unreachable dream of scientists. However, in spite of elusive hopes to build such a theory, contemporary physics possesses a powerful tool which soaked up the knowledge gained by the humanity during many centuries. This tool is conventionally referred to as the Standard Model (SM) that emphasises its monumentality and universality. Nevertheless, from its very first days the SM humbly pre\-tend\-ed to be just an effective theory which can allow one to describe with an acceptable accuracy various phenomena in a limited energy region. The fact that for fifty years since its creation both the predictive power and the applicability domain of the SM have considerably surpassed the most bold expectations of its adherents does not change our understanding that there must exist a more fundamental theory which incorporates the SM as a low-energy approximation. Therefore, either something at odds with the SM will show up at some energy or inaccuracy of its description of the world will become evident thus revealing the desired next-generation theory. These expected deviations from the predictions of the SM are conventionally called New Physics (NP).

The key idea of the SM is to combine several known types of interactions on the basis of some fundamental principles which include relativistic invariance, gauge symmetry, and the spontaneous breaking of its electroweak subgroup through the Higgs mechanism. The first example of successful unification of various types of interactions is the theory of electromagnetism created in the second half of the 19th century and which resulted in the Maxwell system of equations describing electric and magnetic phenomena within a unified approach. The next important milestone in the history of the formation of the theory in it present form was the creation at the beginning of the 20th century of nonrelativistic quantum mechanics and later, already in the middle of the century, the completion of its relativistic generalisation to electromagnetic phenomena which in modern literature is commonly called Quantum Electrodynamics or QED. Thus, it took physicists about a hundred years to proceed from the classical description of electric charges and their interaction to the description of electromagnetic phenomena in terms of field quanta, including photons --- massless carriers of the electromagnetic interaction. At the same time, the physicists fully realised and used an important role of the gauge invariance of electro\-mag\-ne\-tic interactions which grew from a curious artifact to the rank of one of the fundamental principles of the theory. Subsequently, the gauge principle allowed one to describe weak and strong interactions, first as separate theories and then within a unified framework with QED which is eventually known as the SM. It is also difficult to overestimate the role of the Higgs mechanism which gave masses to the gauge bosons of the weak interaction without violating another key property of the theory, namely its renormalisability.

The history of the SM resembles playing Lego bricks: as soon as new ex\-pe\-ri\-men\-tal information arrives, it is naturally integrated into the already existing construction typically leading only to quantitative rather than qualitative changes in the theory. It is sufficient to mention the experimental observation of the neutrino oscillations and, this way, a proof of the existence of a nonzero neutrino mass. Although in the original version of the SM neutrinos were considered massless, the inclusion of their masses did not require serious changes of the theory, although this brought yet another unsolved problem of the SM concerning the origin and hierarchy of the neutrino masses. In the meantime a nonvanishing neutrino mass resolved another fundamental question to the theory since the existence of strictly massless states would imply the presence of additional symmetries not inherent to the SM. In some cases, for example, in the situations with the third generation of quarks, neutral currents or the Higgs boson theoretical predictions of new particles preceded their experimental detection. In the latter case, the experimental confirmation arrived more than forty years later --- the long-awaited discovery of the Higgs boson in 2012 \cite{Chatrchyan:2012ufa,Aad:2012tfa} can without exaggeration be considered a triumph of the SM.

A surprising flexibility of the SM which allows it to stand all experimental challenges has, however, a drawback --- the framework of the model is not regulated intrinsically, within the model itself, and a large number of allowed free parameters and ingredients with an unexplained hierarchy emphasises its nonfundamental nature. This, together with other shortcomings of the SM such as its inability to describe gravity, explain the existence and nature of the dark matter and dark energy or the baryon asymmetry of the Universe, calls for searches for a theory more fundamental than the SM. Despite a large number of theoretical ideas and efforts made to proceed beyond the SM (supersymmetry, extra dimensions, extra generations of particles, technicolour, and so on) to date none of them can be regarded as the mainstream in theoretical physics. A possible way out could be related to an experimental breakthrough with a discovery of a new phenomenon or effect that goes beyond the existing physical paradigm and suggests a direction for the further development of the theory. To this end it is necessary to carry out precision measurements to search either for statistically significant deviations of the experiment from the theory for the processes allowed by the SM or for processes either forbidden by the SM or predicted to be so rare beyond the scope of the contemporary experiment. Measurements of the sides and angles of the unitarity triangle (see chapter \ref{chapter:standmod}) provide an example of the first approach while the second one can be exemplified by searches for the processes with the lepton flavour violation such as
\be 
\mu^-\to\ e^-+\gamma\quad\mbox{or}\quad \mu^-\to e^-+e^++e^-. 
\ee

An interesting and one of the most intriguing properties of our world is the existence of counterparts, heavy and unstable, of all particles of matter. Electron, electron neutrino, $u$- and $d$-quark constitute the first generation of the matter particles. They are sufficient to build the entire world around us and the Nature could have restricted itself to this generation only, however, it has shown a great zeal to create three of them. The branch of the SM devoted to studies of the world beyond the first generation of fermions is called Flavour Physics (FP) which is the main subject of this review. The Nature demonstrates a surprising indifference to these new generations of leptons and quarks: they are neither required nor forbidden by the SM principles\footnote{A disclaimer is in order here that the reproach for the Nature indifference may not be entirely fair --- it has created exactly as many generations of particles as is necessary for our existence. Meanwhile it is unclear which term of the SM Lagrangian encodes the fact of our existence.}. The SM is not able to deduce the number of generations in nature from its basic postulates but it successfully predicted them (sometimes not only the very fact of existence but their properties as well) from experimental measurements through which one can ``smell'' the presence of new heavy particles in the world. So the charmed quark was ``derived'' from a small probability of the decay $K^0\to \ell^+ \ell^-$, the third generation of quarks was predicted based on the violation of the combined $CP$ parity, and it became clear that the mass of the $t$ quark was large, much larger than expected, after the discovery of oscillations of the neutral $B_d$ mesons. The FP contributes most of all to the number of the free parameters of the SM that emphasises once more how vague and nonfundamental this sector of the SM is. This also explains the hope that FP may provide a hint in searches for a more fundamental theory.

We conclude this introductory section with one more important claim. Physics of heavy flavours, in particular physics of $B$-mesons, is closely related to a very mysterious phenomenon in nature, namely the violation of the combined $CP$ invariance which is a subject of many reviews --- see, for example, Refs.~\cite{Danilov:1998yt,Bondar:2007zz}. Let us mention only one problem related to $CP$-violation: in the SM it is not large enough to explain the existence of our Universe. This seems to imply that the SM is incomplete and other sources of $CP$ violation must exist because ... we exist! Most (so far hypothetical) extensions of the SM contain additional sources of the $CP$ violation. If some of them are correct, the first insight into physics beyond the SM can indeed be gained from studies of $CP$ violation.

The Belle II experiment is aimed at precision studies of the FP, in particular at a very accurate measurement of the $CP$ violation in the processes reliably predicted by the SM and thus at revealing a possible disagreement indicating the presence of NP. The aim of this review is to discuss the new opportunities opening up for the physical community with the start of this experiment. Particular attention is paid to those tasks for which Belle II has an advantage over other present-day experiments in the field of High Energy Physics, the Large Hadron Collider (LHC) in the first place. A comprehensive description of the physical program for the Belle II experiment can be found in the Belle II Physics Book \cite{Kou:2018nap}.

The review has the following structure. Chapter \ref{chapter:standmod} contains elementary information about the SM necessary for understanding the rest of the review. Chapter \ref{chapter:bfactories} summarises the history of experiments at electron-positron colliders followed by a detailed discussion of the Belle II detector. Chapters \ref{chapter:measurements} and \ref{chapter:leptons} provide a description of the most interesting and promising measurements related to various SM tests and NP searches which are planned at Belle II and in which it demonstrates an undeniable leadership compared to other experiments. Some aspects of hadronic physics in the Belle II experiment are addressed in chapter \ref{chapter:hadrons}. Concluding remarks are collected in chapter \ref{chapter:conclusions}.

\section{Elementary essentials of the SM}\label{chapter:standmod}

In this chapter we provide basic information on the SM and introduce some concepts and definitions necessary for understanding  the main part of the review. The reader can find more detailed information about the SM and particular theories which constitute it in numerous textbooks and monographs. A discussion of the current situation in the particle physics can be found, for example, in the reviews \cite{Boos:2014vpa,Kazakov:2014ufa,Kazakov:2019fil}. Information about the Standard Model in a compact form convenient for practical applications, the values of its parameters as well as the measured properties of various particles can be found in the regularly updated review of the Particle Data Group (PDG) Collaboration \cite{Zyla:2020zbs}.

The SM is a gauge theory in which the interactions are carried by twelve gauge bosons --- spin-1 particles (massless photon and eight gluons plus massive $W^{\pm}$ and $Z^0$ bosons). In addition to the gauge fields the theory contains fundamental matter fields --- the fermions: leptons and quarks. Neutrinos (neutral leptons) participate in weak interactions only, charged leptons (electron, muon, and $\tau$-lepton) in the electromagnetic and weak ones, and quarks --- in all three types of interactions. The gauge bosons of the electroweak sector and the fermions acquire their masses due to a nonvanishing vacuum expectation value of a scalar field --- this way the Higgs boson appears in the scalar sector of the theory; it participates only in weak interactions. 

The six leptons and six quark fields form three generations, two leptons and two quarks in each, 
\be
\begin{tabular}{lccc}
Generation & I & II & III\\[2mm]
Leptons & 
$\ds\left(\nu'_e \atop e\right)$ & 
$\ds \left(\nu'_\mu\atop\mu\right)$ & 
$\ds \left(\nu'_\tau\atop \tau\right)$
\\[6mm]
Quarks & 
$\ds \left(u'\atop d\right)$ & 
$\ds \left(c'\atop s\right)$ & 
$\ds \left(t'\atop b\right)$,
\end{tabular}
\label{generations}
\ee
where the prime at the upper fermions is related to the mixing of the mass and weak eigenstates, it will be explained below. It is important to note that the (primed) leptons and quarks of all three generations interact identically with the gauge bosons, therefore they are indistinguishable from the point of view of weak, electromagnetic and strong interactions --- they differ only in the interaction with the Higgs field that leads to a hierarchy of the fermion masses. In case of leptons, this phenomenon is known as the lepton universality, the experimental verification of which is one of the most important tests of the SM, and in the case of quarks, this leads to the unitarity of the mixing matrix which is also subject to a thorough experimental verification.

The simplest theory contained in the SM is QED built on the gauge group $U(1)$. This theory is believed (perhaps too arrogantly) to be fully understood.

A distinctive feature of weak interactions, the second gauge theory of the SM, is that in each of the three generations left and right fermion states interact differently with the gauge fields corresponding to the gauge group $SU(2)$: left-handed fermions do interact with the gauge bosons while right-handed ones do not\footnote{It should be noted, however, that after the gauge symmetry breaking by the Higgs field, right-handed fermions begin to interact with the $Z^0$ boson.}. It remains unknown why the nature preferred one chirality to the other. In the SM, this mirror symmetry breaking is simply postulated for consistency with the experiment. This postulate, however, requires the most careful verification since there are many extensions of the SM that predict an admixture of the right-handed currents in the amplitudes of the measured processes that could be detected in this way.

Another important feature of weak interactions is that quark fields interact not by their mass states but by some combinations of them,
\be
\left( \begin{array}{c}
u' \\ c' \\ t'
\end{array}
\right)
=
\left(
\begin{array}{ccc}
V_{ud}&V_{us}&V_{ub}\\
V_{cd}&V_{cs}&V_{tb}\\
V_{td}&V_{cs}&V_{tb}
\end{array}
\right)
\left( \begin{array}{c}
u \\ c \\ t
\end{array}
\right),
\label{CKMmatrix}
\ee
where the mixing matrix of the mass and weak eigenstates $V$ is known as the Cabibbo-Kobayashi-Maskawa (CKM) matrix. The mixing occurs because weak interactions and the Higgs field, which gives masses to the quarks, ``failed to agree'' how the quarks are assigned to different generations. They so to say ``look'' at the quarks from different angles. We do not know why this has happened but we only know that there are no fundamental bans to prevent this.

The CKM-matrix is unitary ($V^\dagger V=VV^\dagger=1$) as a result of the gauge symmetry of weak interactions broken by the Higgs mechanism but initially put at the forefront of the construction of the SM. Gauging with a non-Abelian group ($SU(2)$ in case of weak interactions) requires that the coupling constants of all three generations of quarks with the gauge fields should be the same. This universality of the constants leads to the above unitary rotation. We will dwell in more detail on the experimental checks of this property in chapter \ref{chapter:triangle} below.

An important feature of the CKM-matrix in the case of three or more generations of quarks is its fundamental complexity: different elements of the matrix can contain different complex phases, in contrast to the case of the mixing matrix for two generations when the phase of all four elements is the same and unobservable. It is important to note that ``can contain'' does not necessarily mean ``definitely contain''. Quite surprising is not even that the totalitarian Gell-Mann principle (``Everything that is not forbidden is mandatory'') triumphed in this case and the matrix indeed turned out to be complex, but that this seemingly unrestrained complexity of the CKM-matrix is surprisingly small. Indeed, the Jarlskog determinant (see the definition of various quantities involved in Eq.~(\ref{wolfeinstein}) below) $J = \left| \epsilon^{ijk} {\rm Im} (V_{i1}V_{j2}V_{k3}) \right| \sim A^2\lambda^6 \eta$ \cite{Jarlskog:1985ht}, which is a measure of the ``complexity'' of a unitary matrix, is approximately $10^{-5}$, although for a randomly generated unitary $3\times 3$ matrix it would be uniformly distributed between 0 and 1/8. Thus we can be sure that the CKM-matrix was invented by the nature far from randomly, although we do not know yet what considerations it drew upon in this case. Nevertheless, even such a small complexity leads to one of the most mysterious phenomena of physics, $CP$ violation.

Not less surprising property of the CKM-matrix is its (found experimentally) closeness to the unity matrix \cite{Zyla:2020zbs},
\be
V=
\left(
\begin{array}{ccc}
0.97401 \pm 0.00011 & 0.22650 \pm 0.00048 & 0.00361^{+0.00011}_{-0.00009} \\
0.22636 \pm 0.00048 & 0.97320 \pm 0.00011 & 0.04053^{+0.00083}_{-0.00061} \\
0.00854^{+0.00023}_{-0.00016} & 0.03978^{+0.00082}_{-0.00060} & 0.999172^{+0.000024}_{-0.000035}
\end{array}
\right),
\ee
naturally reflected in the Wolfenstein parametrisation~\cite{Wolfenstein:1983yz}. In this parametrisation, a small deviation from the unity matrix is explicitly emphasised by the expansion in the small parameter $\lambda\equiv\sin\theta = 0.22658\pm 0.00044$, where $\theta$ is the Cabibbo angle. Then, up to the terms of the order $\lambda^4$, the CKM-matrix takes the form
\be
V\!=\!
\left( \! 
\begin{array}{ccc}
1-\frac{1}{2}\lambda^2 & \lambda & A\lambda^3(\rho-i\eta) \\
-\lambda & 1-\frac{1}{2}\lambda^2 & A\lambda^2 \\
A\lambda^3(1-\rho-i\eta) & -A\lambda^2 & 1
\end{array}
\! \right),
\label{wolfeinstein}
\ee
while all other parameters ($A$, $\rho$, $\eta$) are considered to be of the order of unity. 

\begin{figure}[t!]
\centerline{\epsfig{file=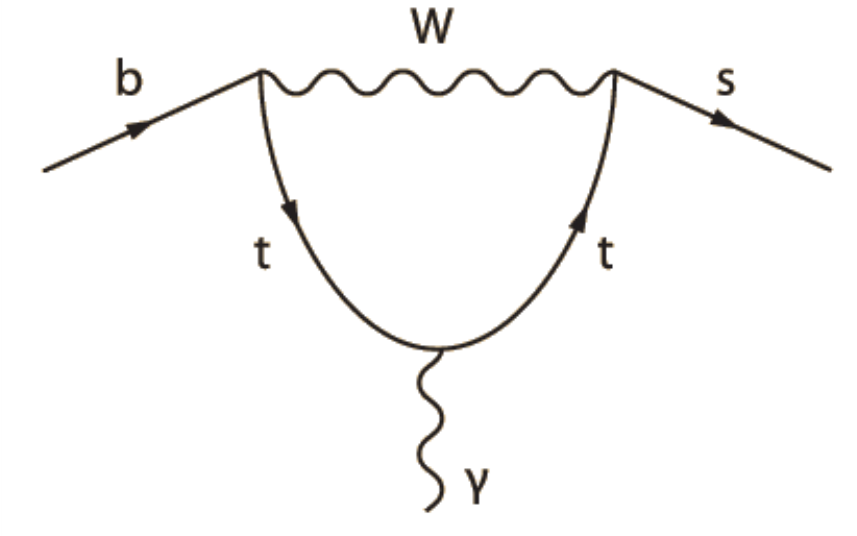, width=0.4\textwidth}
\hspace*{0.01\textwidth}
\epsfig{file=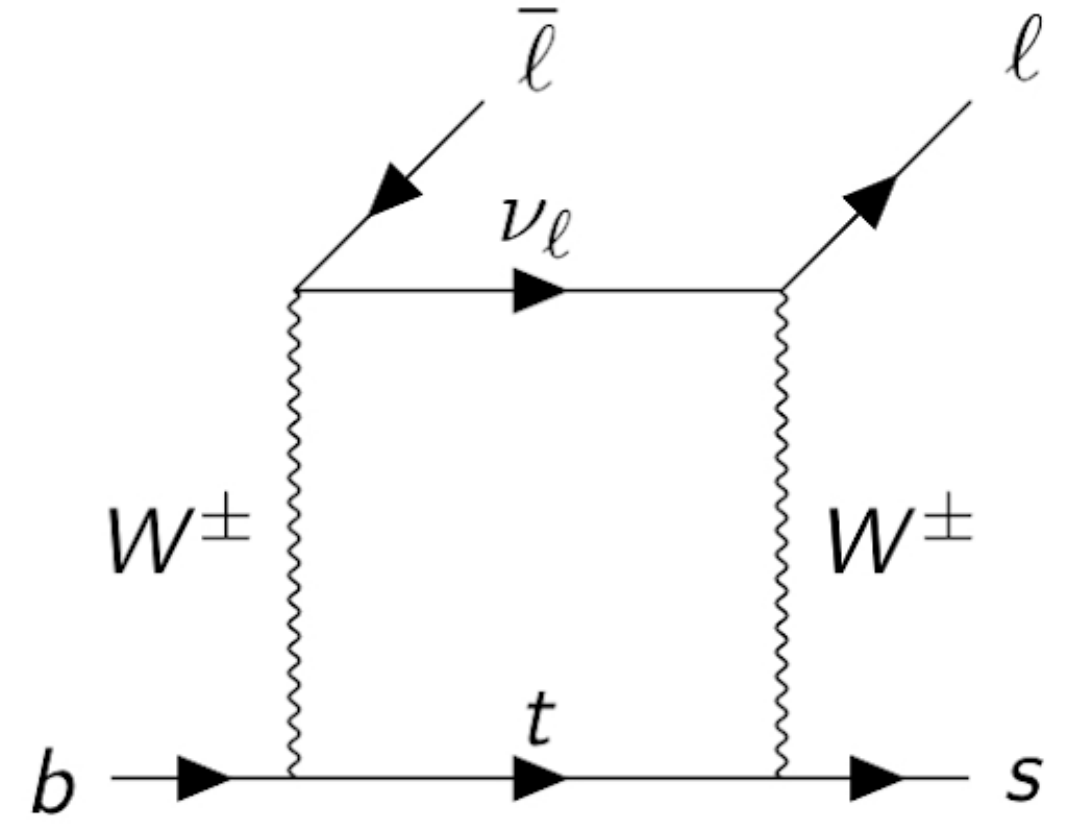, width=0.35\textwidth}} 
\caption{Examples of the loop diagrams for the quark flavour changing processes: ``penguin'' diagram for the transition $b\to s\gamma$ (left plot) and box diagram for the transition $b\to s l\bar{l}$ (right plot). Here $W$ is the $W$-boson, $t$ is the $t$-quark, and $\nu_l$ is the neutrino corresponding to the lepton $l$.}
\label{fig:penguin}
\end{figure}

One of the main tasks of the FP in general and the Belle II experiment in particular is to test self-consistency of the CKM quark mixing mechanism --- this will be discussed in chapter \ref{chapter:measurements}. Not less important are studies of the $b$-quark decays into a lighter quark flavour $q$ without changing the charge, that is into the $d$ or $s$ quark (the abbreviation FCNC for Flavour Changing Neutral Currents is often used in the literature in this respect). In the SM such processes are forbidden at the tree level and in the lowest order of the perturbation theory are described by one-loop diagrams with the $W$-boson exchange (the so-called ``penguin''\footnote{Diagrams of this type were first proposed in Ref.~\cite{Vainshtein:1975sv} and got this name somewhat later due to John Ellis --- see, for example, Ref.~\cite{Shifman:1995hc} and references therein. An example of the Feynman diagram for the transition of $b\to s\gamma$ in the penguin interpretation is shown in Fig.~\ref{fig:penguin}.} and box diagrams --- see Fig.~\ref{fig:penguin}). A consequence of the different interactions of the left- and right-handed fermions with the $W$ boson is, for example, an almost uniquely fixed polarisation of the photon in the decay $b\to s\gamma$ since in the SM the contribution of the operator providing the right polarisation of the photon is suppressed compared to the contribution of the operator responsible for its left polarisation by the ratio of the strange and bottom quark masses, $m_s/m_b\ll 1 $ (see also the discussion below in this chapter). As a result, the photon is left-handed with very high accuracy. The absence of other contributions to the amplitude of the $b\to s\gamma$ transition is very important from the point of view of minimising the role of hadronic effects and, hence, for increasing and better control of the accuracy of theoretical predictions in general. Consequently, experimental detection of a noticeable admixture of the right-handed photon in such a decay would signal a manifestation of NP.

Evaluation of the ``penguin'' processes considerably simplifies if one notices that their typical scale is of the order of the $b$-quark mass ($m_b\simeq 4.3$ GeV \cite{Zyla:2020zbs}) while physics related to the $W$ boson, Higgs boson, and $t$ quark takes place at a much higher electoweak scale of the order of 100 GeV. For this reason, in ``penguin'' diagrams the interactions mediated by such exchanges can be treated as short-ranged and substituted by point-like vertices with the effective couplings known as the Wilson coefficients $C_i$. Then the effective interaction Hamiltonian takes the form \cite{Buchalla:1995vs,Chetyrkin:1996vx}
\be
H_{\rm SM}^{\rm eff}=-\frac{4G_F}{\sqrt{2}}V_{tq}^*V_{tb}\sum_{i}C_iQ_i,
\label{HSMeff}
\ee
where $q=d,~s$, the Fermi constant $G_F=1.166\times 10^{-5}$ GeV$^{-2}$, and $Q_i$ are the corresponding operators. We do not dwell on the explicit form of these operators but will only briefly discuss the physical meaning of the most important of them. Thus, the operators $Q_1$ and $Q_2$ describe the interaction of the quark currents and the operators $Q_7$ and $Q_8$ describe the photonic and gluonic dipole interactions, respectively. To study the decays of the form $b\to q\ell^+ \ell^-$ ($\ell=e,\mu,\tau$) and $b\to q\nu\bar{\nu}$ it is necessary to add the operators $Q_9$, $Q_{10}$, and $Q_L^l$ responsible for the interactions of the quark and lepton currents. 

Let us discuss the operator $Q_7$, which will be needed below. It mixes the left- and right-handed quark components through the insertion of the mass operator in the external $b$-quark line of the penguin diagram (see Fig.~\ref{fig:penguin}). Such an insertion is proportional to the mass of the quark, so that $Q_7\propto m_b$. A similar operator with the mass insertion in the external line of the light quark $Q_7'\propto m_q$ is introduced with the Wilson coefficient $C_7'$. Since $m_b\gg m_q$ ($q=d,s$) then, in the framework of the SM, the operator $Q_7$ provides the dominant contribution while $Q_7'$ gives a small correction only. It is important to note that the photon polarisations in the operators $Q_7$ and $Q_7'$ are different, therefore, as already discussed above in this chapter, with very high accuracy the SM predicts the left-handed polarisation of the photon in weak decays of the form $b\to q\gamma$.

The accuracy gained in calculations of the amplitudes in the effective theory given by the Hamiltonian (\ref{HSMeff}) is determined by the contribution of the neglected terms, that is, by the ratio $m_b^2/m_W^2\sim 10^{-4}$, with $m_W$ for the $W$-boson mass. In addition, since the scale of strong interactions $\Lambda_{\rm QCD} \simeq 350$ MeV is small compared to $m_b$ it is necessary to take into account the gluon exchanges between quarks. In the energy region around $m_W$ the strong coupling constant $\alpha_s$ is sufficiently small that allows one to resort to perturbative theory keeping only the lowest order in $\alpha_s$ compatible with the required precision. The Wilson coefficients at this scale are determined from the requirement of coincidence (up to the terms of the order $m_b^2/m_W^2$) of the amplitudes calculated in the exact and effective theories. Studies of weak decays of $B$-mesons require a knowledge of the Wilson coefficients at the scales of the order $m_b$ that can be achieved by using the renormalisation group analysis.
It is important to note that the Wilson coefficients obtained this way are universal, that is, they do not change from process to process.
All coefficients $C_i$ required for evaluating the processes of the form $b\to s\gamma$ are known in the NNLO (next-to-next-to-leading order) and after the summation of the logarithmically enhanced effects to the order $\alpha_s^2$ \cite{Czakon:2006ss}. At present such a theoretical accuracy generally exceeds that of the available experimental data, however, due to a large statistics, Belle II has good chances to significantly improve the situation from the experimental side (see chapter \ref{chapter:measurements}).
In the given effective approach NP can manifest itself in two ways: through the modification of the Wilson coefficients $C_i$ and through the emergence of new operator structures $Q_i$ absent in the SM.

Unlike the electric charge, the charge of strong interactions (the colour) is not observed in the open state. The corresponding phenomenon is known as confinement of colour. In other words, the observed hadrons (that is, strongly interacting particles) should have wave functions that are singlets in colour. It is easy to see that the simplest hadrons are quark-antiquark mesons and three-quark baryons. However, the non-Abelian nature of the gauge group of strong interactions $SU(3)$ allows one to construct more complex colourless combinations, especially if the studied hadrons contain gluons. An example of such states is provided by hybrid mesons (or simply hybrids), containing not only a quark-antiquark pair but also gluons, or glueballs made of gluons alone. Hadrons with a more complex structure than a quark-antiquark meson or a three-quark baryon are conventionally referred to as exotic. These can also include tetra- and pentaquarks, hadroquarkonia and hadronic molecules, and so on. For a review of the experimental information and theoretical approaches to exotic hadrons containing heavy ($c$ and $b$) quarks see Ref.~\cite{Brambilla:2019esw}. Opportunities related to the Belle II experiment in searches and studies of exotic hadronic states are briefly discussed in chapter \ref{chapter:hadrons}.

The presence of hadronic states in the spectrum of strong interactions results in the fact that the energy dependence of the cross section of the electron-positron annihilation into quark-antiquark pairs can be very much different from a similar annihilation to leptons, in particular, muons. For this reason an important quantity for experimental studies is the so-called $R$ ratio defined as 
\be
R=\frac{\sum_q \sigma(e^+e^-\to q\bar{q})}{\sigma(e^+e^-\to\mu^+\mu^-)}=\sum_q R_q.
\label{ratioR}
\ee
In the absence of hadrons this ratio (up to the higher-order corrections in the coupling constant) should take the form of a step-like function with a constant ``jump'' every time the energy reaches the threshold of a next quark flavour and thus the sum in $q$ acquires a new term. However, the presence of multiquark states in the spectrum of strong interactions results in a quite nontrivial form of the ratio (\ref{ratioR}) as a function of energy. In chapter \ref{chapter:muonanom} possible studies of $R$ in the Belle II experiment are discussed in relation with the problem of measuring the anomalous magnetic moment of the muon. 

\section{Super $B$-factory Belle II}\label{chapter:bfactories}

\subsection{Experiments at electron-positron colliders}\label{chapter:history}

In the framework of the SM, electrons and positrons participate in electromagnetic and weak interactions, and the gauge constants of both theories are very small, so that processes with the smallest possible number of interactions have the largest probability. For this reason, at energies much less than the $Z$-boson mass ($m_Z=91.1876\pm 0.0021$ GeV \cite{Zyla:2020zbs}), the dominating process is the annihilation of the $e^+e^-$ pair into one virtual photon which can then turn to pairs of electrically charged particles: leptons or quarks. Moreover, since particles in the final state are massive, the energy of the colliding beams must exceed a certain critical value (threshold) for the corresponding reaction to take place. Processes with the formation of heavy leptons or hadrons are of primary interest for the tests of the SM and searching for NP. In the latter case, the process proceeds through the conversion of the virtual photon into a vector quarkonium containing a heavy $\bar{Q}Q$ pair and its subsequent decay to an open-flavour final state containing pairs of heavy-light mesons $(\bar{Q}q)+(\bar{q}Q)$, where $q$ denotes the light ($u$, $d$ or $s$) quark.

The first electron-positron collider was constructed at the Frascati National Laboratory (Italy) founded in 1955 and launched in 1959. At the very beginning the laboratory carried out experiments only with an electron beam but in 1960 the Austrian physicist Bruno Tuschek proposed to study collisions of electrons and positrons circulating in a storage ring in opposite directions. This is how the world's first $e^+e^-$ collider AdA (from Italian Anello di Accumulazione) with the beam energies of 250 MeV each appeared. Unfortunately, the luminosity (the number of collisions per unit time per unit of area) of this facility was insufficient to obtain real physical results; therefore, AdA only demonstrated the fundamental possibility of building a collider --- a facility with colliding beams. Independently works on
the construction of such colliders were carried out in No\-vo\-si\-birsk (USSR) under the leadership of G.I. Budker and at the SLAC (from Stanford Linear Accelerator Center) laboratory at Stanford (USA) under the leadership of B. Richter. In 1964, almost simultaneously, both in Novosibirsk and Stanford the first collisions of two electron beams were obtained and already in 1965 physical experiments began at the VEPP-2 electron-positron collider in Novosibirsk. It was realised very quickly that
electron-positron colliders provide a unique opportunity to produce and study the properties of hadrons in very clean conditions. Then since the 70's of the last century such experiments in various laboratories of the world have provided many results which have considerably improved our understanding of elementary particles and their interactions. Eventually physicists came to the idea to build colliders operating in a relatively small energy range but with the maximal possible luminosity --- the so-called factories. Currently there are several such facilities operating all over the world.

For example, the DA$\Phi$NE experimental machine launched in 2002 is still operating in the Frascati National Laboratory. It was created to work in the energy range of the vector $\phi$-meson with the mass of $1019.461\pm 0.016$ MeV \cite{Zyla:2020zbs} that gave the facility its name, the $\phi$ factory. The most famous (though not the only) KLOE experiment \cite{Adinolfi:2002uk,Adinolfi:2002zx} carried out at this facility studied the properties of the 
$\phi$ meson and $CP$ violation in the neutral $K$ mesons. Indeed, since the $\phi$ meson resides very close to and slightly above the $K\bar{K}$ threshold, it decays into a kaon-antikaon pair with a probability of more than 80\%. In 2010, after an upgrade of the accelerator complex, the KLOE-2 \cite{AmelinoCamelia:2010me} experiment started operation.

The VEPP-2000 collider at the Budker Institute of Nuclear Physics of the Siberian Branch of the Russian Academy of Sciences operating in Novosibirsk since 2010 with the energy range in the center-of-mass frame from 320 MeV to 2 GeV is a factory of light vector mesons such as the $\rho$, $\omega$, and $\phi$ mesons as well as their excitations. Two experiments are underway at VEPP-2000: CMD-3 \cite{Khazin:2008zz} and SND \cite{Achasov:2009zza}. The information obtained with them is very important for understanding the nonperturbative regime of strong interactions with the participation of the three light quarks --- $u$, $d$, and $s$.

The next range of energies important for experimental studies at electron-positron colliders are the energies in the region where vector states containing a $\bar{c}c$ pair can be created. The lightest representative of this family of hadrons is the $J/\psi$ meson with the mass $M_{J/\psi}=3096.900\pm 0.006 $~GeV \cite{Zyla:2020zbs} (see Fig.~\ref{fig:hadrons1}). At a slightly higher energy (approximately 3.55 GeV) creation of a lepton pair $\tau^+\tau^-$ becomes possible that also provides new possibilities for studying physics both within the framework of the SM and beyond --- physics of the $\tau$ lepton is discussed in detail in chapter~\ref{chapter:leptons}. For this reason the corresponding accelerators are usually referred to as charm-$\tau$(or $c$-$\tau$)-factories. The most modern operating factory of this type is located at the Institute of High Energy Physics in Beijing (China). An experiment called BES III (from the Beijing Spectrometer III) has been carried out since 2008 using the BEPC II (from the Beijing Electron-Positron Collider) accelerator. The diameter of the ring is 240 m, the energy of the colliding beams can vary from 2 to 4.63 GeV (the collider has recently been upgraded to reach as high maximum energy as 4.9 GeV) and the luminosity is 
$10^{33}$ sm$^{-2}$ sec$^{-1}$. BES III is a successor of the experiments of previous generations (BES  and BES II) as well as the BEPC II accelerator is an upgrade of the BEPC one, the construction of which began in 1984 and the upgrade was carried out in the period from 2004 to 2008. More information about this experiment can be found in Ref.~\cite{Ablikim:2009aa}. At the moment, several projects of further development of $c$-$\tau$-factories --- the construction of Super $c$-$\tau$-factories --- also in Russia on the basis of the Budker Institute of Nuclear Physics in Novosibirsk, are under active discussion. The proposed experimental facility is designed to operate in the energy range from 2 to 6 GeV and provide an unprecedented luminosity of 
$10^{35}$ cm$^{-2}$ s$^{-1}$, which is two orders of magnitude higher than the luminosity reached elsewhere in the world in this energy range \cite{Bondar:2013cja}.

Finally, the electron-positron colliders operating in the energy range of the production of vector bottomonia, that is, hadrons containing a $\bar{b}b$ pair, are commonly called $B$-factories. This name stems from the specific choice of the main operating energy of such a collider --- the experimental data are mainly collected at the summary beam energy near the vector resonance $\Upsilon(4S)$ with the mass $10.5794\pm 0.0012$ GeV \cite{Zyla:2020zbs}. Such a choice of the energy is not accidental since this state lies slightly above the threshold for the production of a pair of bottomed mesons $\BBb$ (see Fig.~\ref{fig:hadrons2}) and, therefore, with almost 100\% probability decays via this channel.

Several factors played their roles in the emergence and successful operation of the $B$-factories. First, the lifetime of the $b$ quark is long enough to allow it to bind with other quarks and form a hadronic state (this is no longer the case for the heaviest $t$ quark). As a result, the $B$-meson, the lightest hadron containing the $b$ quark, also has a rather long lifetime and therefore is available for experimental studies. Moreover, as the ARGUS experiment \cite{Albrecht:1988vy} found in 1987, $B^0$ and $\bar{B}^0$ mesons mix rather strongly \cite{Prentice:1987ap} that opened wide opportunities for studying $CP$ violation with their help. Thus, the $B$-factories were primarily constructed to experimentally measure the parameters of the CKM matrix and study physics associated with it. This was also facilitated by a rapid progress in the technology of the storage rings that made it possible to provide a luminosity sufficient to collect large statistics in a relatively short time --- for 30 years since the launch of the CESR collider in 1980 the number of $B$-mesons produced in $e^+e^-$ collisions has increased by five orders of magnitude! As a result, at the time of the shutdown of the experiments at the $B$-factories of the previous generation --- BaBar \cite{Aubert:2001tu} in 2008 and Belle \cite{Abashian:2000cg} in 2010 --- 530 fb$^{-1}$ and more than 1000 fb$^{-1}$ of data had been collected by them, respectively.

Simultaneous work of at least two competing experiments is an extremely efficient way of timely confirmation or refutation of the observations made in one of them. As mentioned above, until recently, the BaBar experiment at Stanford (USA) and Belle in Tsukuba (Japan) had been working and collecting data. It was expected that both experiments would continue as Super-$B$-factories. So the SuperB \cite{Bona:2007qt} experiment in Tor Vergata (Italy) and Belle II \cite{Abe:2010gxa} in the same place in Japan, were supposed to be the successors to BaBar and Belle, respectively. Unfortunately, because of the economic crisis in Europe, the SuperB project was discontinued in 2012 while Belle II was successfully launched in 2018.

A comprehensive description of the history and development of the previous generations of $B$-factories, their physical programs, and the results obtained by 2014 can be found in the review \cite{Bevan:2014iga}.

\subsection{Experimental facility Belle II}\label{chapter:hardware}

\subsubsection{General information}

In modern experiments on the physics of fundamental interactions two main approaches are used to search for NP. On the one hand, the energy of colliding beams is increased to search for new particles in a wider energy range. In this case a sensitivity to the direct production of new particles depends on the value of the corresponding cross section and the amount of the collected data. On the other hand, one can perform high-precision measurements of already known processes to detect possible deviations from the predictions of the SM. The undoubted world leaders in the first approach are the experiments at the LHC able to detect new particles produced in proton-proton collisions at the centre-of-mass energy up to 14 TeV. On the contrary, the Belle II experiment demonstrates leadership in the second approach --- its main task, as the experiment at a next-generation $B$-factory, is to search for NP by measuring suppressed processes and increase the accuracy of determination of various physical quantities. If discrepancies are found with the predictions of the SM, they can be interpreted from the point of view of various models for NP. In the future new experiments based on the electron-positron colliders can make a significant contribution to such studies. In particular, the CEPC (Circular Electron-Positron Collider) with the centre-of-mass energy up to 240 GeV \cite{cepc} is planned to start operating in China in the 2030's, and a further development of the experiments at CERN under the name FCC (Future Circular Collider) \cite{fcc} is under active discussion. Various possibilities of constructing a linear electron-positron collider are also discussed (see, for example, Ref.~\cite{Behnke:2013xla} and references therein) with a possible energy reach up to 1 TeV. However, at the moment even the location of such an accelerator has not been finally agreed yet.

\begin{figure}[t!]
\begin{center}
\epsfig{file=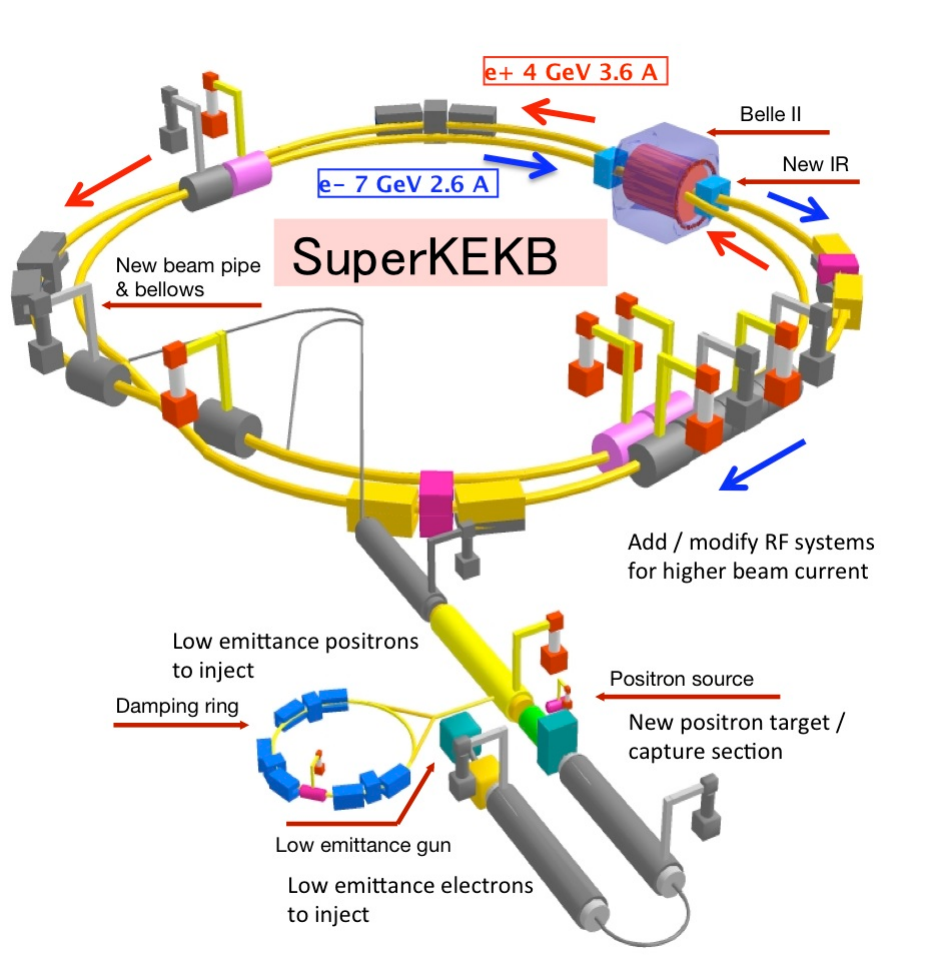, width=0.6\textwidth}
\end{center}
\caption{The design of the Super $B$-factory used in the Belle II experiment. Adapted from the site of the collaboration.}
\label{fig:belle2}
\end{figure}

To meet the goal, the Belle II experiment needs an $e^+ e^-$ collider with asymmetric (unequal energy) beams with a record-breaking luminosity (which exceeds 40 times that of the ac\-ce\-le\-ra\-tor used in the previous-generation experiment Belle) and a detector providing
\begin{itemize}
\item efficient charged tracks reconstruction;
\item high spatial and energy resolution of the photons;
\item identification of charged tracks and clusters of neutral particles;
\item precision measurement of the position in space for charged tracks, 
\end{itemize}
and possessing an efficient trigger and a fast data recording system. 

In Fig.~\ref{fig:belle2} we show the scheme of the experimental facility SuperKEKB employed in the Belle II experiment. The key elements of the design are the storage rings for electrons and positrons intercepting at a single point where the detector is mounted. In chapters~\ref{chapter:superkekb} and \ref{chapter:belle2} the elements of the SuperKEKB collider and Belle II detector are discussed in more detail.

\subsubsection{Accelerator SuperKEKB}\label{chapter:superkekb}

The SuperKEKB experimental complex is the result of a serious upgrade of the previous KEKB project aimed at a significant increase (approximately 40 times) of the instantaneous luminosity. The facility is located at the KEK Research Center (Institute for High Energy Physics) in Tsukuba and is designed to collide beams of electrons and positrons at the centre-of-mass energy in the region of the vector $\Upsilon$-resonances (see chapter \ref{chapter:history}). Typically for a $B$-factory most of the data will be collected at the energy of the bottomonium $\Upsilon(4S)$ which decays into a pair of  $B\bar{B}$ mesons. Such a centre-of-mass energy guarantees collecting very clean $B\bar{B}$ pairs in a quantum-correlated state with the quantum numbers $J^{PC}=1^{--}$. In contrast to hadron-hadron collisions (for example, the proton-proton experiments at the LHC), collisions of the electron and positron beams are characterised by a very low background level that makes possible an effective reconstruction of the final states containing photons from the $\pi^0$ decays as well as the $\rho^{\pm}$, $\eta$, $\eta'$, and $K^0_{\rm L}$ mesons. This property of the Belle II experiment makes it a unique tool for precision tests of the SM.

As already explained in chapter~\ref{chapter:history}, the luminosity of an accelerator is determined by the number of particle collisions per unit time and the transverse size of the particle bunches. The maximum number of particles in a bunch is limited by the so-called collision effect: for a certain charge of the bunch, the influence of the colliding beam becomes so strong that it destabilises the motion of the particles in the storage ring. At the same time, an increase in the number of bunches in the storage ring is also associated with certain technical problems (the need to compensate for the energy losses for synchrotron radiation, to cool the vacuum chamber, to suppress the collective instability of the beams) and therefore is only possible up to a certain limit --- this limit was reached already at KEKB, a previous-generation facility. Thus, since the planned luminosity of the SuperKEKB accelerator is 40 times higher than the luminosity of the KEKB one, a significant upgrade of the entire accelerator complex was necessary \cite{Ohnishi:2013fma}. The key changes which resulted in a radical increase in luminosity are a decrease in the transverse size of the beam at the collision point by about 20 times (from 1 $\mu$m to 50 nm) and a twofold increase in the currents compared to KEKB. This strategy is known as the ``nanobeam'' scheme proposed by P. Raimondi for the design of the Italian SuperB-factory \cite{Bona:2007qt}.
In addition, at SuperKEKB the charged beams collide at a larger angle of 83 mrad as compared to 22 mrad at KEKB. This reduces the number of spurious collisions in the detector and eliminates the need of separating magnets which occupy a useful volume of the detector.

The accelerator is asymmetric, that is, it is constructed with unequal energies of the electron and positron beams to guarantee a boost of the centre of mass of the produced particles in the laboratory frame (this makes an important difference between Belle and Belle II and the experiments at $c$-$\tau$ factories, for example, BES III --- see chapter \ref{chapter:history}). Thus, $B$ and $D$ mesons produced in collisions  have time to travel a large distance in the detector before they decay, thus allowing precise measurements of their lifetime as well as parameters of mixing and $CP$ violation. A slightly lower asymmetry of the beam energies at SuperKEKB as compared to the KEKB (7 and 4 GeV instead of 8 and 3.5 GeV for electrons and positrons, respectively) was chosen to reduce losses because of the Touschek scattering (intrabeam rescattering of particles which can result in their leaving the storage ring) in low-energy beams. Although somewhat reducing the spatial separation of the $B$-mesons, this is advantageous in the solid angle available for studying decays with partially reconstructed final states (missing energy decays). For example, in the analysis of the processes with neutrinos in the final state which require a high hermeticity of the detector.

\subsubsection{Belle II Detector}\label{chapter:belle2}

The Belle II Detector is mounted at the interaction point of the electron and positron beams. Compared to the Belle experiment the new detector should work at a 40-time increase of the luminosity of the accelerator and, therefore, should be able to detect events 40 times faster than before and at a 10-20 times larger counting rate  from the background processes. In particular, the requirements to the new detector can be summarised as follows: 
\begin{itemize}
\item perfect resolution in reconstructing the decay vertices (about 50 $\mu$m); 
\item high efficiency of the charged track reconstruction in a wide momentum range (up to hundreds of MeV);
\item improved reconstruction of the charged tracks with the momenta below 50 MeV;
\item good momentum resolution in the entire studied energy range (up to 8 GeV);
\item precise measurement of the energy and direction of motion for photons with the energy from tens MeV to 8 GeV and an efficient detection of photons with the energy above 30 MeV,
\item high-efficiency identification system for discriminating between pions, kaons, protons, electrons, and muons in the entire kinematical range of the experiment;
\item (nearly) full solid angle covered by the detector;
\item fast and efficient trigger;
\item data acquisition system capable of recording a large amount of information.
\end{itemize}

\begin{figure}[t!]
\centerline{\epsfig{file=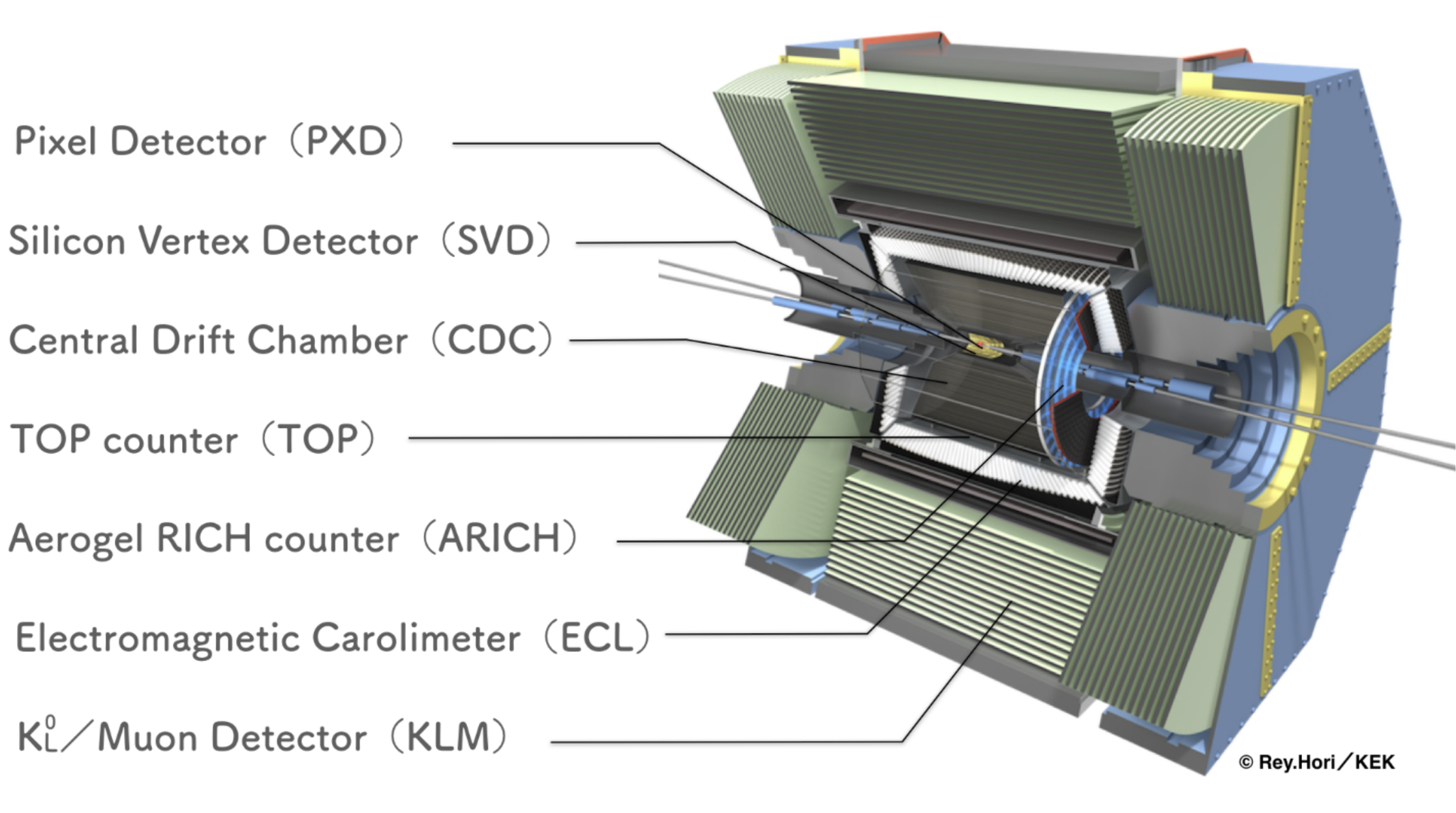, width=0.99\textwidth}} 
\caption{The scheme of the Belle II detector. Adapted from the site of the collaboration.}
\label{fig:detector}
\end{figure}

A schematic picture of the Belle II detector is shown in Fig.~\ref{fig:detector}. Various elements of the detector are located spherically symmetrically around the beam interaction point in the magnetic field of 1.5 T. Although the new detector is very similar to its predecessor and inherited the same superconducting magnet, some of its components are either brand new or highly upgraded~\cite{Abe:2010gxa}. 

The main components of the detector are the vertex detector (VXD), drift chamber (CDC), particle identification system consisting of two subsystems (a time-of-flight system (TOP) and an aerogel-based Cherenkov ring detector (ARICH)), electromagnetic calorimeter (ECL), $K^0_{\rm L}$ and muon detector (KLM), trigger, and data acquisition system (DAQ). Below we provide a brief description of each subsystem separately.

\vspace{0.3cm}\noindent
$\bullet$ {\it Vertex detector}
\vspace{0.3cm}

\noindent The silicon vertex detector allows to reconstruct the spatial position of particle tracks near the interaction point of electron and positron beams. The new vertex detector consists of two subsystems: a silicon pixel detector (PXD) and a silicon vertex detector (SVD) which together form six silicon layers wrapping a beryllium vacuum tube. Pixel and double-sided silicon sensors are used for the first two layers of the PXD and the remaining four layers, related to SVD, respectively. Compared to the vertex detector used in Belle, the beryllium tube and the first two detector layers (pixel detector) are shifted closer to the interaction point, and the outermost layer has now a much larger radius (144 mm at Belle II instead of 88 mm at Belle). As a result, a significant improvement in the determination of the interaction vertex position is anticipated as well as an increase in the reconstruction efficiency of the $K^0_{\rm S} \to \pi^+ \pi^-$ decays with signals in the vertex detector.

\vspace{0.3cm}\noindent
$\bullet$ {\it Drift chamber}
\vspace{0.3cm}

\noindent The central drift chamber (CDC) has three important functions. First, charged tracks are restored in it and their momenta are measured with high accuracy. Second, the CDC provides information for identifying particles with the momenta up to about 1 GeV using measurements of ionisation losses in the gas volume of the chamber. Low-momentum tracks that do not reach the particle identification system can be identified using the CDC only. Finally, the drift chamber provides efficient and reliable signals to start the first-level trigger. Compared to the drift chamber used in the Belle experiment, the new one has a larger radius (1130 mm instead of 880 mm). Besides that, the modern chamber has smaller drift cells than the one used at Belle to be able to handle high event rates and higher background levels. The CDC, described in detail in Ref.~\cite{Abe:2010gxa}, includes 14336 sensitive wires grouped in 56 layers with an axial (along the magnetic field) or stereo (at small angles to the detector axis) orientation. Combining the information from the axial and stereo layers one can reconstruct the three-dimensional position of the spiral particle track. The working gas of the drift chamber, as at Belle, is a mixture of helium and ethane in equal shares. The particular light gas was chosen to reduce multiple scattering of the charged tracks in the chamber. Unlike argon mixtures, this gas has a smaller photoelectron interaction cross section that helps in reducing the effect of the synchrotron radiation on the detector.

\vspace{0.3cm}\noindent
$\bullet$ {\it Particle identification system}
\vspace{0.3cm}

\noindent The time-of-propagation system (TOP) used to identify charged particles moving at large angles to the direction of the beams  \cite{Akatsu:1999hi,Staric:2008zz} is cylindrically symmetrical around the axis of the beams. It is a particular type of the Cherenkov detector in which the information about the Cherenkov ring is extracted from the time of the arrival and the place of the interaction of the Cherenkov photons at the photodetector mounted at one end of a 2.6-m-long quartz rod. The system contains 16 modules, each consisting of a 45-cm-wide and 2-cm-thick quartz rod  with a small wedge (about 10 cm long) at the end. Such a wedge provides an additional point rendering, slightly relaxes the requirements for the synchronisation accuracy, and reduces the occupancy of the photodetector \cite{Staric:2008zz}. Two rows of fast multi-anode photon detectors (16 in total) are installed at the exit window of the wedge.

To identify charged particles moving at small angles to the direction of the beams, an aerogel-based ring Cherenkov detector (ARICH) is used in the front-end region. Such a detector should be capable of detecting low-energy pions and discriminating quite well between the pions and kaons in the momentum range from 0.4 to 4 GeV. The choice of an aerogel as the working medium is dictated by its extremely low (in contrast to other solid materials) refractive index $n$. Since the Cherenkov light is emitted if the velocities of the particles passing through the aerogel exceed the threshold $v>c/n$, for the identification of fast particles $n$ should be close to unity. A key feature of the ARICH detector is the number of the Cherenkov photons which is increased with the help of a new method: to improve focusing, two aerogel layers of the same thickness but with different refractive indices are used. Choosing suitable refractive indices ($n=1.045$ and $n=1.055$ are actually used \cite{Iijima:2005qu,Krizan:2006pc}) one can reduce the contribution of
the uncertainty in determination of the emission point to the overall resolution of the Cherenkov angle of a charged particle. 

A hybrid avalanche photon detector (HAPD) designed in collaboration with Hamamatsu \cite{Nishida:2008zz,Nishida:2014gra} is used as a high-grain single-photon sensor. In such a 73 mm$\times$73 mm unit photoelectrons are accelerated by a potential difference of 8 kV and get detected in the avalanche photodiodes (APD).

\vspace{0.3cm}\noindent
$\bullet$ {\it Electromagnetic calorimeter}
\vspace{0.3cm}

\noindent The electromagnetic calorimeter (ECL) is designed to measure the energy and momentum direction of photons and for electron identification, namely discriminating them from hadrons (in particular, pions) by comparing the cluster energy with the momentum of the corresponding charged
track in the drift chamber. The calorimeter is assembled from 8736 crystal modules (cesium iodide doped with thallium, CsI-Tl, is used) with a typical size of 30 cm$\times$5.5 cm$\times$5.5 cm. The module length (30 cm) corresponds to about 16 radiation lengths. Signals are read off by the silicon
photodiodes (two per each crystal) with a sensitive surface of the size 2 cm$\times$1 cm. The calorimeter covers 90\% of the full solid angle
in the centre-of-mass system. The Belle II experiment uses the same CsI-Tl crystals, preamplifiers, and reference structures previously used by Belle, while the readout electronics and the reconstruction software were completely upgraded.

\vspace{0.3cm}\noindent
$\bullet$ {\it Detector of muons and $K_{\rm L}$ mesons}
\vspace{0.3cm}

\noindent The muon system (KLM, for $K$-long and muons) consisting of 4.7-cm-thick iron plates alternated with the active elements of the detector is located outside the superconducting solenoid \cite{Aushev:2014spa}. The iron plates screen the magnetic field around the detector and at the same time allow to reconstruct the direction of the $K^0_{\rm L}$ mesons which cause hadron showers in iron.

The muon system of the Belle experiment which used resistive plate chambers (RPC) as active elements demonstrated good performance over the entire data taking time. However, at Belle II, as compared to Belle, particular areas of the detector (the ends and inner cylindrical layers) are expected to experience large background loads because of the neutrons formed mainly in the electromagnetic showers caused by background reactions (for example, radiative Bhabha scattering). Long RPC ``dead time'' during the electric field recovery after a strike significantly reduces the reconstruction efficiency at high background counting rate. The resulting misidentification of muons at the end and two inner cylindrical layers of the spectrometer could be so high that such a detector would become useless \cite{Abe:2010gxa}. To solve this problem, the RPC's were replaced by the layers of wavelength-shifting fiber scintillation strips the information from which is read off by the silicon photomultipliers (SiPM) as light detectors \cite{Balagura:2005gh}.

\vspace{0.3cm}\noindent
$\bullet$ {\it Trigger}
\vspace{0.3cm}
 
\noindent The Belle II trigger system plays an important role in the selection of events during data taking. Since in the Belle II experiment there will be much more different tasks in the physical analysis requiring special triggers as compared to Belle, all such triggers should work efficiently at much higher background counting rates expected from the SuperKEKB and in addition meet the requirements of the data acquisition system (DAQ). A well-designed trigger system guarantees a wide range of feasible physical tasks that were unavailable at the previous-generation $B$-factories. At Belle II (like at Belle) the efficiency of the trigger for most decays of  $B$-mesons is close to 100\% for the events reconstructed by stand-alone algorithms. However, in addition to the $B$ physics, Belle II is planned to study processes with the topology similar to the background (for example, $\tau$-lepton or two-photon physics).
In order to meet the requirements of the new experiment, the trigger system previously used by Belle was upgraded and supplied with online algorithms. A detailed description of the trigger and data acquisition system of the Belle II experiment can be found in Ref.~\cite{Abe:2010gxa}.

\section{Precision measurements and searches for NP in the decays of $B$ me\-sons}\label{chapter:measurements}
 
\subsection{Unitarity triangle}\label{chapter:triangle}

An important question to be addressed by the Belle II experiment is whether there are new complex phases in the quark sector in addition to those 
following from the CKM mixing mechanism. To answer this question one needs first to consider the full picture of $CP$ violation within the framework
of the CKM mechanism, that gives an idea of the phases of the CKM matrix, and then search for distortions of this picture provided by possible
contributions from NP. A very illustrative picture, easy for understanding and experimental tests, is provided by the so-called unitarity triangle. A
detailed description of this triangle is given in the review \cite{Bondar:2007zz}, so here we only briefly repeat the main ideas concerning its origin
and methods for measuring its angles and sides; then we dwell in detail on the accuracy achieved by the $B$-factory since the work
\cite{Bondar:2007zz} was published and what new results in this field can be expected from the new Belle II experiment.

\begin{figure}[t!]
\centerline{\epsfig{file=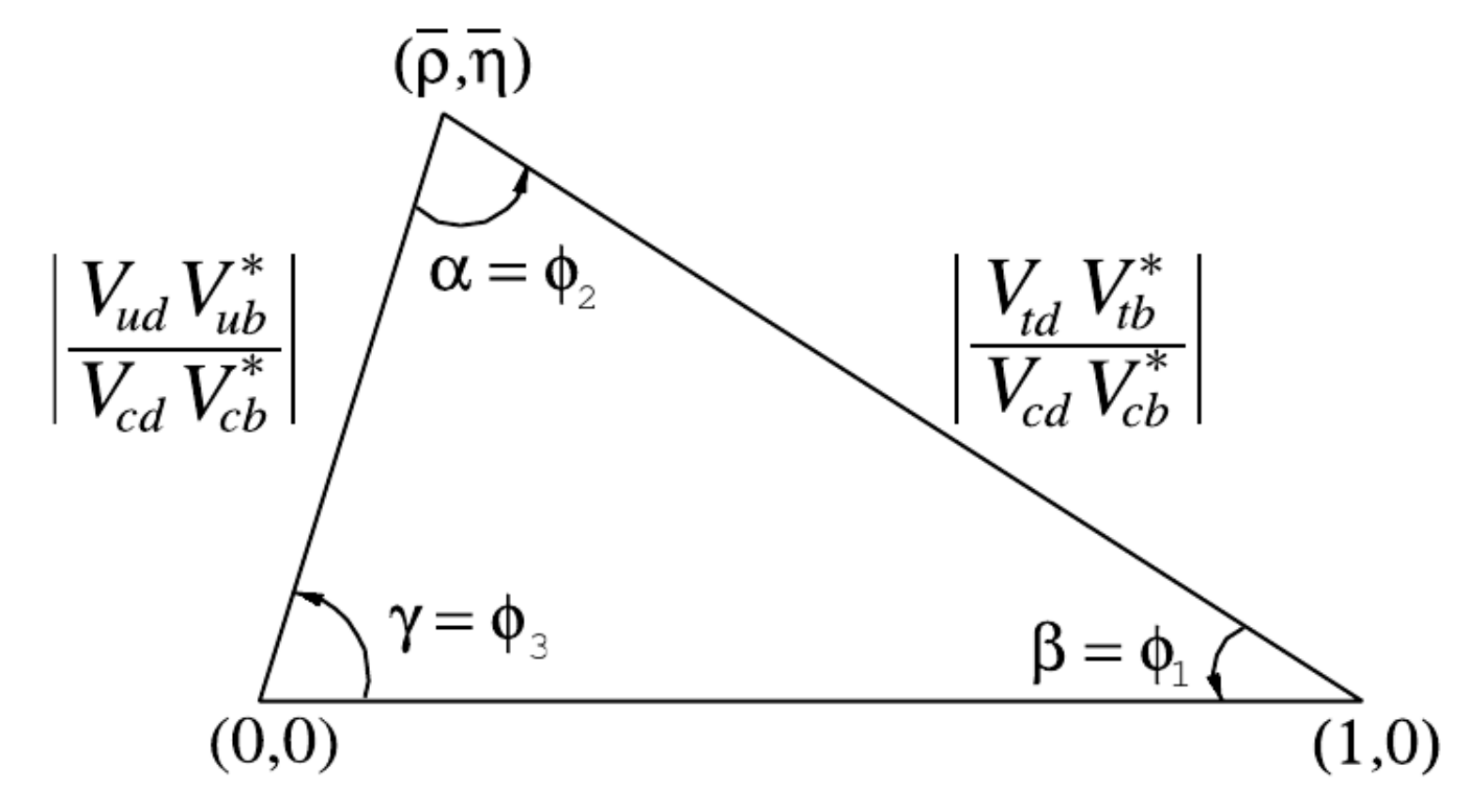, width=0.5\textwidth}} 
\caption{A schematic picture of the unitarity triangle defined in Eq.~(\ref{triangle}) after its base normalisation to 1. Adapted from the PDG review \cite{Zyla:2020zbs}.}
\label{fig:triangle0}
\end{figure}

The unitarity condition for the CKM matrix (see the definition (\ref{CKMmatrix})) $VV^\dagger=1$ gives rise to nine conditions imposed on its elements, $\sum_i V_{ij}^* V_{ik} = \delta_{jk}$, each being a sum, equal either to zero or one, of three complex numbers naturally visualised as vectors in a complex plane. Then three conditions (for $j=k$) correspond to squares and six (for $j\ne k$) --- to triangles. All squares and four of the six triangles are of little use for experimental verification because of their very different sides: in fact, they all nearly degenerate into line segments and, therefore, allow one to check something only if the measurements are extremely accurate. Meanwhile the two remaining triangles look really like triangles of a general form from the school textbooks in geometry with all sides unequal to each other but with the length of the same order, and with no corners collapsed to 0. Consider one of them ($j=u$, $k=b$),
\be
V_{ud}V_{ub}^*+V_{cd}V_{cb}^*+V_{td}V_{tb}^*=0,
\label{triangle}
\ee
which is closely related to the decays of the \Bn and $B^+$ mesons. It is convenient to divide relation (\ref{triangle}) by $\left|V_{cd}V_{cb}^*\right|$, so that the length of one side of the triangle turns to 1 while the coordinates of its opposite apex are denoted as $(\bar{\rho},\bar{\eta})$ --- see a graphical representation of the unitarity triangle in Fig.~\ref{fig:triangle0}. 

Measurements of various parameters in the $B$-meson decays allow to calculate the sides and angles of the unitarity triangle independently of each other. If such decays of the $B$-mesons do not contain contributions from NP, then all the relations known from the school geometry course, such as the cosine theorem or that about the sum of the angles of a triangle, will hold for the unitarity triangle. This verification was an important task for the $B$-factories which have already stopped data taking and still remains a major task for the new Belle II experiment. Since the length of the base of the triangle is normalised to unity, measurements of any other two elements (angles or sides) uniquely determine the entire triangle. In this case, each next measurement provides a check of the Euclidean nature of the complex plane of the quarks' mixing constants that is tantamount to checking the validity of the SM since NP would lead to a deformation of this plane.

\subsubsection{Angles of the unitarity triangle}\label{chapter:angles}

In this chapter we discuss measurements of the angles of the unitarity triangle. In the Wolfenstein parametrisation (\ref{wolfeinstein}), the angles of the unitarity triangle are the phases of the two elements of the CKM matrix ($V_{ub} = \left|V_{ub} \right| e^{i\gamma}$, $V_{td} = \left| V_{td} \right| e^{i\beta}$) while all other elements are real. To measure a particular angle one needs to pick up such decays of the $B$-mesons which receive contributions from two diagrams, with the difference of the phases being precisely this angle, and measure $CP$ violation in these decays. For such measurements one can use the indirect $CP$ violation through the oscillations of the neutral $B$-mesons in which the second diagram containing \BBb oscillations acquires an additional phase from them ($\beta$ in the case of \BBd\ and 0 in the case of \BBs). A detailed derivation of the relation between the angles of the unitarity triangle and the $CP$-violation parameter at the example of the angle $\beta$ is given in the review \cite{Bondar:2007zz}. For simplicity, in what follows we will refer to the amplitude of the first diagram as the amplitude of the process while the additional phase acquired by the second diagram will be called the phase of oscillations.

The original idea was based on using the indirect $CP$ violation that allows one to avoid theoretical uncertainties related to hadronic corrections (strong phases and decay form factors not calculated in QCD). The following measurements were suggested for these purposes: 
\begin{itemize}
\item angle $\beta$ in the transition $b\to c\bar{c}s$, for example, in the decay $B^0 \to \jpks$ with a real amplitude and the phase of oscillations $\beta$.
\item angle $\alpha$ в in the transition $b\to u \bar{u} d$, for example, in the decay $B^0 \to \pi^+ \pi^-$ with the decay amplitude phase $\gamma$ and the oscillation phase $\beta$. This way the angle $180^\circ -\gamma-\beta$ is measured that equals the angle $\alpha$ in the SM. 
\item angle $\gamma$ in the transition $b \to u \bar{c} s$, for example, in the decay $\Bs \to D_s^+ K^-$ with the phase of the decay amplitude $\gamma$ and a vanishing phase of the \BBs oscillations. 
\end{itemize}

Such measurements would indeed be free of theoretical uncertainties but for the con\-tri\-bu\-ti\-on from penguin diagrams that can distort the measurement strongly and in an uncontrollable way as pointed out in Refs.~\cite{London:1989ph,Gronau:1989ia}. It will be demonstrated below how, for each of the angles, one can avoid model errors stemming from the impossibility to accurately calculate the amplitudes incorporating strong interactions of hadrons. In addition to the theoretical uncertainties, it is also important to discuss a way to control experimental errors. Thus, when assessing the accuracy achieved in the Belle II experiment one should take into account that the systematic error does not always improve proportionally to the integrated luminosity. Part of it is determined on control data samples which, like the signal samples, grow with the increase in the data statistics but sometimes systematic errors may occur which can not be reduced. Fortunately, the latter are almost never met for the measurements discussed below.

\vspace{0.3cm}\noindent
$\bullet$ {\it The angle $\beta$}
\vspace{0.3cm}

The angle $\beta$ is so far the most precisely measured parameter of the unitarity triangle. On the one hand, this measurement makes a very important contribution to the global approximation of the quark mixing parameters fixing the remaining parameters under the assumption of the validity of the SM with a better accuracy than can be achieved in direct measurements. On the other hand, verification of the self-consistency of the unitarity triangle requires at least two more measurements of the same accuracy (which, as will be discussed later in this chapter, unfortunately, are not yet available). Nevertheless, it is convenient to start the construction of the triangle from this angle extracted from the time-dependent (indirect) $CP$-violation parameter in the $\Bn\to\jp\Kn$ decay, which gives the relative phase of the \BBd mixing and the decay amplitude. The lifetime-dependent asymmetry generated in this case is
\be
A_{CP} (t) \equiv \frac{N(\Bn(t)\to f) - N(\Bb(t)\to f)}{N(\Bn(t)\to f) + N(\Bb(t)\to f)}
= S_{f} \sin(\Delta m_d t) + A_{f} \cos(\Delta m_d t),
\label{ASdef}
\ee
where $S_{f}$ and $A_{f}$ are the parameters of the indirect and direct $CP$ violation in the decay channel $B^0\to f$, respectively; $\Delta m_d$ is the mass difference for the two mass eigenstates of the \Bd mesons.

As demonstrated in the review \cite{Bondar:2007zz}, the magnitude of the indirect $CP$ violation $S_{\jpks}$ turns out to be, with good accuracy, equal to $\sin{2\beta}$. A small correction to this equality coming from an additional contribution to this decay from the penguin diagram ($S_{\jpks} = \sin{2\beta} +\Delta S_{\jpks} = \sin{2(\beta + \delta_{\jpks})}$) is still much smaller than the present-day experimental error because the penguin contribution has almost the same phase as the tree diagram. Modern estimates of $\delta_{\jpks}$ give a hope that even with a further increase in the ex\-pe\-ri\-men\-tal accuracy by about a factor of 5 the penguin contribution will not introduce too large uncertainty. Meanwhile this leaves the open question how reliable the existing theoretical estimates are.

An important parameter governing the phase of the penguin contribution is the direct $CP$-violation parameter $A_{\jpks}$ which should vanish if the penguin contribution does not bring any additional phase. The current value of $A_{\jpks}$ obtained in the Belle experiment is indeed consistent with zero ($0.015\pm 0.021 \pm 0.045$~\cite{Adachi:2012et}), and the error should decrease 4 times as a result of the Belle II work.

In the Belle II experiment, the systematic uncertainty is mainly controlled on the data and remains smaller than the statistical one. For this reason, the
accuracy of the angle $\beta$ measurements will improve proportionally to the statistics collected in the experiment. It should be noted that
experiments at the LHC, in particular the LHCb, will achieve perhaps even better accuracy in measurements of $\sin 2\beta$. Meanwhile Belle II will
make its unique contribution to the elimination of the ambiguity of extracting the value of this angle from the measured sine of the double angle by
measuring with a high accuracy also $\cos 2\beta$ from the parameter of the $CP$ asymmetry in decays $\Bn\to\Dn h^0$, where $h^0$ is a neutral light
hadron, $\pi^0$, $\eta$, $\omega$, and so on. The first such measurement has already been done in the joint work by Belle and
BaBar~\cite{Adachi:2018itz} where for the first time a statistically significant $CP$ violation was found and it was shown that $\cos 2\beta>0$. In
the future,  to check the self-consistency of the SM, it will be possible to confront the values of $\sin{2\beta}$ and $\cos 2\beta$ as independent measurements of the parameter $\beta$ in
two processes with different tree diagrams.

\vspace{0.3cm}\noindent
$\bullet$ {\it The angle $\alpha$}
\vspace{0.3cm}

\begin{figure}[t!]
\centerline{\epsfig{file=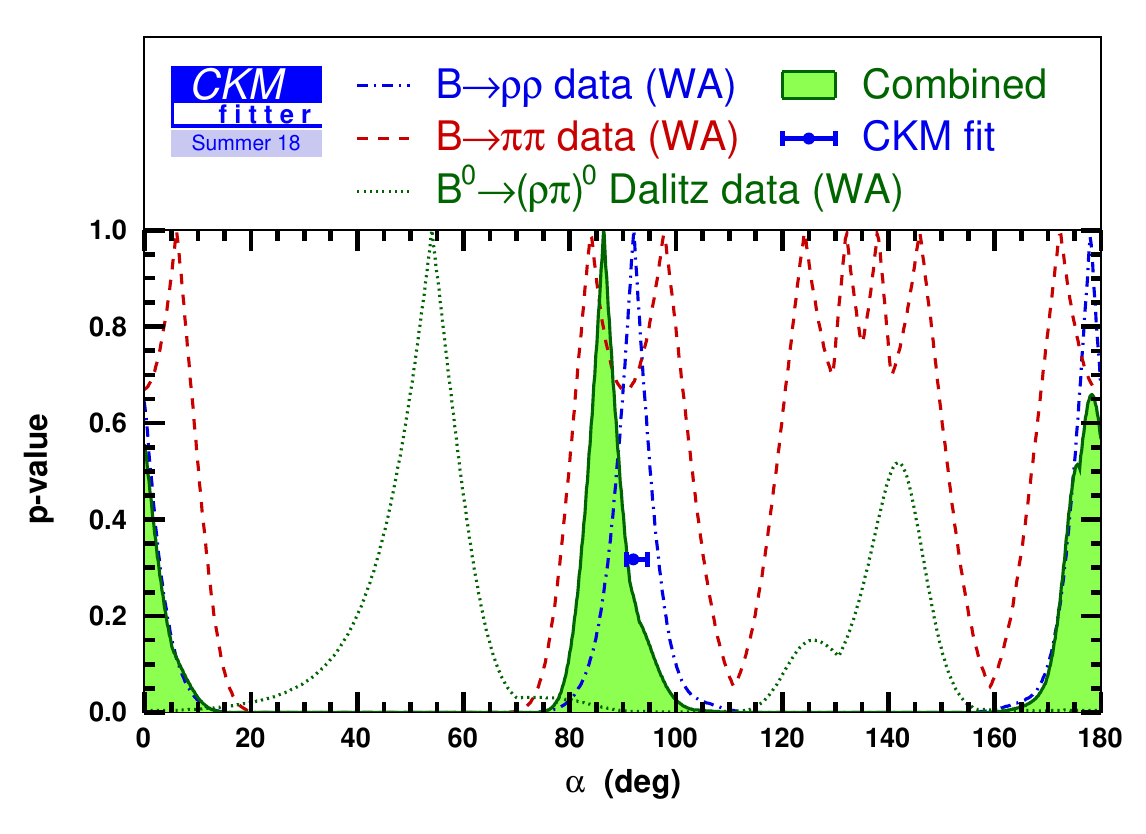, width=0.8\textwidth}} 
\caption{Scan of the confidence level for the values of the angle $\alpha$ obtained from the isospin analysis of the $b \to u$ decays and the allowed intervals for the values of $\alpha$. Dashed, dotted, and dash-dotted lines correspond to the constraints from the analysis of the $B\to \pi\pi$,  $B\to\rho\rho$ and $B\to \rho\pi$ decays, respectively. Adapted from the site of the CKMfitter Collaboration.}
\label{fig:alpha}
\end{figure}

Measuring the angle $\alpha$ turned out to be much more difficult. The parameter of the indirect $CP$ asymmetry in the decay $\Bn\to 
\pi^+\pi^-$, $S_{+-}$ (see Eq.~(\ref{ASdef}); the subscripts correspond to the charges of two $\pi$ mesons in the final state), would equal $\sin {2\alpha}$ if the penguin amplitude with an additional phase associated with the angle $\beta$ did not mix with
the tree-level decay diagram. However, it was found experimentally that this admixture is substantial, as follows from a large value of the direct
$CP$ violation and a high probability of the decay $\Bn \to \pi^0 \pi^0$. Extracting the angle $\alpha$ from a large number of possible
measurements of the decay $B\to \pi \pi$ required a lot of efforts. The underlying idea is based on the difference in the isotopic structures of the
tree-level and penguin transitions: the former may result in the isospin change by $3/2$ and $1/2$ while only $1/2$ is possible for the latter. It
was shown in Ref.~\cite{Gronau:1990ka} that six decay amplitudes ($\Bn \to \pi^+ \pi^-$, $\Bn \to \pi^0\pi^0$, $B^+ \to \pi^+ \pi^0$ and their
charge-conjugate counterparts) are subject to isospin relations, and a recipe is provided how to extract the angle $\alpha$ from the analysis of
such relations. The full isospin analysis allows one to calculate this angle with an eightfold ambiguity but relatively small model uncertainty which
does not cause troubles for the measurements. The method is described in detail in the review \cite{Bondar:2007zz}, so here we only discuss the
results achieved by the two $B$-factories, Belle and BaBar, for the ten years of work. The input variables of the method are $B_{+-}$, $B_{+0}$,
$B_{00}$, $S_{+-}$, $A_{+-}$, $A_{00}$, where the quantities $B$ denote the averaged probabilities of the decay itself and its $CP$-conjugate
counterpart, $S$ and $A$ are the values of the parameter of the indirect and direct $CP$ violation, respectively (see formula (\ref{ASdef})). The list of variables does not contain $S_{00}$ which is responsible for
the indirect $CP$ violation in the $\Bn \to \pi^0 \pi^0$ decay. Its measurement is impossible in the existing experiments since the restoration of the decay vertex for a completely neutral final state is very difficult. The inclusion of this variable should play an important role
for the future measurement on the Belle II data. The present-day world average values of all parameters are given in Table~2 of
Ref.~\cite{Gronau:2016idx} where, in addition, solutions for the angle $\alpha$ are provided as obtained with the help of the Monte Carlo code which
generates the above six observables under the assumption that they obey Gaussian distributions and evaluating $\chi^2$ for different values of the
studied angle. The minimum value of $\chi^2$ is reached for four values of $\alpha$: $95.0^\circ$, $128.9^\circ$, $141.1^\circ$, $175^\circ$ (see
Fig.~\ref{fig:alpha}), although initially the method assumed an eightfold ambiguity in the solution. This is a result of some ``luck'' since one of
the phases in the analysis turned out to be close to zero thereby having partially reduced the ambiguity.

It is noteworthy that one can improve the precision of $\alpha$ only by reducing the uncertainty of all six variables while reducing it for just one of them has almost no effect on the result. This is where the contribution of the Belle II experiment will be decisive. Although the experiments at the LHC can provide very accurate measurements of the parameters of the charged mode $\Bn \to \pi^+ \pi^-$ and, possibly, will do it even better than Belle II, the modes with $\pi^0$ are practically unattainable for study in the experiments at the LHC because of a huge neutral background and impossibility to associate the $\pi^0$ with a particular decay vertex of the $B$-meson. Moreover, Belle II is able to measure $S_{00}$ using conversion photons. Here weighting of the detector (generally speaking, undesirable) played a positive role since, in order to recover tracks at high counting rates, more layers of the vertex detector were required that increased the amount of the material and, consequently, the probability of the photon conversion. The additional variable $S_{00}$, although not so accurately measured, will discriminate solutions near $\alpha=129^\circ$ and $141^\circ$ giving $S_{00}\simeq -0.70$, from the solutions near $\alpha=95^\circ$, or $175^\circ$ where $S_{00}$ changes the sign and equals approximately 0.67. The expected accuracy of measuring the main parameters for extracting $\alpha$ in the $B\to\pi\pi$ decays on the Belle II data will be improved by 3-10 times, where a relatively modest threefold improvement is caused by an unavoidable systematic error. Thus, in the Belle II experiment, the variable $S_{00}$ will be used for the first time and measured with an accuracy of $\pm 0.28(\text{stat}) \pm0.03(\text{sys})$~\cite{Kou:2018nap}. 

For measuring the angle $\alpha$ several more final states due to the $b\to u$ transitions are suitable such as $B\to\rho\rho$ and $B\to\pi\rho$. They also contain penguin contributions, so everything said about the difficult isospin analysis to extract the value of $\alpha$ remains valid here, too. Without going into detail we present only the results obtained from these modes, $\alpha= (86.4^{+4.5}_{-4.3})^\circ$ and the second solution $(-1.8^{+4.3}_{-5.1})^\circ$ (see Fig.~\ref{fig:alpha}). As can be seen from the figure, in spite of more complicated final states, the accuracy of the measurement sometimes exceeds that achieved in the $B\to\pi\pi$ decays and the ambiguities in the solution do not always overlap with those in $B\to\pi\pi$. In this case, the better accuracy and lower ambiguity in the $\rho\rho$ and $\pi\rho$ modes result from the fact that, due to accidental reasons, the penguin contributions in these final states are smaller than in the $\pi\pi$ mode.

In conclusion we emphasise once again that all final states, including those containing neutral pions, are extremely important to further improve the accuracy of measuring the angle $\alpha$ that means that Belle II, with its ability to cleanly restore them, is absolutely necessary. The accuracy of the  determination of the angle $\alpha$, expected upon reaching the planned integrated luminosity at Belle II, will be approximately $0.6^\circ$~\cite{Kou:2018nap}, and the ambiguity of the solution will be completely eliminated by using all possible decay modes and the new variable 
$S_{00}$.

\vspace{0.3cm}\noindent
$\bullet$ {\it The angle $\gamma$}
\vspace{0.3cm}

To measure the last angle of the unitarity triangle, $\gamma$, the decay $B^+\to D^0 K^+$ without penguin corrections was picked up, so the method is theoretically clean. Recall that to measure an angle, it is necessary to consider the interference of two diagrams with different CKM phases (and different strong phases) that can be gained only by ``entangling'' \Dn\ and \Db\ in the final state (see Fig.~5 of the review \cite{Bondar:2007zz}). The price for the absence of theoretical uncertainties is twofold. On the one hand, a direct $CP$ violation has to be used for the measurement that introduces an additional unknown parameter --- an uninteresting strong phase difference which has to be measured together with the angle of interest $\gamma$. An extra unknown quantity obviously decreases the precision of the measurement. The second problem is that the decays used for this measurement are rare. Therefore, although the angle $\gamma$ is determined with a smaller theoretical ambiguity than $\alpha$, from a statistical point of view it remains the worst measured angle of the unitarity triangle.

The easiest way to entangle \Dn\ and \Db\ is to use flavour-blind decay channels of the \Dn meson, for example, $\Dn\to K^+ K^-$. This method is relatively simple but suffers from small statistics. A more promising method discussed in Ref.~\cite{Bondar:2007zz} is based on three-body \Dn\ decays such as $\Dn\to K^0_S\pi^+\pi^-$. In this final state, the entanglement of the \Dn\ and \Db\ is not complete but depends on the masses of the intermediate resonances in the $K^0_S \pi^+ \pi^-$ system. However, the need to correctly take into account the contribution of all intermediate resonances results in a rise of the model uncertainty which is not dominant with the existing statistics but will become such with a further decrease of the statistical error.

For the ten years of data taking at $B$-factories for measuring the $\gamma$ angle an accuracy of $5^\circ$ has been achieved without ambiguities and with a minimal model error. In the future, Belle II is not only to reduce the statistical and systematic error by 4-5 times but also to reduce the model error by fixing the $\Dn_{CP}\to K^0_S\pi^+\pi^-$ decay model from the experimental data from BES III or a planned experiment at the Super $c$-$\tau$-factory \cite{Bondar:2013cja}. Despite a significant improvement in accuracy one has to admit that Belle II may not become the leader in this measurement. Since the final state is fully charged, the LHCb not only does not lose out to Belle II in accuracy in this measurement but even surpasses it due to the ability to extract the same angle from the time-dependent $CP$ asymmetry in the decay $B_s \to D_s^{\pm} K^{\mp}$. 

\subsubsection{Sides of the unitarity triangle}\label{chapter:sides}

Measurements of the sides of the unitarity triangle are associated with extracting the absolute values of the CKM-matrix elements from the decay probabilities or \BBb-oscillation pa\-ra\-me\-ters. In solving this problem not everything depends on the experimentalists' skills. A lot depends on the ability of the theory to accurately predict the relations between the fundamental constants and measurements. In contrast to the angles of the triangle, measurements of its sides require knowledge of the decay form factors or hadronic constants of mesons which cannot be calculated from the first principles of QCD. Very often various tricks are needed to reduce theoretical uncertainties. However, even having succeeded in these tricks to keep theoretical errors at the level of the statistical accuracy of $B$-factories, to improve accuracy at Belle II in the future theoretical progress is mandatory. Fortunately, in the past decade, lattice calculations have provided the theory with a powerful tool for studying various nonperturbative aspects of QCD with a controlled and systematically improved accuracy. Calculations within the framework of lattice QCD are performed by independent international collaborations which not only carry out the actual calculations but can also extrapolate their expected accuracy to the future. Note that in recent years significant progress in calculations has been achieved not only due to a fantastic increase in the computer performance but --- and this is perhaps the most important issue --- due to the success of numerous tests of lattice calculations in the experiment. This allows one to hope that the efforts of experimentalists to improve their accuracy will not go in vain and the theoretical accuracy will not be an obstacle in reaching the goal.

\vspace{0.3cm}\noindent
$\bullet$ {\it The side opposite to the angle $\beta$}
\vspace{0.3cm}

The side of the unitarity triangle opposite to the angle $\beta$ is equal to $\left| V_{ud} V_{ub}/V_{cd} V_{cb}\right|$. To measure it one needs to know the absolute values of these four CKM-matrix elements. Two of them are known with a high precision \cite{Zyla:2020zbs},
\be
\Vud=0.97370 \pm 0.00014,\quad \Vcd= 0.221 \pm 0.004,
\ee
so it remains only to improve the accuracy of measuring the other two elements, \Vcb\ and \Vub. They are best measured in the semileptonic transitions $b \to c \ell^-\bar{\nu}$ and $b \to u \ell^-\bar{\nu}$ since in such decays half of the particles in the final state do not participate in strong interactions and therefore reduce (although do not remove completely) the necessity to involve theoretical calculations. For this reason Belle II, rather than the experiments at the LHC where final states involving neutrinos can not be reliably reconstructed, allows an accurate measurement of these sides. 

The value of the \Vcb\ has already been measured at the $B$-factories with a reasonable (about one per cent) accuracy using the exclusive decays $B\to D^{(*)} \ell^-\bar{\nu}$ \cite{Zyla:2020zbs},
\bea
\Vcb_{D^*\ell\nu} & = & (39.05 \pm 0.47 (\text{exp}) \pm 0.58(\text{theor})) \times 10^{-3}, \nonumber \\[-2mm]
\\[-2mm]
\Vcb_{D\ell\nu\phantom{.}} & = & (39.18 \pm 0.94 (\text{exp}) \pm 0.36(\text{theor})) \times 10^{-3},\nonumber
\eea
where the first error is defined by the experiment and the other one comes from theoretical uncertainties. The two measurements perfectly agree with each other, however the exclusive one (from the probability of the decay $B\to X_c \ell^-\bar{\nu}$, where $X_c$ is a sum over all possible charmed hadrons in the final state) lies three standard deviations higher,
\be
\Vcb_{\text{exclusive}} = (42.19 \pm 0.78) \times 10^{-3},
\ee
that looks somewhat worrisome.

In this regard, a task for Belle II is to re-check all above measurements with a high accuracy and make sure that the predictions of the theory used to extract \Vcb\ agree with all kinematical constraints for the decays. In particular, for the exclusive method it is necessary to make sure that the shapes of the lepton momentum spectra and the $X_c$ mass spectrum predicted by the lattice QCD coincide with the data and, if so, perform a reliable measurement.

Determination of \Vub\ from the inclusive decays $B\to X_u \ell^-\bar{\nu}$ is very difficult because of a large background from the decays $B\to X_c \ell^-\bar{\nu}$. In the small region of the phase space where this background is kinematically forbidden, the theoretical uncertainties are maximal and do not allow extracting \Vub\ with a reasonable theoretical accuracy. Alternatively one can use exclusive decays, such as $B\to \pi (\rho) \ell^-\bar{\nu}$ with the full reconstruction of the second $B$-meson in the event, but this requires a truly huge statistics, and at the moment the statistical uncertainty dominates in such measurements. Averaging everything that has been done at $B$-factories to date, PDG gives the following value for \Vub\ with an accuracy of slightly better than 10\%:
\be
\Vub= (3.82\pm 0.24) \times 10^{-3}.
\ee

The goal of Belle II is to achieve a one per cent accuracy for \Vub\ (that is, to improve accuracy by a factor of 10!) using all available experimental and theoretical approaches. The main role will be played by the method that uses the exclusive decays $B \to \pi \ell^+ \nu$ and $B_s \to K \ell^+ \nu$ since the statistical experimental error rather than the accuracy of the theory dominates in them. Although the second decay has not even been observed yet it must be possible to measure using the data collected at the energy of the $\Upsilon(10860)$-resonance (see Fig.~\ref{fig:hadrons2}), so that the heavier meson in the final state ($K$ instead of $\pi$) will allow a more accurate theoretical prediction. Using this method, at Belle II both \Vub\ and its expected accuracy will be extracted from a simultaneous fit to the dependence of the differential width of the semileptonic decay $B_{(s)} \to \pi (K) \ell^+ \nu$ on the momentum transfer, $q^2= (P_B-P_{\pi(K)})^2$, for the experimental data and simulations (with their statistical and systematic errors) of QCD on the lattice. Such a simultaneous fit will balance the fact that the experimental measurements of the width are most accurate for large hadron momenta in the final state while the form factors are best calculated for small $q^2$. If the shapes of the $q^2$ dependence in the experiment and lattice QCD coincide, one can be sure that the measurement of the \Vub\ is correct.

\vspace{0.3cm}\noindent
$\bullet$ {\it The side opposite to the angle $\alpha$}
\vspace{0.3cm}

To measure the side opposite to the angle $\alpha$ one needs to measure the ratio $\left| V_{td}/V_{ts} \right|$ in which the theoretical uncertainties are partially cancelled. Measuring this ratio directly in $t$-quark decays would be theoretically very clean but not feasible experimentally given that it is almost impossible to distinguish between the jets formed by the $s$ and $d$ quark. The only available processes with $B$-mesons in which these matrix elements are present should obviously include loop or box diagrams since the $t$ quark can be present only virtually. This ratio can be extracted from the mixing parameters of the \BBd\ and \BBs, $\Delta m_d$ and $\Delta m_s$, measured by the $B$-factories and LHCb very precisely \cite{Zyla:2020zbs}, 
\be
\Delta m_d= (0.5065 \pm 0.0019)~\text{ps}^{-1},\quad \Delta m_s=(17.749 \pm 0.020)~\text{ps}^{-1}.
\ee

To extract the values of the sought matrix elements it is also necessary to know the vacuum constants ($f_{d,s}$) and ``bag''-parameters ($B_{d,s}$) responsible for the probability of finding the quark-antiquark pair at one point inside the meson. Their values are calculated in lattice QCD with the accuracy of about 4\%, however, the uncertainties are partially cancelled in the ratio~\cite{Zyla:2020zbs},
\be
f_s\sqrt{B_s}/f_d\sqrt{B_d}= 1.206 \pm 0.038,
\ee
that allows one to extract the studied ratio with a relatively good accuracy,
\be
\left| V_{td}/V_{ts} \right| = 0.205 \pm 0.001 (\text{exp}) \pm 0.006 (\text{theor}).
\ee
Therefore, at the moment everything depends on the theoretical uncertainties while the experiment can only help to keep them under control by providing various precision me\-a\-su\-re\-ments for comparison with the theoretical calculations.  

An alternative method of evaluating $\left| V_{td}/V_{ts} \right|$ is based on measuring the ratio of the probabilities of the penguin radiative decays $B \to \rho \gamma$ and $B \to K^* \gamma$. This method, however, is much less accurate~\cite{Zyla:2020zbs},
\be
\left| V_{td}/V_{ts} \right| = 0.220 \pm 0.016 (\text{exp}) \pm 0.037 (\text{theor}).
\ee
From the experimental side this is caused by a relatively poor identification of hadrons at $B$-factories that does not allow one to resolve a weak $B\to\rho\gamma$ signal against the background of the dominant $B\to K^*\gamma$ one, and in the theoretical part this is because the neutral and charged $B$-mesons were averaged to compare with the experiment while the theoretical uncertainty in the latter is much larger than in the former as a result of the contribution of an additional diagram with the $W$ annihilation. In the future, an excellent identification of charged hadrons in the Belle II experiment will allow to measure $\Br(B\to \rho(\omega)\gamma)$ on the full statistics of $50~\text{ab}^{-1}$ with the accuracy of about $4\%$ \cite{Kou:2018nap}. At the same time, a progress in the theoretical calculations is also expected. This makes the given method also competitive compared to the first one described above.

\subsection{The unitarity triangle at present and in ten years} 

The discussion of various measurements of the angles and sides of the unitarity triangle presented above in this chapter allow to overview the whole picture of the achievements in metrology of the triangle for more than ten years of work of the $B$-factories later joined by the LHCb experiment (measurements of the parameters of the $CP$ violation were also made in the CMS \cite{Khachatryan:2015nza} and ATLAS \cite{Aaboud:2016bro} experiments). Since the base of the triangle was set equal to unity, its coordinates are $(0,0)$ and $(1,0)$ (see Fig.~\ref{fig:triangle0}). Each of the five measurements described above in this chapter, had it been absolutely accurate, would have set the locus for the triangle apex (it is appropriate to recall here school problems of plotting triangles with a compass and ruler). For example, the position of the apex can be totally defined by the circles centred 
at $(0,0)$ and $(1,0)$ if the corresponding side is known exactly, by a ray if the angle adjacent to the base is known, or by the arc if one knows the angle opposite to the base. Five independent measurements imply that the triangle is overdetermined. This is precisely the goal: not to find out the exact position of the apex which, almost certainly, will tell us nothing but to check if these five measurements do not contradict each other.

\begin{figure}[t!]
\centerline{\epsfig{file=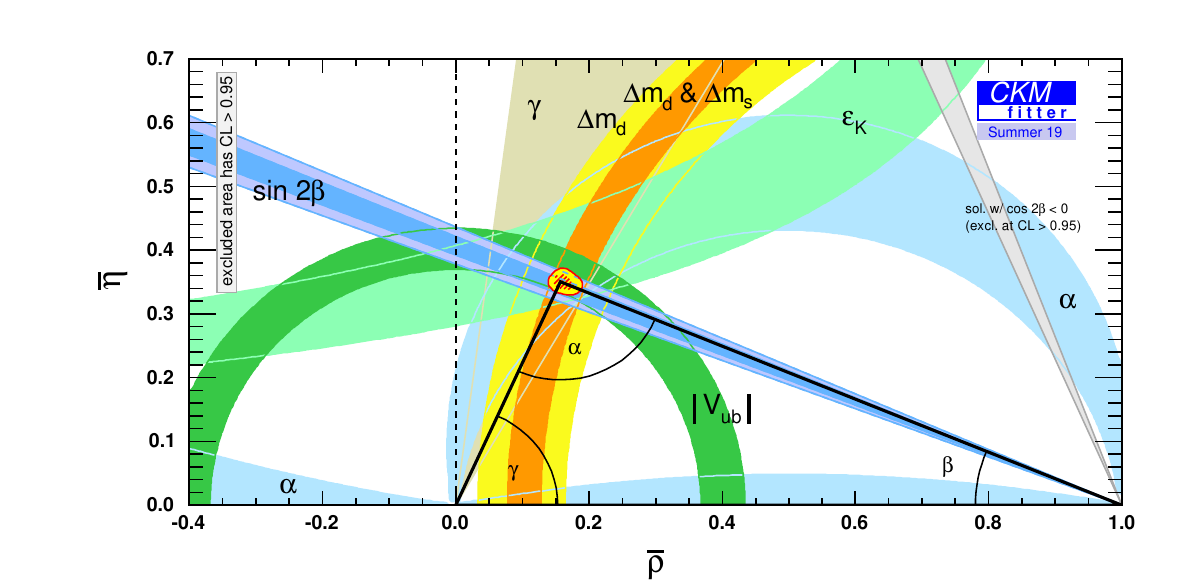, width=0.8\textwidth}} 
\caption{The fit for the untarity triangle apex from all measurements of its parameters. Coloured bands show the allowed positions of the apex of the triangle within one standard deviation constrained from different measurements: the blue and grey sectors and the blue region between the two arcs come from the measurements of the angles $\beta$, $\gamma$, $\alpha$, respectively; the dark green and orange rings are the allowed regions from the measurements of the sides $|V_{ub}|$ and $|V_{td}|$, respectively; the light green region bounded by the hyperbolas is obtained from the measurements of the parameter $\varepsilon_K$.
Adapted from the site of the CKMfitter Collaboration.}
\label{fig:ckmfitter}
\end{figure}

The resulting picture of the triangle can be characterised most clearly with the drawing of the CKMfitter Collaboration \cite{Charles:2004jd} which is engaged in averaging and simultaneous presentation of the results of experimentalists and theorists. This is a highly nontrivial task because of the statistical, systematic, and theoretical uncertainties inherent to all measurements, the correct account for which requires a considerable effort. Uncertainties lead to the fact that the sharp boundaries (rays, circles, ellipse) are blurred and become bands within which the position of the apex is likely, say, at a confidence level of 67\% that corresponds to one standard deviation. In particular, it is instructive to see if all the bands intersect at one point. The result of CKMfitter is shown in Fig.~\ref{fig:ckmfitter} from which one can see that all five measurements indeed agree with each other rather well.

\begin{figure}[t!]
\centerline{\epsfig{file=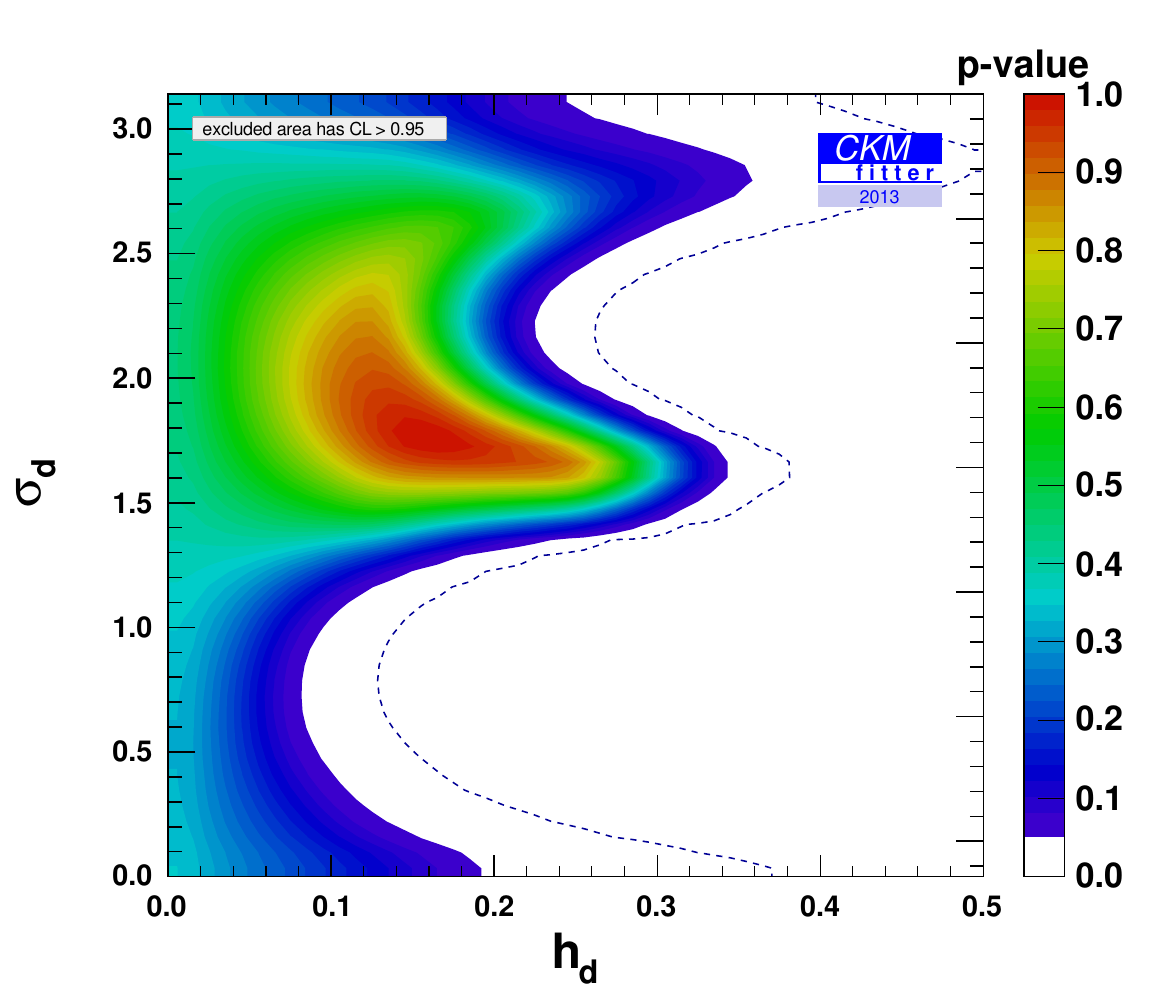, width=0.4\textwidth}
\epsfig{file=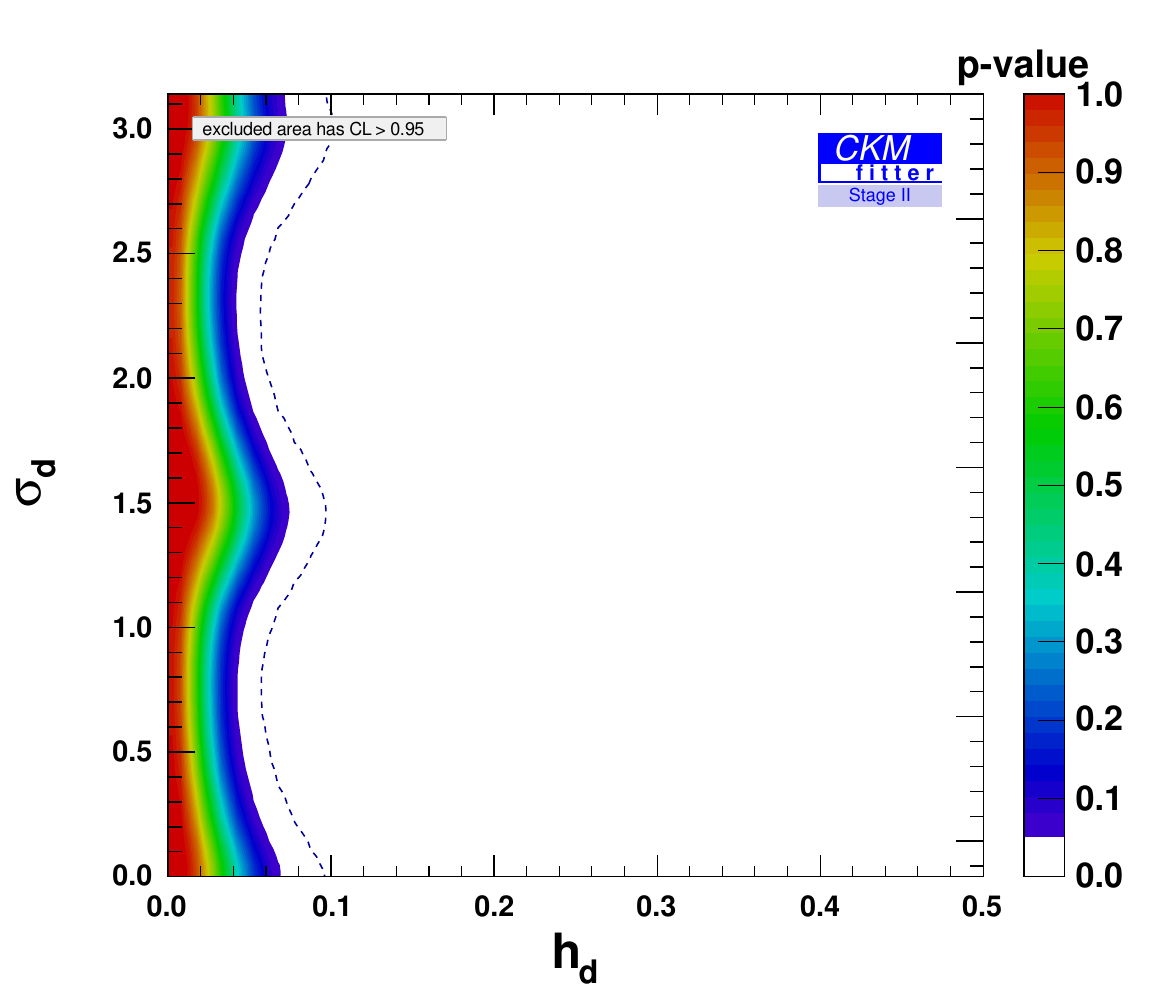, width=0.4\textwidth}} 
\caption{The upper limit on the amplitude of the NP contribution to the \Bn\Bb mixing: at present (left plot) and expected in 10 years of the Belle II work (right plot). Different colours correspond to different confidence levels (CL), from the highest (red) to the lowest (blue). The upper limits with CL=95\% are shown by the dashed line. The estimate is made by the CKMfitter Collaboration. Adapted from its site.}
\label{fig:NP_mix}
\end{figure}

The bands that clearly ``miss'' the apex should not mislead the reader since they just show the second solutions if they exist (for example, due to the ambiguity in extracting the angles $\alpha$ or $\beta$). Let us also pay attention to the band of the hyperbola shape extracted from the parameter of indirect $CP$ violation in the \Kn\Kb\ system $\varepsilon_K$ and thus demonstrating an external constraint for the $B$ physics. The observed agreement implies that the SM has once again successfully overcome a challenging test. In addition, it is obvious that the contribution of NP, if any, is significantly less than that of the Standard Physics. Assuming that NP contributes only to the box diagrams, one can describe it by an additional term in the expression for the amplitude of the \Bn\Bb oscillations,
\be
M_{12}^d=(M_{12}^d)_\text{SM}+(M_{12}^d)_\text{NP}=(M_{12}^d)_\text{SM} \times (1+h e^{i\sigma_d}),
\ee
where $h$ is the ratio of the amplitudes of NP and the SM and $\sigma_d$ is a relative phase of NP. Constraints obtained from the above results are shown at the left plot in Fig.~\ref{fig:NP_mix}. As one can see from the plot, at the moment the amplitude of the NP contribution to the processes with the $B$ s is constrained at the level of 0.1-0.3 (note that this constraint had exceeded 1 before the $B$-factories started operation).

The Belle II experiment will improve the accuracy of individual measurements by 4-10 times. It is possible to extrapolate how this will affect the
exclusion of NP: constraints on $\left|h\right|$ will be at the level of 0.04-0.07 (see the right plot in Fig.~\ref{fig:NP_mix}). One can go even further and estimate the restriction on
the energy scale of NP $\Lambda_\text{NP}$ we can get in 10 years. The amplitudes of the processes with loops behave like
$1/\Lambda_\text{NP}^2$, so that now experiments studying $B$-mesons are sensitive to the processes at the energies about 10 times higher than the
scale of the electroweak theory, that is above the scale achieved at the LHC, and Belle II will surpass this achievement by at least an order of
magnitude.

\subsection{$B$-meson decays}\label{chapter:decays}

As discussed in chapter~\ref{chapter:standmod}, processes with flavour-changing neutral currents, $b\to q\gamma$ and $b\to q l^+l^-$ (as in chapter~\ref{chapter:standmod}, hereinafter a light quark $q$ is understood to be the $d$- or $s$-quark), are of a particular interest in view of their sensitivity to NP. The SM predicts the probability of transitions with the quark flavour change at the level $10^{-6}$, however, it is very sensitive to the type and number of particles involved in these processes. For this reason, rare decays of the $B$-mesons are a powerful tool in searches for NP since new particles appearing beyond the SM can significantly change the decay probabilities and angular distributions in the final state. In addition, radiative
electroweak penguin decays of the $B$-mesons provide a perfect testground for NP searches because the final states containing photons or pairs of leptons can be both
calculated theoretically and measured experimentally with a high accuracy.

Inclusive measurements of the transitions $b\to q\gamma$ and $b\to q l^+ l^-$ as well as the decays of the $B$-mesons into final states containing a pair of photons, neutrinos or $\tau$-leptons are difficult to access in the LHCb experiment. So Belle II turns out to be the only experiment that in the near future can provide detailed new information on the processes that change the quark flavour.

The physical program of the Belle II experiment in this field of research is focused at measurements of the inclusive decays 
$B \to X_{d,s}\gamma$ and $B \to X_{d,s} l^+ l^-$ (hereinafter $X_{d,s}$ stands for a meson containing the $d$- or $s$-quark) as well as rare decays 
$B_{d,s}\to \gamma \gamma$, $B \to K^* \nu \bar{\nu}$, $B_{d,s} \to \tau^+ \tau^-$, and $B \to K^{(*)} \tau^+ \tau^-$. Below in this chapter these processes are considered one by one.

\subsubsection{The process $B \to K^*\gamma$}\label{chapter:B2Kgamma}

The radiative transition $b \to s \gamma$ was first observed by the CLEO Collaboration in 1993 in studies of the process $B \to
K^* \gamma$ \cite{Ammar:1993sh} which is the most experimentally clean decay of the type $B \to X_s \gamma$. At present this decay is of large interest for searches for NP since it allows to measure such important quantities as the photon polarisation and the isospin and $CP$-asymmetry. 

As discussed in chapter~\ref{chapter:standmod}, the SM predicts that the polarisation of the photon emitted in the electroweak penguin loop 
$b\to s\gamma$ is predominantly left-handed (up to corrections suppressed by a small mass ratio $m_s/m_b\sim 0.03$). This is because the
$W$-bosons interact only with left-handed quarks. However, some extensions of the SM compatible with all experimental data available to date \cite{Bertolini:1987pk,Baer:1996kv,Hewett:1996ct,Carena:2000uj,Fujikawa:1993zu,Babu:1993hx,Cho:1993zb} predict that the right-hand polarisation of the photon can be enhanced due to the presence of a heavy fermion in the loop. Thus studies of the photon polarisation in the given process at the expected large statistics in the Belle II experiment will allow either to detect a deviation from the SM or to set a strong upper limit on it.

It should be mentioned that measurements of the partial branchings of the transitions $B \to K^*\gamma$ and $\bar{B} \to \bar{K^*} \gamma$ provide only weak constraints on NP since the predictions of the SM suffer from large uncertainties in the hadronic form factors. In the meantime, the isospin asymmetry,
\be
\Delta_{0+}=\frac{\Gamma(B^0 \to K^{*0}\gamma)-\Gamma(B^+ \to K^{*+}\gamma)}{\Gamma(B^0 \to K^{*0}\gamma)+\Gamma(B^+ \to K^{*+}\gamma)},
\label{D0plus}
\ee
and the direct $CP$-asymmetry, 
\be 
A_{\rm CP}=\frac{\Gamma(\bar{B} \to \bar{K^*} \gamma)-\Gamma(B \to K^*\gamma)}{\Gamma(\bar{B} \to \bar{K^*} \gamma)+\Gamma(B \to K^*\gamma)},
\label{ACP}
\ee 
where $\Gamma$'s are the widths of the corresponding decays, appear to be more theoretically clean observables since the aforementioned theoretical un\-cer\-ta\-in\-ti\-es cancel against each other in them. 

The decay $B \to K^* \gamma$ was studied in several experiments, in particular, by CLEO \cite{Coan:1999kh},
Belle \cite{Nakao:2004th}, BaBar \cite{Aubert:2009ak}, and LHCb \cite{Aaij:2012ita}. In 2017 Belle announced \cite{Horiguchi:2017ntw} the first statistically significant (3.1 standard deviations) experimental observation of the isospin breaking in the decay
$B \to K^* \gamma$,
\be
\Delta_{0+}(K^*\gamma)=(+6.2\pm 1.5\mbox{(stat)}\pm 0.6\mbox{(sys)}\pm 1.2(f_{+-}/f_{00}))\%,
\label{D0plusBelle}
\ee
where the last error stems from the uncertainty in the ratio of the number of the $B^+B^-$ and $B^0 \bar{B^0}$ pairs produced in the decays of the vector bottomonium $\Upsilon(4S)$ ($f_{+-}$ and $f_{00}$, respectively), and the first measurement of the $CP$-asymmetry difference for the processes 
$B^0 \to K^{*0}\gamma$ and $B^+ \to K^{*+}\gamma$,
\be
\Delta A_{CP}=(+2.4\pm 2.8\mbox{(stat)}\pm 0.5\mbox{(sys)})\%,
\label{ACPBelle}
\ee
which is consistent with zero.

So far all measured values of the asymmetries (\ref{D0plus}) and (\ref{ACP}) agreed well within errors with the predictions of the SM. However, the present 
accuracy of the experimental values is worse than that of the theoretical calculations. For example, the prediction of the SM for the isospin asymmetry vary from 2\% to 8\% with a typical uncertainty of the order of 2\%
\cite{Keum:2004is,Lyon:2013gba,Beneke:2004dp,Ball:2006eu,Kagan:2001zk,Ahmady:2013cva}. For the $CP$-asymmetry there is only one theoretical prediction \cite{Paul:2016urs},
\be
A^{\rm SM}_{CP}(B^0 \to K^{ 0}\gamma)=(0.3\pm 0.1)\%.
\label{ACPSM}
\ee

Note also that at the moment the statistical uncertainty dominates in the experimental results, and more data are needed to reduce it. Thus,
the Belle II experiment, with the projected luminosity exceeding the one achieved by its predecessor Belle about 40 times, as no one else is suitable for accomplishing the task of a radical reduction of the statistical error of the measurements.

In the Belle II experiment $K^*$ mesons can be reconstructed in the following decay channels: $K^- \pi^0$, $K^0_{\rm S} \pi^-$, $K^-
\pi^+$, and $K^0_{\rm S}\pi^0$. Combining the $K^*$ meson with a hard photon one can reconstruct a $B$-meson candidate. The final states $K^- \pi^0$, $K^0_{\rm S}\pi^-$, and $K^- \pi^+$ are eigenstates in the basis of flavours that allows one to use them for measurements of the $CP$-asymmetry while the $K^0_{\rm S} \pi^0$ channel with the tagged flavour of the second $B$-meson allows to measure the time dependence of the $CP$ asymmetry.
In the latter case the decay rate for the neutral $B$-meson to the $CP$ eigenstate depends on time as
\be
P(\Delta{t},q)=\frac{e^{-|{\Delta{t}}|/\tau_{B^0}}}{4\tau_{B^0}}\Bigl(1+q[S\sin(\Delta
m_d\Delta t)+A\cos(\Delta m_d\Delta t)]\Bigr),
\label{Ptq}
\ee 
where $S$ and $A$ are the parameters defining the $CP$ asymmetry, $\tau_{B^0}$ is the $B^0$-meson lifetime, $\Delta{t}$ is the difference of the
decay times of the $B^0$ and $\bar{B^0}$ meson, $\Delta m_d$ is the mass difference for the $B^0$ and $\bar{B^0}$ meson, $q=1$ and $q=-1$ correspond to the decay of the $\bar{B^0}$ and $B^0$ meson, respectively. The quantity (\ref{Ptq}) depends on the Wilson coefficients $C_7$ and $C_7'$ (see chapter \ref{chapter:standmod}) and, for this reason, is sensitive to the polarisation of the photon in the final state. 

If the isospin asymmetry $\Delta_{0 +}(K^*\gamma)$ is measured in the Belle II experiment with 5 ab$^{-1}$ of collected data (that is, approximately
one tenth of the total statistics expected for the entire duration of the experiment) the systematic error of 0.5\%, which comes from the uncertainty in determining the number of the produced $B^+B^-$ and $B^0 \bar {B^0}$ pairs (here also a significant progress has been achieved in comparison with the Belle experiment --- see formula (\ref{D0plusBelle})), will still dominate. However, this uncertainty is already five times smaller than the most up-to-date theoretical prediction of the SM.

In the measurements of the direct $CP$ asymmetry at Belle II the statistical error will still prevail. 
The corresponding uncertainties are estimated as 0.2\% and 0.3\% for the $A_{\rm CP} (B^0\to K^{ 0} \gamma)$ and $A_{\rm CP}(B^+ \to K^{ +} \gamma)$, respectively, that eight times exceeds the accuracy of the previous result by Belle \cite{Horiguchi:2017ntw}. The theoretical estimate (\ref{ACPSM}) still has a smaller uncertainty than the result accessible in the Belle II experiment, however, as accurate experimental measurement of the quantity  $A_{\rm CP} (B^0 \to K^{0} \gamma)$ as possible is an important task by itself.

\subsubsection{The process $B \to K^* l^+ l^-$}\label{chapter:B2Kll}

The inclusive decays $B \to X_q l^+ l^-$ provide information on the $b$-quark sector in addition to that extracted from the inclusive decays $B \to X_q \gamma$.

The transition $b \to s l^+l^-$ was first observed in the Belle experiment in 2001 in the $B \to K l^+l^-$ decay~\cite{Abe:2001dh}. Two years later Belle announced the discovery of the $B \to K^*l^+l^-$ decay~\cite{Ishikawa:2003cp}. Experimental investigation of these two processes opened a promising way for NP searches in the electroweak penguin decays of the $B$-mesons. 

In particular, the angular analysis of the $B^0 \to K^{*0}l^+l^-$ decay is of interest. Of the twelve variables describing it, the angular distribution of the products is determined by three angles (the angle between the direction of the $l^+$ in the dilepton  rest frame and the direction of the $l^+l^-$ dilepton motion in the $B$-meson rest frame, the angle between the direction of the kaon $K$ in the $K^*$ rest frame and the direction
of $K^*$ in the $B$-meson rest frame, and the angle between the  $l^+l^-$ plane and the $K^*$ decay plane) and the squared dilepton mass $q^2$. At small $q^2$ the result is sensitive to the contribution of the right-handed penguin operator $Q_7'$ (see chapter \ref{chapter:standmod}) \cite{Jager:2014rwa,Jager:2012uw,Becirevic:2011bp,Grossman:2000rk} and therefore important for assessing the impact of NP. At present the main uncertainties in the theoretical predictions of the angular variables stem from the unknown corrections of the order ${\cal O}(\Lambda_{\rm QCD}/m_b)$ and
uncertainties in the calculation of the hadronic form factors. Attempts to reduce these uncertainties led to the introduction of the so-called ``optimised''
variables defined as ratios in which the dependence on the form factors is severely reduced, so that the optimised variables are less sensitive to theoretical uncertainties (see the discussion in chapter \ref{chapter:triangle}).

In 2013 the LHCb Collaboration announced a deviation from the predictions of the SM ~\cite{Aaij:2013qta} for one of the optimised variables in the angular analysis of the process $B^0 \to K^{*0} \mu^+ \mu^-$ obtained with 1~fb$^{-1}$ of collected data. Two years later using the full data collected in Run I that correspond to 3~fb$^{-1}$, LHCb repeated the angular analysis of the same process and found a discrepancy with the SM at the level of 3.3 standard deviations~\cite{Aaij:2015oid}. Soon after that the Belle Collaboration performed its angular analysis of the processes $B^0 \to K^{*0} \mu^+ \mu^-$ and
$B^0 \to K^{*0} e^+ e^-$ with the full statistics (about 711~fb$^{-1}$)~\cite{Wehle:2016yoi}. The result obtained agrees with that from LHCb~\cite{Altmannshofer:2017fio}.

The declared deviation from the SM emphasises the necessity of further detailed studies of the process $B^0 \to K^{*0}\mu^+ \mu^-$ to reduce the experimental uncertainty. In particular, the Belle II experiment can considerably reduce the systematic uncertainty by using the results obtained previously at Belle. For example, the difference between the simulations and data can be estimated directly from the $B\to J/\psi K^*$ decay studied in detail in the Belle experiment. The accuracy of measuring both (electron and muon) modes with just 2.8~ab$^{-1}$ of the data collected at Belle II 
is expected to be comparable with that for the results from LHCb obtained in the analysis of the muon channel only with the 3~fb$^{-1}$ data sample. 
It is obvious that the analysis of the entire bulk of the data (about 50~ab$^{-1}$) planned to be collected in the Belle II experiment will allow one to achieve the results which will substantially surpass the precision of those from LHCb and, thus, either to resolve the conflict with the SM or confirm it at a qualitatively new level.

Other promising quantities to be explored are the ratio of the probabilities of the electron and muon modes and the forward-backward asymmetry as a function of $q^2$.

\subsubsection{The processes $B \to K^{(*)} \nu \bar{\nu}$}\label{B2Knunu}

Decays $B \to K^{(*)} \nu \bar{\nu}$ related to the transition $b\to s$ are also quite promising in searches for NP~  
\cite{Buras:2014fpa,Altmannshofer:2009ma,Kamenik:2009kc}. From the point of view of the theory, rare $B$-meson decays to the final states containing a pair of neutrinos belong to the most clean processes with the flavour-changing neutral currents (see chapter \ref{chapter:standmod}). Since neutrino is electrically neutral,  here, on the contrary to other $B$-meson decays, factorisation of the hadron and lepton currents is exact.  For this reason, precision measurements of the processes $B \to K^{(*)} \nu \bar{\nu}$ should allow one to extract the form factors of the transition $B \to K^{(*)}$ with a high accuracy. 

The processes $B \to K^{(*)} \nu \bar{\nu}$ are also closely related to other $B$-meson decays proceeding through the formation of an exotic state which, in turn, decays into a pair of neutrinos. Studies of such signals are very promising in searches for the dark matter and can allow one to investigate the interrelation between the SM and so-called dark sector of the Universe~\cite{Kamenik:2011vy}.

Searches for the processes $B \to K^{(*)} \nu \bar{\nu}$ with $K^+$, $K^{*+}$, and $K^{*0}$ in the final state were performed by the Belle and BaBar Collaborations using the method of the hadron \cite{Lutz:2013ftz,Lees:2013kla} or semileptonic \cite{Grygier:2017tzo,delAmoSanchez:2010bk} tagging. The established upper limits on the probabilities of such processes exceed the predictions of the SM by two to five times \cite{Buras:2014fpa}. 
Therefore, improving the accuracy of measuring the processes $B \to K^{(*)} \nu \bar{\nu}$ is an extremely important task. Then, even if NP does not contribute anything new to the transitions $b \to s \nu \bar{\nu}$, all three decays of the type $B\to K^{(*)} \nu \bar{\nu}$ mentioned above will be observed at Belle II already with the statistics of 10 ab$^{-1}$ while the accuracy of measuring the probabilities of such processes with the full statistics of 50 ab$^{-1}$ will be about 10\%, that is already comparable with the uncertainty of the theoretical predictions in the SM. Once the decays $B \to K^{(*)+} \nu \bar{\nu}$ and $B \to K^{(*)0} \nu \bar{\nu}$ are observed experimentally, important quantities to measure are the ratio of the probabilities for the two above decays and the polarisation of the $K^*$-meson. In particular, the expected precision of the measurement of the longitudinal component of the $K^*$ polarisation with the full Belle II statistics is about 8\% for both charged and neutral kaons that is compatible with the uncertainty of the SM predictions of about 3\%.

\subsubsection{The process $B\to \nu\bar{\nu}$}\label{chapter:b2nunu}

\begin{figure}[t!]
\begin{center}
\epsfig{file=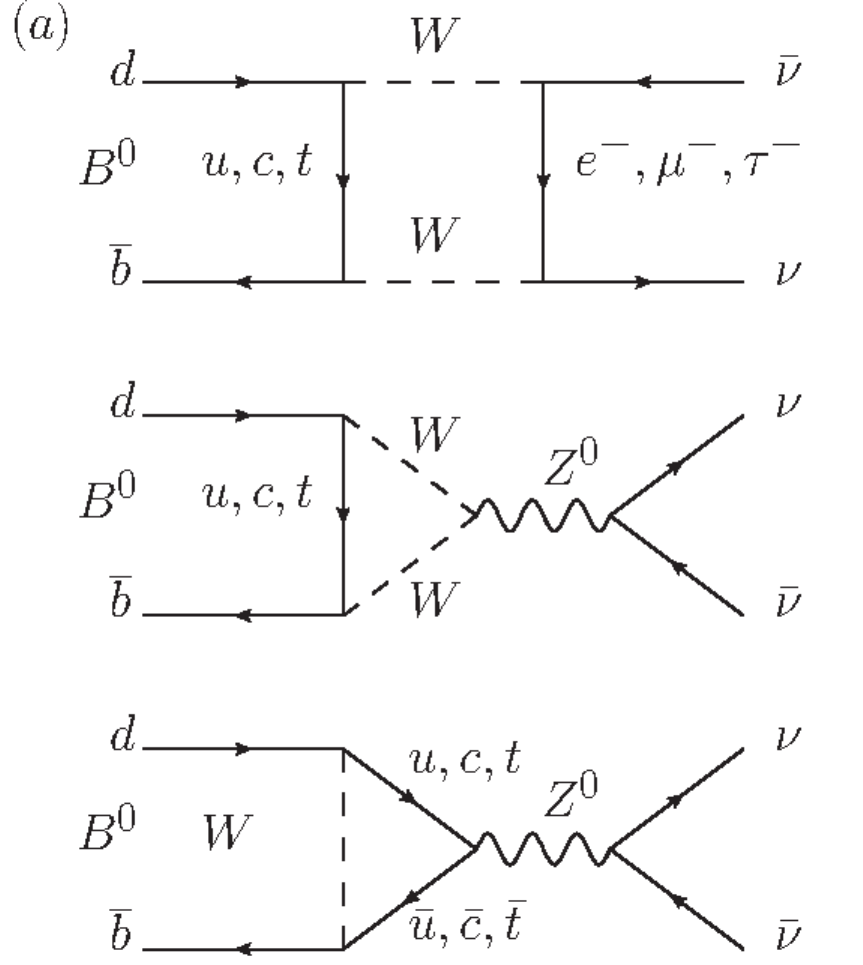, width=0.49\textwidth}
\epsfig{file=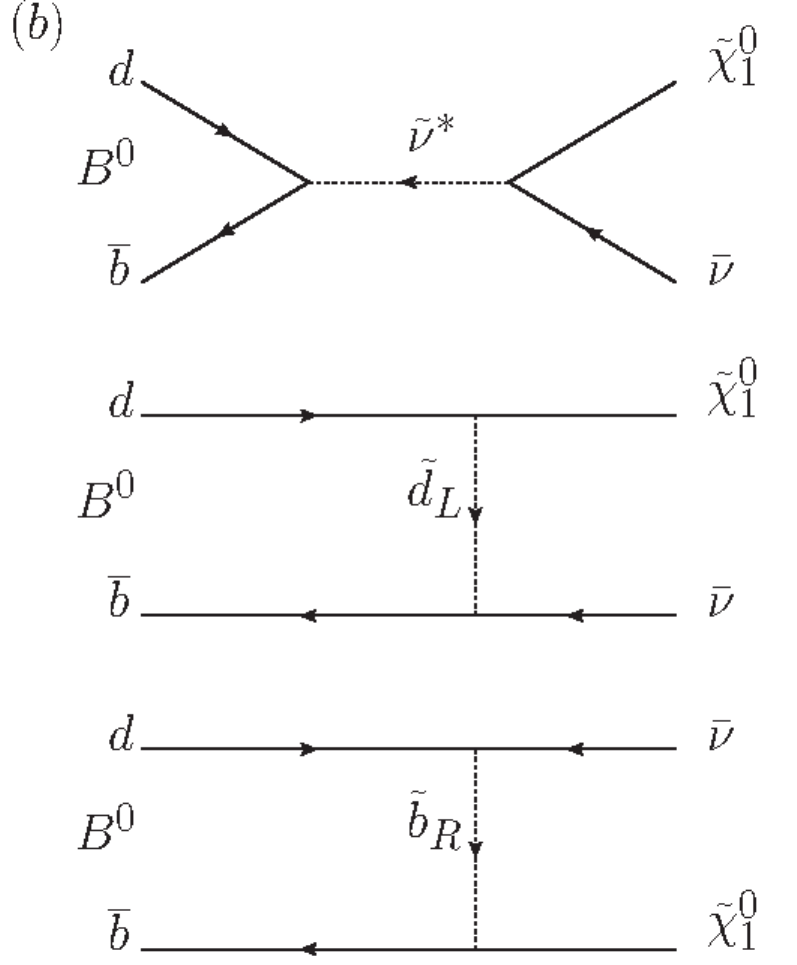, width=0.45\textwidth}
\end{center}
\caption{Feynman diagrams describing the processes (a) $B^0 \to \nu \bar{\nu}$ in the SM and (b) $B^0 \to \tilde{\chi}_1^0 \bar{\nu}$ with hypothetical particles of NP. Adapted from Ref.~\cite{Lees:2012wv}.}
\label{fig:b-nunu}
\end{figure}

In the SM the $B^0 \to \nu \bar{\nu}$ decay can proceed in three possible ways --- the corresponding Feynman diagrams are shown in
Fig.~\ref{fig:b-nunu}(a). Theoretical calculations demonstrate that the amplitude of such a process is suppressed by a tiny factor
$(m_{\nu}/m_{B})^2$ (here $m_\nu$ and $m_B$ are the neutrino and $B$-meson mass, respectively), so that the probability of this process turns out to
be much lower than the values achievable in the present-day experiment. However, this probability can be enhanced by a contribution of NP if the
corresponding new states also participate in the process resulting in the final states with weakly interacting particles which provide the same
signature as the $B^0 \to \nu \bar{\nu}$  decay (see the diagrams in Fig.~\ref{fig:b-nunu}(b)). It should be noted that experimentally the decay $B^0 \to \nu \bar{\nu}$ is identified through the missing daughter particles from one of the $B$ mesons and thus it is indistinguishable from other decays with a larger number of neutrinos or other invisible particles (for example, hypothetical Dark Matter particles). 
It was demonstrated in Ref.~\cite{Bhattacharya:2018msv} that the probability of the $B$-meson decays into four-neutrino final states exceeds substantially that of the decays into two-neutrino final states, so that such transitions should also be taken into account when interpreting the results of the searches for the decays $B^0 \to \mbox{\em ``nothing''}$ from the point of view of the NP signals. 

Decays of the $B^0$ into $\nu \bar{\nu}$ and invisible final states were studied by the BaBar~\cite{Lees:2012wv} and Belle~\cite{Hsu:2012uh} Collaborations using the method of the hadron tagging. None of the two collaborations managed to observe a significant signal and, therefore, upper limits were established for the probabilities of such processes at the confidence level of 90\% which constitute $2.4 \times 10^{-5}$ and $1.3 \times 10^{-4}$, respectively.

The Belle analysis \cite{Hsu:2012uh} was performed using an old method of the hadron tagging which later was improved. Besides that making use of additional requirements for suppressing the continuum $\tau^+ \tau^-$ backgrounds also facilitated further improving the accuracy of the measurements. The efficiency of the reconstruction with the help of the hadron tagging is expected to improve by a factor of five  given the large statistics planned to be collected in the Belle II experiment. It is also anticipated that the analysis using the semileptonic tagging will allow to establish  upper limits for the probabilities of the $B^0$-meson decays to invisible final states with the accuracy three times better than the one achieved with the help of the hadron tagging. Using both the hadron and semilepronic tagging methods in the analysis with the full statistics of the Belle II experiment the upper limit $\Br(B^0 \to \nu \bar{\nu})\sim 1.5 \times 10^{-6}$ can be established. In addition, the combined analysis using both these tagging methods will also allow one to investigate the process $B_{\rm s} \to \nu \bar{\nu}$ which has not been yet experimentally studied and establish the upper limit on the partial branching of this process at the level $\Br(B_s \to \nu \bar{\nu})\sim 1.5 \times 10^{-6}$.

\section{Tests of the SM and searches for NP in the lepton sector}\label{chapter:leptons}
In this chapter we discuss some tasks in the lepton sector of the SM related to the muon and $\tau$-lepton which look rather promising for studies in the Belle II experiment.

\subsection{Measurements of $R$ and the anomalous magnetic moment of the muon}\label{chapter:muonanom}

The most accurate measurement of the anomalous magnetic moment of the muon (hereinafter denoted as $a_\mu$) performed at the Brookhaven National
laboratory \cite{Bennett:2006fi} is known to give the result exceeding by about 3.7 standard deviations \cite{Davier:2019can} the theoretical prediction in the SM 
based on the experimental data on $e^+e^-$ annihilation into hadrons. In order to understand whether such a discrepancy is a statistical fluctuation or manifestation of NP, a new measurement of $a_\mu$ is currently in progress in the Enrico Fermi laboratory (USA) \cite{Logashenko:2015xab} and another experiment is being prepared in the J-PARC laboratory (Japan) \cite{Abe:2019thb}; in each of them it is planned to improve the accuracy of our knowledge of the value of $a_\mu$ four times. This leads to the necessity to improve the accuracy of the theoretical predictions as well, with the largest contribution to the uncertainty coming from the hadronic vacuum polarisation. Evaluation of this effect is possible from first principles (lattice calculations) or using precision measurements of the total cross section of the $e^+e^-$ annihilation into hadrons ($R$ ratio --- see formula (\ref{ratioR})) \cite{Aoyama:2020ynm}. Belle II has a unique opportunity to significantly increase the accuracy of measurements of $R$ in the energy range in the centre-of-mass system from the threshold of the hadronic production up to about 5 GeV using the method of the initial-state radiation (ISR) \cite{Druzhinin:2011qd}. More specifically, this method allows one to measure with high accuracy the cross section of $e^+e^-$ annihilation into
all kinds of hadronic final states. Because of the emission of an energetic $\gamma$-quantum by the initial electron (positron), annihilation occurs at a noticeably lower energy, so that the resulting hadronic system has a large momentum and moves inside a cone with a small span angle that allows to detect hadrons with a high probability which weakly depends on the production mechanism. This also allows to strongly suppress systematic effects
related to limited knowledge of the mechanism of hadron production. High efficiency of this method for measuring $R$, which is complementary to the traditional method of scanning, has been proven in recent years by successful experiments at the $e^+e^-$-factories --- the experiments BaBar, Belle, KLOE, and BES III \cite{Aoyama:2020ynm}. The integrated luminosity planned in the Belle II experiment will exceed the previously collected one by two orders of magnitude that will allow one not only to refine the cross sections of the main processes in this energy range but also increase the sensitivity
to poorly studied or not yet observed final states. Improvement in the accuracy of the hadronic vacuum polarisation at low energies is important not only
to refine the prediction of $a_\mu$ but also to determine such fundamental quantities as the running fine structure constant $\alpha(s)$, light quark masses \cite{Shifman:1978by}, quark and gluon condensates \cite{Eidelman:1978xy}, and so on.

\subsection{$\tau$-lepton studies and tests of the leptonic universality}\label{chapter:tau}

The $\tau$ lepton and its neutrino $\nu_\tau$ are two of the six fundamental leptons of the SM (see chapter \ref{chapter:standmod} and, in 
particular, formula (\ref{generations})). Being the heaviest lepton, $\tau$ decays into other leptons and hadrons --- the Particle Data Group in its latest
edition mentions over 200 different decay modes of $\tau$ \cite{Zyla:2020zbs}. Although semileptonic and leptonic decays of hadrons with the
$\tau$ lepton in the final state are widely used in the experiments at the LHC \cite{Sirunyan:2017khh,Aaboud:2018pen}, the presence of the corresponding neutrino, $\nu_\tau$,
in the final state of any $\tau$-lepton makes detailed studies of $\tau$ decays in the experiments at hadronic
colliders very difficult. Therefore, the leadership in providing the sources of $\tau$ leptons belongs to $e^+e^-$ colliders where $\tau$ leptons are
produced in the reaction $e^+e^-\to\tau^+\tau^-$ with a large cross section of 0.92 nb at the energy of the $\Upsilon(4S)$-resonance that makes each
$B$-factory also a $\tau$-lepton factory that produces $0.92 \times 10^6$ $\tau^+\tau^-$ pairs per fb$^{-1}$ of its integrated luminosity. Produced in
the above electromagnetic process the $\tau$ lepton decays through the weak interaction with hadrons in the final state in 65\% of cases, so the
description of such decays requires taking into account effects of strong interactions through the introduction of hadronic form factors. Thus,
studies of the production and decays of $\tau$ allow to investigate all interactions contained in the SM and search for NP effects. It
should also be noted that in the $\tau$ decays the final states with only one charged particle (85\% of cases) prevail, so that without exaggeration
investigation of $\tau$ leptons provides a clean laboratory without hadrons in the initial and just a small number of them in the final state. Therefore,
it should not go as a surprise that considerable progress in studies of the $\tau$-lepton properties achieved after 2005 is associated with
$B$-factories. Table~\ref{tab:lepton} contains the values of the integrated luminosity and the total numbers of $\tau^+\tau^-$ pairs produced in
various experiments. One can see that the integrated luminosity of 50 ab$^{-1}$ planned in the Belle II experiment corresponds to production of
46$\times 10^9$ $\tau$-lepton pairs that will allow one to search for and study rare decays with the  branching
fractions at the level $10^{-9}$-$10^{-10}$.

\begin{table}[t!]
\begin{center}
\begin{tabular}{|c|c|c|}
\hline
Experiment & Integrated luninosity, fb$^{-1}$ & Number of $\tau$-lepton pairs, $10^6$ \\
\hline
LEP (Z-boson peak) & 0.34 & 0.33 \\
\hline
CLEO (10.6~GeV) & 13.8 & 12.6 \\
\hline
BaBar (10.6~GeV) & 469 & 431 \\
\hline
Belle (10.6~GeV) &980 & 902 \\
\hline
Belle II &$5\times 10^4$ & $4.6\times 10^4$ \\
\hline
\end{tabular}
\end{center}
\caption{Experimental studies of the $\tau$ lepton}
\label{tab:lepton}
\end{table}

To test the leptonic universality of the SM (see chapter \ref{chapter:standmod}) the mass, lifetime, and the  branching fraction of the decay
$\tau^- \to e^-\bar{\nu}_e\nu_\tau$ or $\tau^- \to \mu^-\bar{\nu}_\mu\nu_\tau$ need to be measured with an extremely high precision. Let us focus on each of these measurements separately. 

\subsubsection{Measurements of the $\tau$-lepton mass}

Precise knowledge of the mass of the $\tau$ lepton plays a crucial role since its decay width is proportional to the fifth power of the mass, so that any tests of the SM critically depend on the value and the accuracy of the $\tau$ mass. Its most accurate measurements are made using the so-called threshold method in which the energy dependence of the cross section of the production of a pair of $\tau$-leptons is measured near the production threshold. The most accurate to-date measurement of the mass made at the BES III detector \cite{Ablikim:2014uzh} gives $m_\tau = 1776.91 \pm 0.12^{+0.10}_{-0.13}$ MeV.

$B$-factories employ the pseudomass method developed and used for the first time at the ARGUS detector \cite{Albrecht:1992td}. In this method
the mass is estimated at the edge of the invariant mass spectrum for all detected decay products, and the masses of both positively and negatively charged $\tau$-lepton are measured independently that allows one to test the $CPT$ invariance. In the most accurate measurement made at the Belle detector using this method the result is $m_\tau = 1776.61 \pm 0.13 \pm 0.35$ MeV \cite{Abe:2006vf}. Analysis of the systematic uncertainties in this measurement demonstrates that, if the traditional for this method decay mode $\tau^- \to \pi^-\pi^+\pi^-\nu_\tau$ is used, one can optimistically hope to achieve the systematic uncertainty of the order of 0.15 MeV at Belle II (that is, to reduce it by more than two times). Meanwhile improvements in the adopted methodology is required in order to further increase the accuracy, for example, using other decay modes with heavier final particles, such as the decay
$\tau^- \to K^-K^0_S\nu_\tau$, in which the edge of the pseudomass spectrum is significantly shifted towards the $\tau^-$ mass.

\subsubsection{Measurements of the $\tau$-lepton lifetime}

At an asymmetric collider, the angle between $\tau$ leptons produced in the process $e^+e^- \to \tau^+\tau^-$ is not 180$^\circ$ in the laboratory frame, so the point of their production is determined by the intersection of the two trajectories fixed by the decay points and the directions of the $\tau$-lepton momenta. Determining the direction of each $\tau$ in the laboratory frame amounts to solving a quadratic equation, so that there are two solutions. This feature of the experiments at an asymmetric $B$-factory allowed one to measure the $\tau$-lepton lifetime with a high precision \cite{Belous:2013dba}.
Using over 630 million generated $\tau^+\tau^-$ pairs and the fully kinematically reconstructed events with $\tau$ decaying into $3\pi\nu_\tau$, the Belle Collaboration obtained for the lifetime of the $\tau$ lepton the value $(290.17 \pm 0.53 \pm 0.33) \times 10^{-15}$~s that is 1.6 times more accurate than the world average value mainly based on the measurements at LEP \cite{Zyla:2020zbs}. In addition, the difference in lifetimes between the positively and negatively charged $\tau$'s was also measured for the first time to be $|\langle\tau_{\tau^+}\rangle-\langle\tau_{\tau^-}\rangle|/\langle\tau_\tau\rangle<7.0 \times 10^{-3}$ at a 90\% confidence level. One can see that with the expected increase in statistics at Belle II the statistical uncertainty can be considerably reduced. Besides, the systematic uncertainty of the Belle measurement is dominated by the contribution coming from the accuracy of the vertex detector alignment, so one can hopefully avoid or at least noticeably suppress the contribution of this effect in the corresponding brand new subsystem of the new detector (see chapter~\ref{chapter:belle2}).

\subsubsection{Branching ratio of the decay $\tau^- \to e^-(\mu^-)\bar{\nu}_{e(\mu)}\nu_\tau$}

Measurements of the probability of this decay, similarly to other decays with just one charged particle in the final state, is a very difficult task since the probabilities of $\tau^-$ decays into $e^-\bar{\nu}_{e}\nu_\tau$, $\mu^-\bar{\nu}_\mu\nu_\tau$, and $h^-\nu_\tau$ (here $h=\pi,K$) are strongly correlated. Among the experiments at the energy of the $\Upsilon(4S)$ and the corresponding kinematics, the most accurate measurements of these probabilities   were performed by CLEO \cite{Anastassov:1996tc} and BaBar \cite{Aubert:2009qj} (with the integrated luminosity 3.56 fb$^{-1}$ and 467 fb$^{-1}$, respectively). Interestingly, the systematic uncertainties of these measurements are comparable despite a 130-times larger integrated luminosity in the BaBar experiment. It is worth mentioning that a precision measurement of the probability of a lepton decay requires a good understanding of higher-order effects, in particular, a separate measurement of the radiative decays with a real $\gamma$-quantum in the final state. Such a measurement was made recently with the BaBar detector \cite{Lees:2015gea} in which both radiative decays $\tau^- \to e^-\bar{\nu}_e \nu_\tau \gamma$ and $\tau^- \to \mu^-\bar{\nu}_\mu \nu_\tau \gamma$ were measured with the integrated luminosity of 431 fb$^{-1}$ with a many times higher accuracy than before. While for the muon decay the result is
in good agreement with the theory, for the electronic decay the measured probability differs from the theoretical prediction by 3.5 standard deviations \cite{Fael:2015gua}. Similarly to the study of the structure of the electroweak interaction, in the leptonic decay of the muon \cite{Fetscher:1986uj} a detailed study of the spin-spin correlations between $\tau^+$ and $\tau^-$ generated in the process $e^+e^- \to \tau^+\tau^-$ allows one to determine the so-called Michel parameters expressed in terms of the constants of the electroweak Lagrangian and determine the differential cross section of this process \cite{Tamai:2003he}. Studies of the Michel parameters in the leptonic decays of $\tau$ using the integrated luminosity of 485 fb$^{-1}$ collected at Belle demonstrated that their statistical errors were already at the level of $10^{-3}$ and the total systematic error from the physical and detector corrections was below 1\% \cite{Epifanov:2017kly}. An important role is played by the systematic effects related to the corrections for the detection and trigger efficiency. In the Belle II experiment the expected statistical error is $10^{-4}$ so that the systematic uncertainties dominate which require a high and uniform efficiency of the two-track trigger to reduce. The general conclusion regarding the tests of the lepton universality is clear: a detailed analysis of the systematic effects should allow a noticeable increase in accuracy which does not seem impossible given the huge expected statistics in the Belle II experiment.

\subsection{Searches for NP in $\tau$-lepton decays}

\subsubsection{Electric dipole moment of the $\tau$ lepton}

A nonzero value of the electric dipole moment (EDM) is prohibited both by $T$ and $P$ invariance. The strongest direct experimental restriction on 
the $\tau$-lepton EDM $d_\tau$ at the level of $10^{-17}e$ sm was obtained in the Belle experiment using only a small part of the total statistics 
\cite{Inami:2002ah}. A possible increase in accuracy on the larger integrated luminosity will be based on the same method of optimised
observables \cite{Bernreuther:1993nd,Atwood:1991ka} in which the sensitivity to the $d_\tau$ is maximum. The density matrix squared for the
process $e^+e^- \to \tau^+\tau^-$ is given by the sum of the SM contribution, ${\cal M}^2_{\rm SM}$, the EDM  term $|d_\tau|^2{\cal M}^2_d$, and the interference term,
\begin{equation}
{\cal M}_{\rm prod}^2={\cal M}_{\rm SM}^2
+{\cal M}_{\rm Re}^2\mbox{Re}(d_\tau) 
+{\cal M}_{\rm Im}^2\mbox{Im}(d_\tau)
+{\cal M}_{d^2}^2|d_\tau|^2, 
\end{equation}
where $\mbox{Re}(d_\tau)(\mbox{Im}(d_\tau))$ is the real (imaginary) part of the EDM. The optimised observables are defined as
\begin{equation}
{\cal O}_{\rm Re} = \frac{{\cal M}_{\rm Re}^2}{{\cal M}_{\rm SM}^2},\quad
{\cal O}_{\rm Im} = \frac{{\cal M}_{\rm Im}^2}{{\cal M}_{\rm SM}^2},
\end{equation}
and their numerical values are extracted using the most probable values of the spin and momentum directions of the $\tau$-leptons, $\vec{S}_\pm$ and $\hat{\vec{k}}$, respectively. The averaged values of ${\cal O}_{\rm Re}$ and ${\cal O}_{\rm Im}$ are proportional to the EDM and, therefore, have the strongest sensitivity to it which can be further increased by measuring as many decay modes as possible, like, for example, eight modes in Ref.~\cite{Inami:2002ah}. The analysis of the systematic uncertainties gives a hope to achieve $|\mbox{Re},\mbox{Im}(d_\tau)|\leqslant 10^{-18}$-$10^{-19}$. 

\subsubsection{Anomalous magnetic moment of the $\tau$-lepton}

The SM prediction for the $\tau$-lepton anomalous magnetic moment or, more precisely, for the anomalous part of its magnetic moment, is
$a_\tau=(1.17721 \pm 0.00005) \times 10^{-3}$ \cite{Eidelman:2007sb}. Any significant deviation of the measured value from this prediction may
indicate a manifestation of NP. In most models for NP the effect of its influence on $a_l$ is proportional to the square of the lepton mass, so that
the $\tau$-lepton is $(m_\tau/m_\mu)^2\approx$ 283 times more sensitive to the effects of NP than the muon. This explains the great interest to
measurements of $a_\tau$. The experimental resolution achieved to date for the anomalous magnetic moment is about $10^{-2}$ which is an order of
magnitude larger than the value predicted by the SM. Unlike the case of the electron and muon, $a_\tau$ can not be measured through the spin
precession in a magnetic field because of a short $\tau$-lepton lifetime. The existing limits, $ -0.052 <a_\tau <0.013 $ at the 95\% confidence level
and $a_\tau=-0.018\pm 0.017$, were obtained in the DELPHI experiment at LEP2 in the measurement of the total cross section of the process $e^+e^- \to
e^+e^-\tau^+\tau^-$ at the energies between 183 and 208 GeV \cite{Abdallah:2003xd}. Possibility of alternative methods of extracting $a_\tau$ was
analysed in Ref.~\cite{Eidelman:2016aih}. To this end the process $e^+e^- \to \tau^+\tau^-$ with a consequent decay of both $\tau$'s was simulated for
the setup of the Belle experiment. First, an old idea was tested  of using the so-called radiation zeros \cite{Laursen:1983sm}, that is, vanishing matrix element squared of the radiative decay $\tau^- \to l^-\nu_\tau\bar{\nu}_l\gamma$ taking place in specific kinematic cases. Simulations performed
demonstrated that this method practically did not allow to improve the sensitivity to $a_\tau$. More promising is the use of the full information
about the events in the entire phase space (using the unbinned maximum-likelihood method) when the $\tau$ lepton from the signal side
decays radiatively. The results of simulations demonstrate that the method does not allow to improve measurements of the EDM, however, given the huge
statistics expected at Belle II, it allows one to increase sensitivity to the anomalous magnetic moment.

\subsubsection{Searches for lepton-flavour-violating $\tau$ decays}

In the SM the lepton flavour is conserved and the neutrinos are massless. Observation of the neutrino oscillations has demonstrated that this is not the case in nature and that the lepton flavour is not conserved in the neutrino sector. However, this does not imply that it is necessarily violated in 
charged leptons as well and processes with its violation will be observed in the near future. Even if the SM is extended to incorporate the neutrino mass generated by the Higgs mechanism, processes with the lepton-flavour violation in charged leptons are suppressed by the fourth power of this mass,
so that the resulting probabilities turn out to be negligible (for example, the decay probability  $\tau^- \to \mu^-\gamma$ is less than $10^{-53}$).

\begin{figure}
\centering
\includegraphics[width=0.8\textwidth]{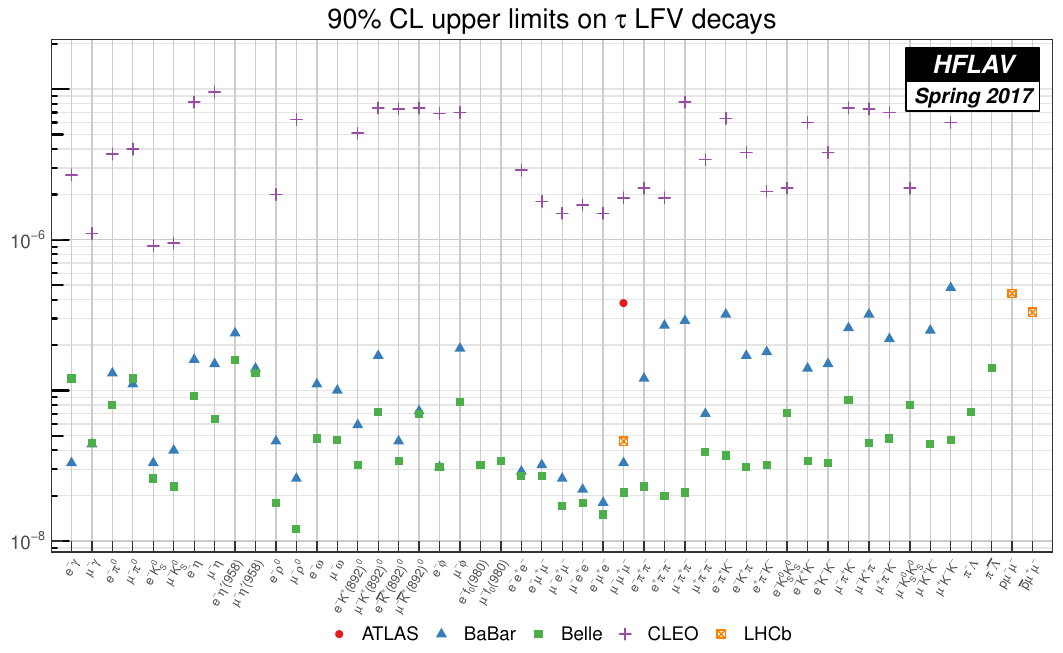}
\caption{Experimental upper limits on the probabilities of the lepton-flavour-violating $\tau$ decays. Adapted from Ref.~\cite{Amhis:2016xyh}.}
\label{fig:lfv}
\end{figure}

In nature, symmetries associated with the lepton flavour are not precise but (to the best of our present understanding) are only accidental. In the meantime, many theories beyond the SM with a typical mass scale of the order of TeV predict, at the level experimentally achievable in the near future, 
the existence of lepton-flavour-violating interactions in the charged lepton sector \cite{Kou:2018nap}.
Given the $\tau$'s large mass, studies of its lepton-flavour-violating decays are much more promising compared 
to the muon decays.  
For example, there appear a large number of final states with one or two mesons that allows one to check the coupling constants between the quarks and leptons for the lepton-flavour-violating interactions. Moreover, one can search for exotic decays like $\tau^+ \to \mu^-e^+e^+$, with violation of all possible
lepton symmetries, or $\tau^- \to \Lambda \pi^-$ in which, additionally, the baryon number is not conserved. Figure~\ref{fig:lfv} \cite{Amhis:2016xyh} shows the upper limits on the probabilities of the lepton-flavour-violating decays thus providing the full picture of such searches for the $\tau$ decays in various experiments.

Proceeding to the Belle II experiment with the integrated luminosity two orders of magnitude larger than before one may hope that for the decay modes with only charged particles in the final state, and therefore with almost no background, such as $\tau^- \to \mu^-e^+e^-$, the upper limits will be better than at Belle proportionally to the luminosity thus reaching the level of ${\cal O}(10^{-10})$. For the decays with a background, like $\tau^- \to \mu^-\gamma$, reaching the level better than $10^{-9}$ looks feasible but can require specific and more complex methods of the analysis. 
 
\subsubsection{$CP$ violation in $\tau$ decays}

In the SM with three generations, $CP$ violation is explained by the Kobayashi-Maskawa mechanism according to which $CP$ violation occurs in the quark sector and is absent in the lepton one (see chapter \ref{chapter:standmod}). For this reason, studies of $CP$ violation in hadronic decays of 
$\tau$ provide an interesting opportunity to search for NP by detecting a new source of $CP$ violation that is not present in the
Kobayashi-Maskawa mechanism. In particular, observation of $CP$ violation in $\tau$ decays will mean that a new type of interaction, besides the $K^0$-$\bar{K}^0$ mixing induced in the SM, is present in the $\tau$-$\nu_\tau$ and/or $d$-$u$ or $s$-$u$ current. So far there are two measurements of the 
$CP$ asymmetry in the BaBar and Belle experiments.

The BaBar Collaboration made an attempt to measure the $CP$ asymmetry by considering the difference of the widths of the decays $\tau^+ \to \pi^+K^0_S\bar{\nu}_\tau$ and $\tau^- \to \pi^-K^0_S\nu_\tau$~\cite{BABAR:2011aa},
\be
{\cal A}_\tau=\frac{\Gamma(\tau^+ \to \pi^+K^0_S\bar{\nu}_\tau)-\Gamma(\tau^- \to \pi^-K^0_S\nu_\tau)}
{\Gamma(\tau^+ \to \pi^+K^0_S\bar{\nu}_\tau)+\Gamma(\tau^- \to \pi^-K^0_S\nu_\tau)}.
\ee
In the SM this asymmetry differs from zero because of the $K^0$-$\bar{K}^0$ mixing and, under the assumption of  $CPT$ invariance, can be expressed through the kaon mixing parameter $\epsilon$ \cite{Bigi:2005ts,Grossman:2011zk},
\be
{\cal A}^{\rm SM}_\tau \approx 2\mbox{Re}(\epsilon)\approx (0.36\pm0.01)\%.
\ee
The BaBar result reads
\be
{\cal A}_\tau=(-0.36 \pm 0.23 \pm 0.11)\%,
\ee
that is, it disagrees with the prediction of the SM by 2.8$\sigma$. A comparable level of $CP$ violation also caused by the kaon mixing should be observed in the decays of the $D$ meson,
\be
{\cal A}_D=\frac{\Gamma(D^+ \to \pi^+K^0_S)-\Gamma(D^- \to \pi^-K^0_S)}
{\Gamma(D^+ \to \pi^+K^0_S)+\Gamma(D^- \to \pi^-K^0_S)}.
\ee
This asymmetry is related to the $\tau$-lepton $CP$ asymmetry as ${\cal A}_\tau=-{\cal A}_D$~\cite{Grossman:2011zk}. From the experimental data the mean $CP$ asymmetry of the $D$ meson is ${\cal A}_D=(-0.41 \pm 0.09)\%$, that is, it is known much better than ${\cal A}_\tau$. The Belle Collaboration 
searched for the $CP$-violation using a quite different method with the help of the angular observables in the decay $\tau^- \to \pi^- K^0_S \nu_\tau$ to extract the forward-backward asymmetry ${\cal A}^{CP}_{FB}$ on the data which correspond to the integrated luminosity of 699 fb$^{-1}$ \cite{Bischofberger:2011pw}. It is well known that there are several overlapping resonances with different quantum numbers in the mass region $M(\pi^- K^0_S)<m_\tau$ which need to be analysed to determine ${\cal A}^{CP}$. It should be noted that in the Belle experiment almost all contributions to the systematic uncertainty of ${\cal A}^{CP}$ are related to the detector effects and could be estimated on a test data set in which one $\tau$ decayed into three charged pions and a neutrino. Once this source of the systematic uncertainty depends on the statistics of the control set, it is expected that at Belle II the uncertainty will be diminished proportionally to the growth in the integrated luminosity as compared to the Belle experiment. Then, for the full luminosity of 50 ab$^{-1}$ one can expect an improvement in both the statistical and systematic uncertainties by $\sqrt{70}$ times, that should give $|{\cal A}^{CP}|<(0.4\mbox{-}2.6) \times 10^{-4}$ at the 90\% confidence level under the assumption of the central value of ${\cal A}^{CP}$ equal to 0. For such a high accuracy, extraction of the parameters of $CP$ violation needs to be performed simultaneously with the determination of the form factors. 

Thus, a more precise measurement of ${\cal A}_\tau$ и ${\cal A}^{CP}$ is definitely one of the high-priority tasks for Belle II. Other ways of searching for $CP$ violation in $\tau$ decays are discussed in the review~\cite{Kou:2018nap}.

\subsection{Semileptonic $\tau$-lepton decays}

In this chapter we briefly discuss semileptonic (in what follows for simplicity referred to as hadronic)  $\tau$-lepton decays and perspectives of their studies in the Belle II experiment. 

\subsubsection{$\tau$-lepton decays to final states with hidden strangeness}

Hadronic decays of the $\tau$ lepton with hidden strangeness in the final state, that is Cabibbo-unsuppressed ones, constitute about 62\% of all $\tau$ decays \cite{Zyla:2020zbs}. Detecting various final states of this type one can extract the so-called spectral functions which correspond to different quantum numbers of the produced hadronic system. This complicated task was accomplished in the OPAL \cite{Ackerstaff:1998yj} and ALEPH \cite{Davier:2013sfa} experiments at the LEP $e^+e^-$ collider. Despite a relatively low precision of these measurements they allow one to extract fundamental parameters of QCD, in particular to obtain one of the most accurate values of the strong coupling constant $\alpha_s(m^2_\tau)$ \cite{Pich:2016bdg}. A precision measurement of the spectral functions is an important task for Belle II. The main hadronic decays of the $\tau$ lepton, Cabibbo-unsuppressed decays to the final states with several pions ($\tau^- \to \pi^-\pi^0\nu_\tau$, $\tau^- \to (3\pi)^-\nu_\tau$, $\tau^- \to (4\pi)^-\nu_\tau,~\ldots$) in the first place, have been measured with a high precision \cite{Zyla:2020zbs}. Decays with the pions in the isovector state, in particular, those with an even number of pions (as well as the decays to $\omega\pi^-$, $\eta\pi^+\pi^-$, $\ldots$) are subject to a relation, based on the conservation of the vector current (CVC), between the total cross section $e^+e^- \to H^0$ at the energy $\sqrt{q^2}$ and the differential cross section of the decay $\tau^- \to H^-\nu_\tau$, where $H$ is a hadronic system of the mass $q$ \cite{Tsai:1971vv,Thacker:1971hy},
\be
\frac{d\Gamma}{dq^2}=\frac{G_F|V_{ud}|^2S_{\rm EW}}{32\pi^2 m_{\tau}^3}
(m_{\tau}^2-q^2)^2(m_{\tau}^2+2q^2)v_1(q^2),\quad 
v_1(q^2)=\frac{q^2\sigma^{I=1}_{e^+e^-}(q^2)}{4\pi\alpha^2},
\ee
where $S_{\rm EW}=1.0201 \pm 0.0003$ is the electroweak correction~\cite{Marciano:1988vm}. In Ref.~\cite{Eidelman:1990pb} this relation was systematically verified for different final hadronic states on the basis of the entire bulk of the experimental data on the $e^+e^-$-annihilation and $\tau$ decays available at that time and the above relations were demonstrated to be accurate up to 5-10\%, in agreement with the expected accuracy of the isospin symmetry. As new experimental data arrived and their precision increased, the picture started to change. The analysis performed in 2002-2003 demonstrated that the spectral functions obtained from $\tau$ decays systematically exceeded the corresponding functions extracted from the $e^+e^-$-annihilation \cite{Davier:2002dy,Davier:2003pw}. In a latter analysis from Ref.~\cite{Jegerlehner:2011ti} the discrepancy between the spectral functions is declared to almost vanish with a correct account for the $\rho$-$\gamma$-mixing. Nevertheless the problem of a precise evaluation of the corrections due to isospin breaking persists and, in order to be solved, will require both new tests of the CVC relations, based on more accurate measurements of the hadronic spectra in $\tau$ decays to become possible in the Belle II experiment, and new theoretical efforts. The experience gained from the CLEO experiment
\cite{Asner:1999kj} has shown that even a relatively low data sample in the decay $\tau^- \to \pi^-\pi^0\pi^0\nu_\tau$ allows one to extract interesting information on the structure of the $a_1(1260)$ meson. This gives hope that a precision measurement of the three-pion decays of the $\tau$ with
two possible charge combinations in the final state ($\pi^-\pi^+\pi^-$ and $\pi^-\pi^0\pi^0$) will considerably improve our understanding of the decays and excitations of the $a_1(1260)$ meson.
 
\subsubsection{Cabibbo-suppressed decays and kaon spectroscopy}

Cabibbo-suppressed decays with open strangeness constitute slightly more than 3\% of all $\tau$-lepton decays \cite{Zyla:2020zbs}. These are the decays to the final states with one kaon ($K^-$ or $\bar{K}^0$) plus several pions and/or $\eta$-mesons and (with a small probability\footnote{The branching ratio of the decay mode $\tau^- \to K^-K^+K^-\nu_\tau$ is approximately $2 \times 10^{-5}$.}) to three kaons. Detection of all possible final states and measuring the mass distribution in the hadronic system (the stranged spectral function) allows one to estimate the mass of the strange quark and the element  $V_{us}$ of the CKM matrix \cite{Antonelli:2013usa}. Similarly to the decays with hidden strangeness, this problem has not yet been solved at $B$-factories and calculations rely on the old results on the spectral functions from ALEPH \cite{Barate:1999hj} and OPAL \cite{Abbiendi:2004xa}. In many cases the produced hadronic system is experimentally known to form a resonant state --- an excited kaon. At present the spectroscopy of such states is poorly understood. The last systematic study of excited kaons was carried out over thirty years ago in $K^-p$ collisions in the LASS experiment (see the review \cite{Aston:1988wf} and references therein). Since then, except for considerable efforts to prove the existence of $K^*_0(700)$ or $\kappa$ \cite{Zyla:2020zbs}, new experimental information on excited states of the $K$ mesons arrived mainly from a few experiments with the amplitude analysis in the decays of the $D$ \cite{Aaij:2017kbo} and $B$-mesons \cite{Guler:2010if}. In all these experiments, $K^*$'s are produced accompanied by other hadrons, and besides that hadrons are also present in the initial state, that leads to poorly controlled effects of strong interactions and potentially
unaccounted systematic errors. These serious drawbacks are absent if the decay products of the $K^*$ meson are produced in a $\tau$-lepton decay. Then there are no other hadrons involved in the process that facilitates the interpretation of the results and reduces systematic effects. An example of such a study is the decay $\tau^- \to K^-\pi^+\pi^-\nu_\tau$ for which the analysis of statistics of about 7000 events allowed to observe the $K_1(1270)$ and $K_1(1400)$ meson \cite{Asner:2000nx}. Unfortunately, this data sample is insufficient for more accurate measurements because of the proximity of the studied states to each other and their large widths. High-statistics experiments like, for example, studies of the decay $\tau^- \to K^0_S\pi^-\nu_\tau$ at Belle using over 53 thousand events, allowed to measure the mass and width of the $K^*(892)$-meson with the world best
precision \cite{Epifanov:2007rf}. There are currently 12 known $K^*$-states with the mass below the mass of the $\tau$ lepton. A significant increase in statistics at Belle II together with the improved methods of the amplitude analysis developed in recent years \cite{Grube:2019hoa} will allow a detailed study of these mesons and their decays and, thus, will significantly improve our understanding of the strong interactions involving the $s$-quark.

\subsubsection{Searches for second-class currents in $\tau$-lepton decays}

Hadronic currents can be classified according to their quantum numbers $J^{PG}$, spin, parity, and $G$ parity as the first-class currents (FCC) with the quantum numbers $J^{PG}=0^{++} (\sigma)$, $0^{--}(\pi)$, $1^{+-}(a_1)$, $1^{-+}(\rho)$~\cite{Weinberg:1958ut} and the second-class currents (SCC) with 
$J^{PG}=0^{+-} (a_0)$, $0^{-+}(\eta),$ $1^{++}(b_1)$, $1^{--}(\omega)$ which have not yet been discovered.
The $G$ parity combines the charge and isospin symmetries. The latter is broken since $m_u \neq m_d$ и $q_u \neq q_d$. However, once this breaking is relatively small, then the $G$ parity is a good approximate symmetry of strong interactions. For this reason, in the framework of the SM and for some quantum numbers $J^P$, hadronic systems with the $G$ parity corresponding to the left-handed weak (light) quark current are allowed and can be easily produced. Meanwhile hadronic systems with a ``wrong'' $G$ parity possess SCC quantum numbers and are suppressed. In the SM a small violation is induced by the isospin breaking that results in {\it induced} SCC. In addition to this suppressed effect, ``generic'' weak SCC may exist which are caused by NP and reveal themselves in somewhat enhanced probabilities compared to what is expected from the isospin-symmetry breaking or a background evaluated in the SM. Observation of the decay $\tau^- \to b^-_1\nu_\tau$ or $\tau^- \to a^-_0\nu_\tau$ would be a clear manifestation of the SCC~\cite{Leroy:1977pq}.
For the most frequently discussed SCC decay $\tau^- \to \eta\pi^-\nu_\tau$ theory predicts the partial branching at the level of $10^{-5}$-$10^{-6}$. Smallness makes this quantity sensitive to various background processes. To better understand the latter Belle made an attempt to search for various exclusive decays with the $\eta$ meson in the final state~\cite{Inami:2008ar}. 
In the BaBar searches, the decay mode $\eta \to \pi^+\pi^-\pi^0$ was used which, in addition to the dominating background from the decay 
$\tau^- \to \eta\pi^-\pi^0\nu_\tau$, is contributed by other processes. In the future the decay $\eta \to 2\gamma$ may appear quite promising despite a considerable background contribution from the decays $\tau^- \to \eta\pi^-\pi^0\nu_\tau$ and $\tau^- \to \pi^-\pi^0\nu_\tau$.
Nevertheless, for the branching ratio ${\cal B}(\tau^- \to \eta\pi^-\nu_\tau) \sim 1 \times 10^{-5}$ predicted by  theory, the statistics at Belle II should definitely guarantee the discovery of the SCC. In this case NP can reveal itself through an anomalously large probability of such a decay and allow one to set the upper limit on the possible charged Higgs boson exchange~\cite{Descotes-Genon:2014tla}.
 
The decay $\tau^- \to \omega\pi^-\nu_\tau$ proceeding through the vector hadronic current unduced by the $\rho$-, $\rho'$-, $\rho''$-,\ldots mesons has the probability of the order of 2\%. However, the SCC may contribute to it, for example, through the $b_1(1235)$ resonance, that will result in a modification of the angular distribution for the angle between the normal to the decay plane of the $\omega$ meson and the direction of the fourth pion in the $\omega$ rest frame. The expected forms of such distributions are given in Ref.~\cite{Chung:1968zz}. BaBar used the statistics of 347.3 fb$^{-1}$ to search for the  SCC contribution and set the upper limit at the level of 1.4$\times 10^{-4}$~\cite{Aubert:2009an} while the theoretical calculation gives 2.5$\times 10^{-5}$~\cite{Paver:2012tq}. A simple estimate demonstrates that at Belle II, with the approximately 1400 times larger luminosity, one can expect to observe SCC in this $\tau$ decay, too. 

\section{Hadronic physics}\label{chapter:hadrons}

\begin{figure}[t!]
\begin{center}
\begin{tabular}{cc}
$B$-meson decays & Two-photon production \\
(any quantum numbers) & ($J^{PC}=0^{\pm+}$ $2^{\pm+}$)\\
\includegraphics[width=0.45\textwidth]{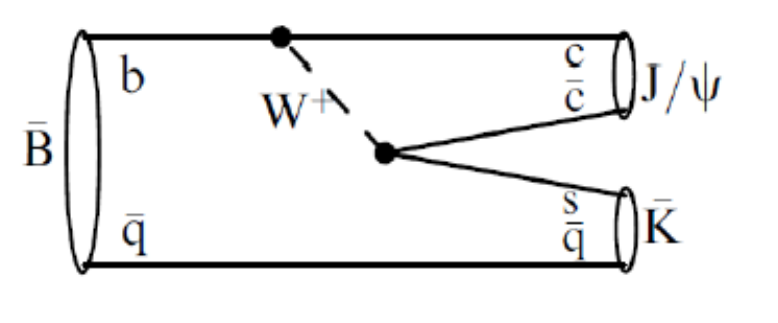} & \includegraphics[width=0.45\textwidth]{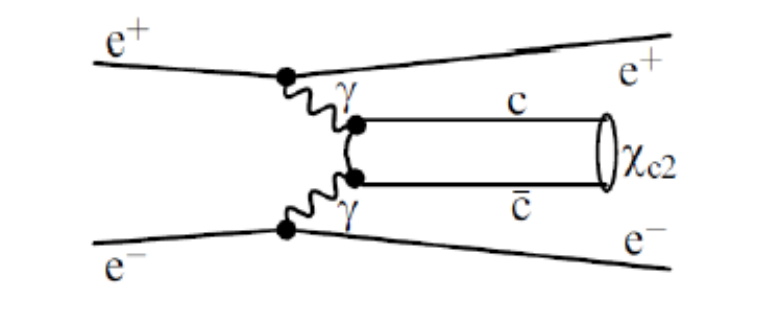} \\
Double production & Initial-state radiation \\
($J^{PC}=0^{\pm+}$) & ($J^{PC}=1^{--}$) \\
\includegraphics[width=0.45\textwidth]{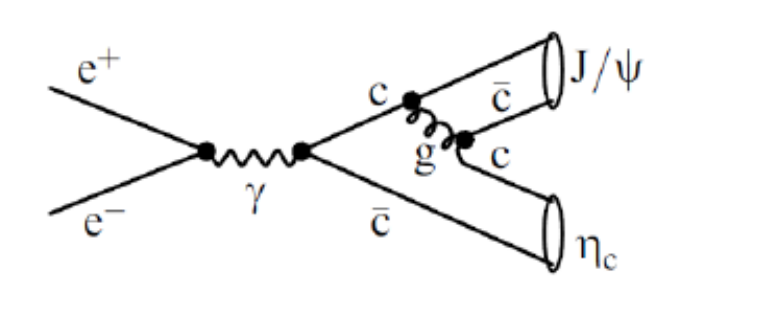} & \includegraphics[width=0.45\textwidth]{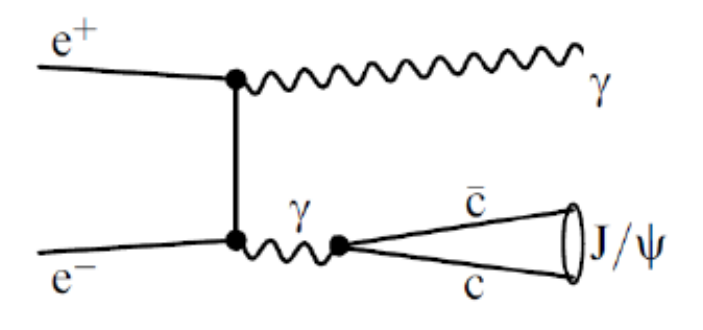}
\end{tabular}
\end{center}
\caption{Production mechanisms for charmonia at $B$-factories. Adapted from Ref.~\cite{Pakhlova:2010zza}.}
\label{fig:charmprod}
\end{figure}

Experiment Belle II, like the preceding experiments at $B$-factories, provides new unique opportunities in searches and studies of hadronic states containing heavy quarks. The most up-to-date classification of such states according to PDG \cite{Zyla:2020zbs} is given in Table~\ref{tab:PDGnames}. 

\begin{table}[t!]
\begin{center}
\begin{tabular}{|c|c|cccc|}
\hline
\multicolumn{2}{|r|}{$PC$}&${-+}$&${+-}$&${--}$&${++}$\\
\hline
Isospin & \vphantom{$\Bigl($} $Q\bar{Q}$ pair & & & & \\
\strut $I=0$ & $c\overline c$&$\eta_c$&$h_c$&$\psi$&$\chi_c$\\
\strut $I=0$ & $b\overline b$&$\eta_b$&$h_b$&$\Upsilon$&$\chi_b$\\
\strut $I=1$ & $c\overline{c}$&$\Pi_c$&$Z_c$&$R_c$&$W_c$\\
\strut $I=1$ & $b\overline{b}$&$\Pi_b$&$Z_b$&$R_b$&$W_b$\\
 \hline
\end{tabular}
\caption{Classification of states containing a heavy $Q\bar{Q}$ pair adopted by PDG \cite{Zyla:2020zbs} (adapted from the review \cite{Brambilla:2019esw}). For isovector states $C$ implies the charge parity of their electrically neutral components. An additional lower index may indicate the total spin of the state $J$ --- see Figs.~\ref{fig:hadrons1} and \ref{fig:hadrons2}. If the quantum numbers of a state are not determined, it is temporarily denoted as $X$.}
\label{tab:PDGnames}
\end{center}
\end{table}

Vector bottomonia $\Upsilon$ can be produced in $e^+e^-$ collisions directly through the annihilation of the electron-positron pair into a virtual photon. It is sufficient just to set the invariant energy of the $e^+e^-$ pair near the mass of the corresponding resonance, $m_\Upsilon$, and collect statistics sufficient for the analysis. Bottomonia with other quantum numbers are produced in the decays of the vector $\Upsilon$ with emission of light hadrons (for example, pions or $\eta$ mesons) and photons. The ability to detect such particles, especially neutral, makes the Belle II experiment a unique tool for studies of hadronic states containing heavy quarks. 

\begin{figure}[t!]
\centerline{\epsfig{file=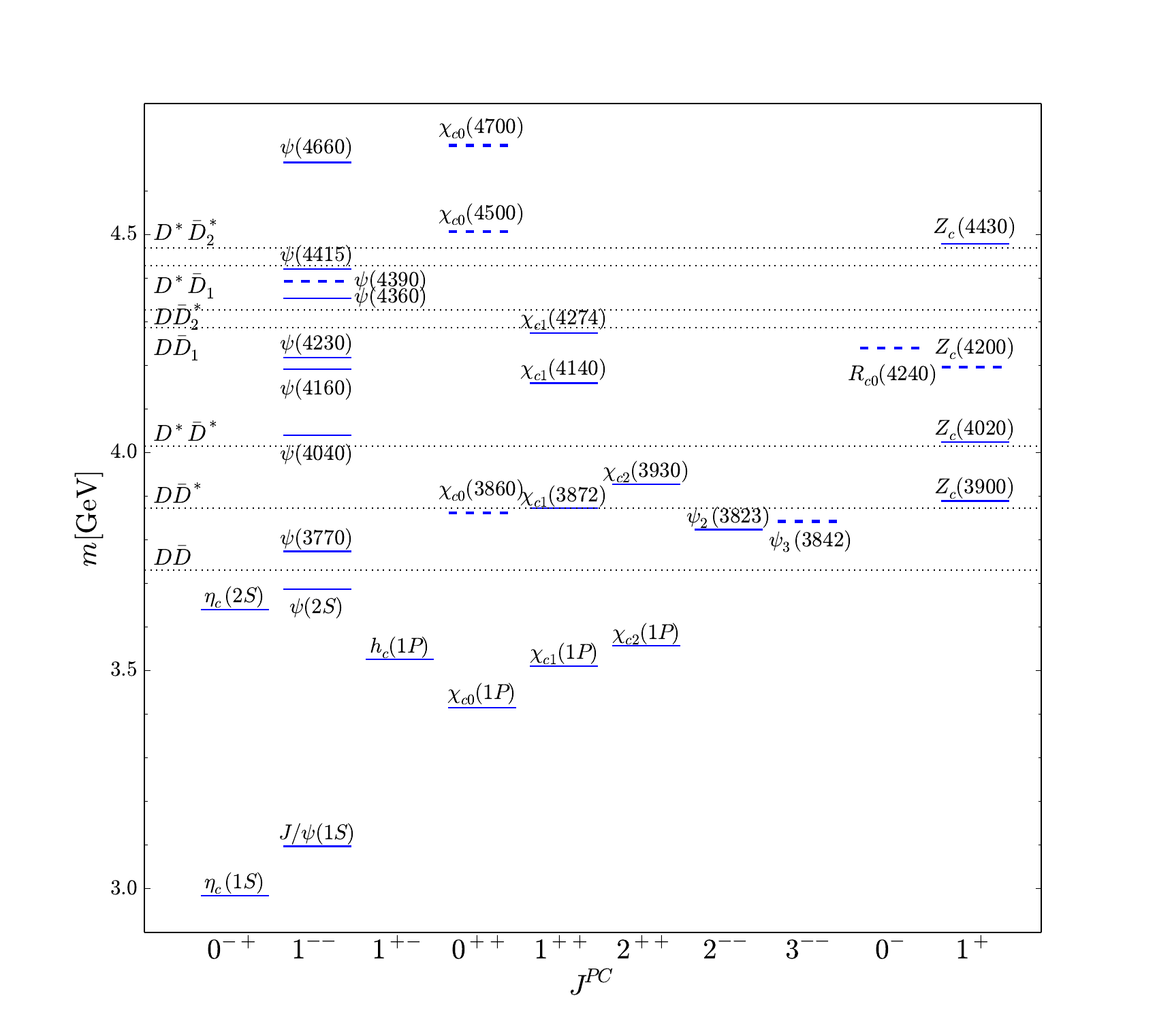, width=0.9\textwidth}}
\caption{The spectrum of experimentally observed charmonium. States regarded as well etsablished (according to PDG) are shown with the solid line and those which still need a confirmation --- with the dashed line. The name of each state (see Table~\ref{tab:PDGnames}) is followed, in parentheses, either by its measured mass or the quantum numbers of the $Q\bar{Q}$ pair if this state is well described by the quark model as a generic quark-antiquark meson. Adapted from the review \cite{Brambilla:2019esw}. }
\label{fig:hadrons1}
\end{figure}

\begin{figure}[t!]
\centerline{\epsfig{file=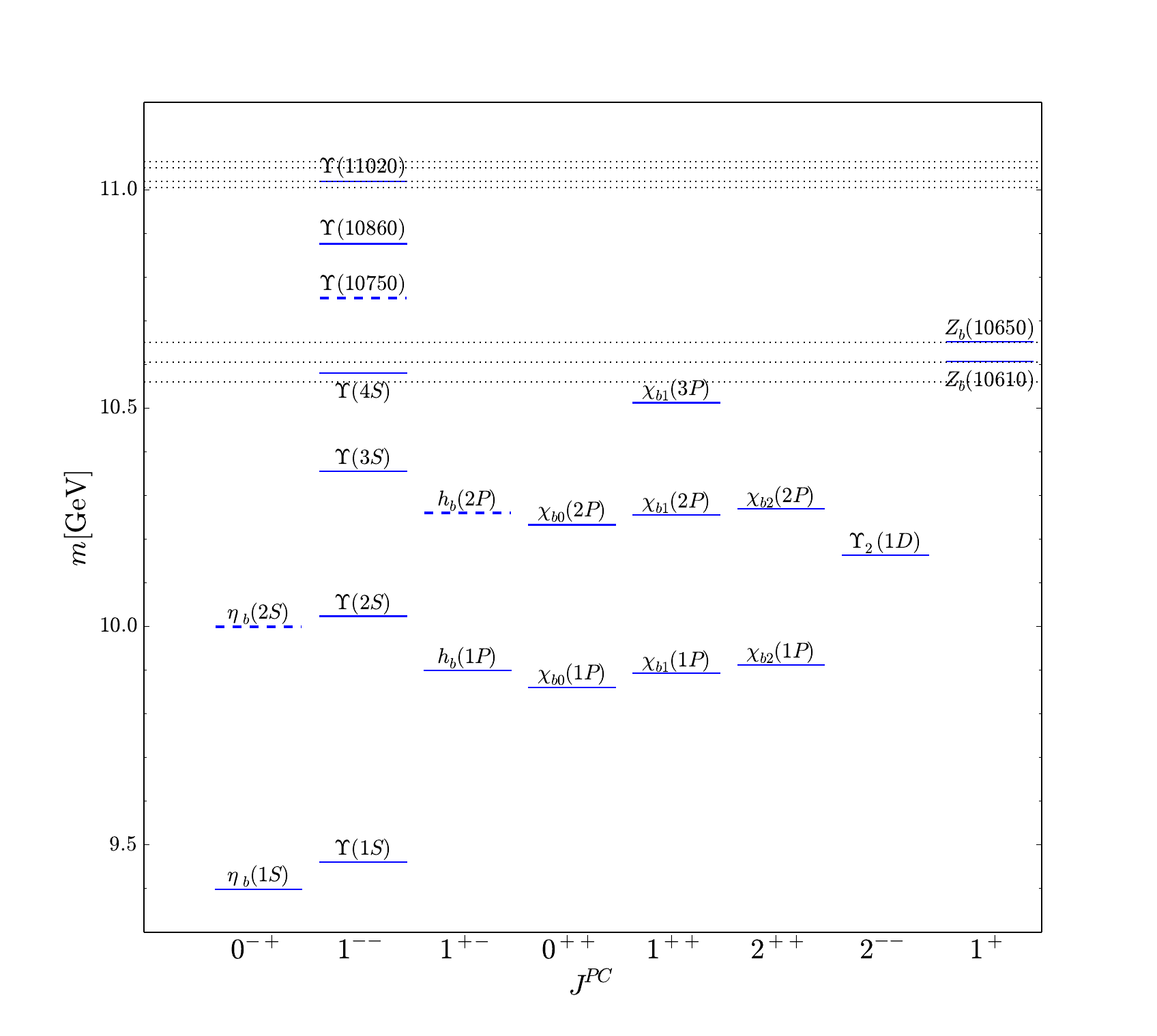, width=0.9\textwidth}}
\caption{Same as in Fig.~\ref{fig:hadrons1} but for the spectrum of bottomonium.}
\label{fig:hadrons2}
\end{figure}

Charmonia with various quantum numbers can be produced at $B$-factories through several mechanisms --- the most important of them are illustrated in Fig.~\ref{fig:charmprod}. Like in case of bottomonia, vector charmonia are produced most copiously. Figures~\ref{fig:hadrons1} and \ref{fig:hadrons2} demonstrate experimentally observed to date states containing a heavy quark-antiquark ($c\bar{c}$ or $b\bar{b}$) pair. Many of them were observed and studied in the experiments at $B$-factories, and the Belle Colaboration contributed a lot to these studies. As examples one could mention in\-ves\-ti\-ga\-ti\-ons of the low-lying (and therefore the most difficult for the experimental studies) states in the spectrum of charmonium and bottomonium --- the pseudoscalars 
$\eta_c(1S)$, $\eta_c(2S)$ \cite{Xu:2018uye} and $\eta_b(1S)$ \cite{Fulsom:2018hpf}, $\eta_b(2P)$ \cite{Mizuk:2012pb} or axial vectors  $h_b(1P)$ and $h_b(2P)$ \cite{Mizuk:2012pb,Tamponi:2015xzb}. On the other hand, the Belle experiment has contributed a lot to the observation of the candidates to previously unknown charmonia (for example, $\chi_{c0}(2P)$ \cite{Chilikin:2017evr}) and 
bottomonia (for example, $\Upsilon(10750)$ \cite{Abdesselam:2019gth}). Recently Belle provided the first measurement of the radiative decay from a vector bottomonium to an axial vector charmonium $\Upsilon(1S)\to\gamma\chi_{c1}$ \cite{Katrenko:2019vdd}. Thus an important task for Belle II is the observaion of the remaining states predicted by the quark model below the open-flavour threshold, for instance, the tensor charmonium $\eta_{c2}(1D)$. The quark model predicts it at around 3.74 GeV (see, for example, Ref.~\cite{Ferretti:2013faa}). Experimental searches for this state can be performed in the $B$-meson decay chain $B\to \eta_{c2}(1D)K\to h_c(1P)\gamma K\to \eta_c(1S)\gamma \gamma K$ with a consequent $\eta_c(1S)$ decay to hadronic channels. Such a task is not feasible for the experiment LHCb while it could be accomplished at Belle II. A detailed information on the perspectives of the experiment Belle II in various studies of bottomonia and bottomonium-like hadrons can be found in the review \cite{Bondar:2016hva}.

An important result obtained in the Belle experiment and still open for further im\-pro\-ve\-ments is the measurement of the exclusive cross sections of $e^+e^-$ annihilation to open-charm states \cite{Zhukova:2017pen} (the accuracy of the most recent Belle data exceeds that of all previous studies performed by CLEO, BABAR as well as Belle itself about ten years earlier). A theoretical analysis of these data using unitary coupled-channel approaches should allow to extract the parameters of the vector states in the spectrum of charmonium and bottomonium in a model-independent way \cite{Uglov:2016orr}. An alternative source of information on such exclusive cross sections is provided by the BES III experiment. 

The expected potential of Belle II will exceed that of Belle proportionally to the luminosity growth and, therefore, the collected data. In particular, this will allow one to establish the quantum numbers of the states which can not be analysed at the statistics of Belle. Also, this will allow to perform an energy scan in a wider range and with a smaller step than it was done before at the $B$-factories of the previous generation. 

The list of achievements of the Belle experiment and hopes related to Belle II concerning hadronic physics would not be complete without mentioning exotic hadrons. From Figs.~\ref{fig:hadrons1} and \ref{fig:hadrons2} one can see that some hadronic states (for example, isovectors --- see the notations in Table~\ref{tab:PDGnames}) definitely contain not only a heavy $\bar{Q}Q$ pair but also light quarks that automatically classifies such hadrons as exotic (not quark-antiquark states --- see chapter~\ref{chapter:standmod}) and makes them very interesting objects for both experimental and theoretical investigations. Let us discuss in some more detail the potential of Belle II in studies of such hadrons. 

After the November 1974 revolution (the experimental discovery of the charmed quark) followed in 1977 by the observation of the $b$-quark the spectroscopy of hadrons containing heavy quarks developed extensively and predictably. New states of charmonium and bottomonium were found experimentally and quite well described by the quark model as quark-antiquark mesons. Furthermore, given that the $c$ and $b$ quark are quite heavy compared to the typical QCD scale ($m_c,m_b\gg\Lambda_{\rm QCD}$), a good description of heavy quarkonia can be already achieved  in the framework of the simplest nonrelativistic quark model, like the Cornell one  \cite{Eichten:1978tg}. Proceeding to excited charmonia and bottomonia implied only an inclusion of the relativistic corrections, however, no surprises were expected on this way. 

This \emph{status quo} was challenged in 2003 when Belle discovered the state $X(3872)$ in the spectrum of charmonium with the properties at odds with the predictions of the quark model \cite{Choi:2003ue}. It is interesting to notice that despite a very wide spectrum of studies performed at Belle, so far  this publication of the collaboration remains the most cited. A review of the current status of this state can be found in Ref.~\cite{Brambilla:2019esw} while its description in the framework of one of the most successful approaches to exotic hadrons --- the molecular model --- is contained in the work \cite{Kalashnikova:2018vkv}.

For nearly twenty years that passed since the first observation of the $X(3872)$ about twenty new exotic charmonium-like and bottomonium-like states have been discovered. To make certain of the considerable progress gained in this field of hadronic spectroscopy one can check the review \cite{Pakhlova:2010zza} where the state of the art of the charmonium spectroscopy is outlined for the end of the first decade of the 21 century. In particular, at that moment, the number of observed exotic states was less than 10 while now the number of the declared exotic states in the spectrum of charmonium exceeds 20, and approximately half of them are regarded as confirmed and well established --- see Figs.~\ref{fig:hadrons1} and \ref{fig:hadrons2}\footnote{In this figure, the states are labelled according to Table~\ref{tab:PDGnames}, so that the $X(3872)$ is denoted as $\chi_{c1}(3872)$ as per its quantum numbers $J^{PC}=1^{++}$ \cite{Zyla:2020zbs}.}. It is easy to see that one of the key features of the $X(3872)$ is that it resides fantastically close to the  neutral threshold $D\bar{D}^*$ that should strongly affect the process of the formation and further properties of this charmonium-like state (the latest and the most precise measurement of its properties can be found in Ref.~\cite{Aaij:2020xjx}). Furthermore, it is legitimate to claim that exotic states could be observed due to the ability of the contemporary experiment, at $B$-factories in the first place, to work at the energies above the open-flavour threshold where decays of a heavy quarkonium to pairs of heavy-light mesons in the relative $S$ wave become kinematically allowed. 

\begin{figure}[t!]
\centerline{\epsfig{file=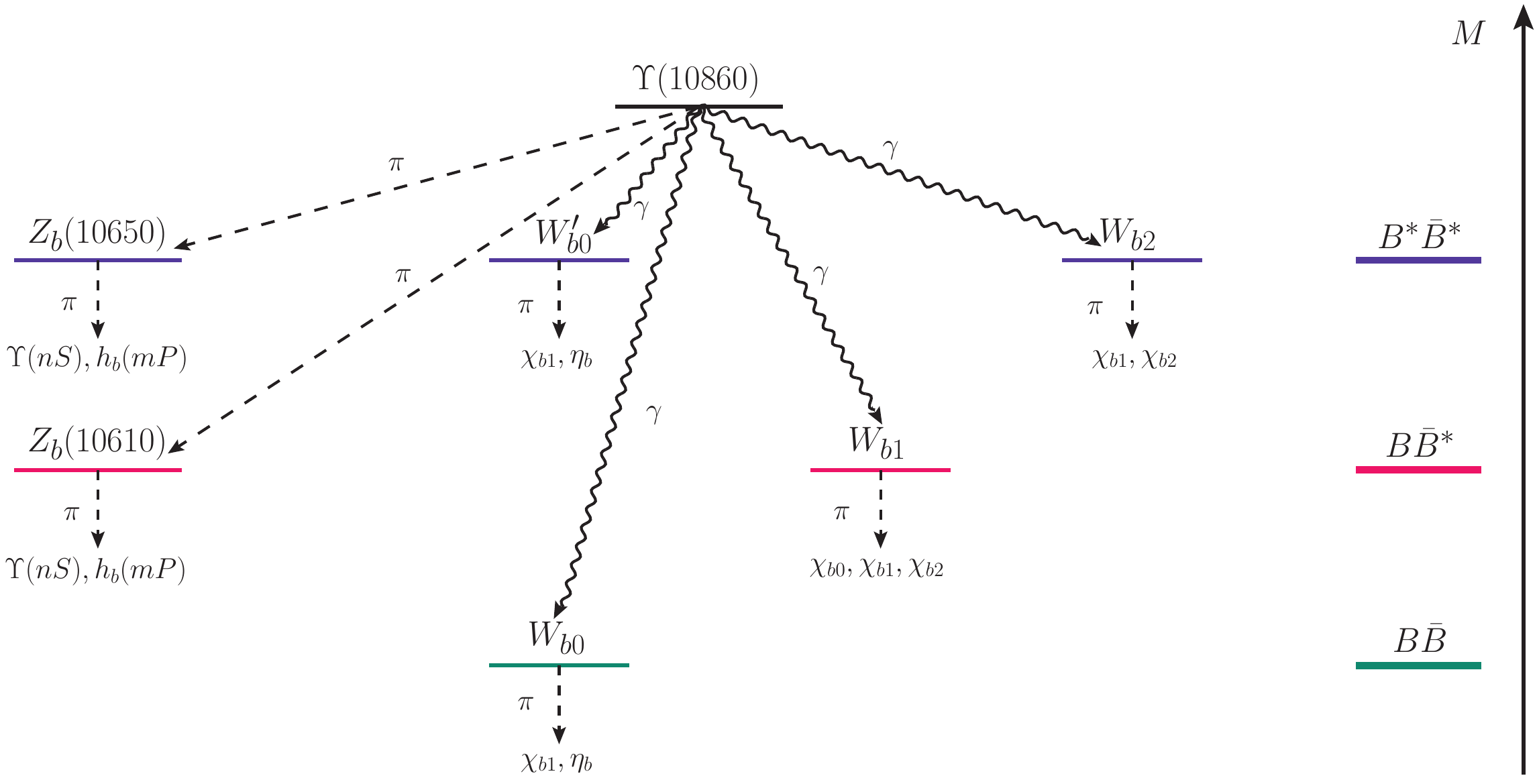, width=0.95\textwidth}} 
\caption{Production and decays of the $Z_b$ states and their predicted spin partners $W_{bJ}$. Adapted from Ref.~\cite{Baru:2019xnh}.}
\label{fig:wbj}
\end{figure}

Let us illustrate the potential of the experiment Belle II in studies of exotic hadrons at the example of the  $Z_b(10610)$ and $Z_b(10650)$ states in the spectrum of bottomonium --- see Fig.~\ref{fig:hadrons2}. In 2011 these states were found by the Belle Collaboration in dipion transitions from the $\Upsilon(10860)$ to lower-lying vector bottomonia $\Upsilon(nS)$ ($n=1,2,3$) and axial vector bottomonia $h_b(mP)$ ($m=1,2$) \cite{Belle:2011aa}. Somewhat later Belle observed the same resonances in the decay channels of the $\Upsilon(10860)$ to the open-bottom final states $\pi B^{(*)}\bar{B}^{*}$ \cite{Collaboration:2011gja,Adachi:2012cx,Garmash:2015rfd}. Exotic nature of the $Z_b$ bottomonia  is obvious since they are seen as peaks in the mass distributions $\Upsilon(nS)\pi^\pm$ and $h_b(mP)\pi^\pm$, that is on the one hand they contain a $b\bar{b}$ pair but on the other hand are charged (have isospin 1) --- the feature accessible only if at least one pair of light quarks is contained in the system. Thus the minimal quark content of these states is four-quark. The main competing theoretical approaches which pretend to describe the data on the production and decay modes of the $Z_b$ states are the molecular and tetraquark models. A detailed description of these approaches and related references can be found in the reviews 
\cite{Guo:2017jvc,Bondar:2016hva} on the molecular model and in the review \cite{Esposito:2014rxa} on the tetraquark model. It should be noted that the molecular model for the $Z_b$'s suggested shortly after their experimental discovery \cite{Bondar:2011ev} allowed one to naturally resolve one of the main puzzles related to the dipion decays of the bottomonium $\Upsilon(10860)$. The problem is that in order to build a vector state ($J^{PC}=1^{--}$) from a quark and an antiquark their spins need to be aligned so that the total spin of the quarks is  $S_{b\bar{b}}=1$ while for an axial vector state ($J^{PC}=1^{+-}$) it should be $S_{b\bar{b}}=0$. Therefore, the transition from a vector bottomonium to another vector bottomonium proceeds without the heavy quark spin flip while transitions to 
axial-vector bottomonia 
require such a flip. Then since the spin-dependent operators in the Hamiltonian contain the mass of the corresponding particle in the denominator the probability of the spin flip for the heavy $b$ quark is suppressed by a small ratio $\Lambda_{\rm QCD}/m_b\ll 1$. 
If applied to the decays $\Upsilon(10860)\to\pi\pi\Upsilon(nS)$ ($n=1,2,3$) and $\Upsilon(10860)\to\pi\pi h_b(mP)$ ($m=1,2$) this implies a considerable (up to two orders of magnitude) suppression of the latter decays compared to the former ones. Contrary to this expectation, both types of decays were experimentally found to have similar probabilities \cite{Belle:2011aa}. To explain this result it is conjectured in the work \cite{Bondar:2011ev} that studied decays proceed not directly but as cascades through the formation of intermediate isovector bottomonia $Z_b$ and $Z_b'$ with the quantum numbers $J^{PC}=1^{+-}$ (as before, we quote the $C$ parity of the neutral component of the isovector) residing near the thresholds $B\bar{B}^*$ and $B^*\bar{B}^*$, respectively. If in addition the wave functions of the $Z_b$'s are assumed to be two orthogonal superpositions of the states with different orientations of the heavy quark spins, $S_{b\bar{b}}=0$ and $S_{b\bar{b}}=1$, then transitions to the final states with $\Upsilon$ and $h_b$ have similar probabilities since they proceed via different components of the wave functions which.
however, have the same weight. This conclusion is an example of predictions of the the so-called Heavy Quark Spin Symmetry (HQSS) which establishes relations between partial probabilities of the decays of the $Z_b$ molecular states (see, for example, Refs.~\cite{Voloshin:2011qa,Mehen:2011yh}). Furthermore, this symmetry predicts the existence of molecular states with a different orientation of the quark spins and as a result with different quantum numbers. Such states are known as spin partners and, for a given total spin $J$, (only $S$-wave molecules are considered, so that the angular momentum equals zero) their spin wave functions are composed from the combinations of the form $[S_{Q\bar{Q}}\otimes S_{q\bar{q}}]_J$, where $S_{Q\bar{Q}}$ and $S_{q\bar{q}}$ are the total spins of the heavy and light quark-antiquark pair, respectively \cite{Voloshin:2011qa}. The spin partners of the $Z_b$ states are traditionally denoted as $W_{bJ}$, have the quantum numbers  $J^{PC}=0^{++}$, $1^{++}$, $2^{++}$ (there are two $0^{++}$ states) and reside near the thresholds $B\bar{B}$, $B\bar{B}^*$, and $B^*\bar{B}^*$ \cite{Bondar:2011ev,Voloshin:2011qa,Mehen:2011yh,Bondar:2016hva,Baru:2017gwo,Baru:2019xnh}. Since the bottomonia $W_{bJ}$ have a negative $G$ parity, they can be produced in the radiative decays of the vector $\Upsilon(10860)$ --- the corresponding scheme is depicted in Fig.~\ref{fig:wbj}. Model-independent predictions for the line shapes and branching ratios of the spin partners $W_{bJ}$ in different channels can be found in Ref.~\cite{Baru:2019xnh}. 

It is important to mention that the observation of the $Z_b$'s became possible after the Belle Collaboration raised the energy of the colliding beams up to approximately 11 GeV (data taking was performed at the energies close to the masses of $\Upsilon(10860)$ and $\Upsilon(11020)$) that exceeds the ``standard'' energy of the $B$-factories working near the $\Upsilon(4S)$ resonance for approximately 1 GeV. Similar measurements but with a substantially higher statistics are expected in the Belle II experiment that is very important for the discovery of the $Z_b$'s spin partners. Indeed, because of a different production mechanism of the $W_{bJ}$'s as compared to the $Z_b$'s (radiative decays rather than pionic ones) the corresponding probabilities are expected to be about two orders of magnitude lower than for the $Z_b$'s. For this reason the data collected at Belle seem insufficient for searches of their spin partners. In the meantime, a considerable increase of the luminosity and, as a result, a much higher statistics expected at Belle II should allow one to perform data analysis to search for the $W_{bJ}$'s. 

In conclusion of this chapter let us note that the experiments at LHC also provide a lot of quite valuable information on hadrons with heavy quarks. Here we quote only a few interesting and important observations made by the LHCb Collaboration in this field quite recently. In Ref.~\cite{Aaij:2019vzc} they report an evidence of the existence of narrow pentaquark states (for only a year after publication these results have been already cited hundreds of times) and in Ref.~\cite{Aaij:2019evc} a candidate is found for a genuine charmonium state $\psi_3(3842)$ with the quantum numbers $3^{--}$ --- see Fig.~\ref{fig:hadrons1}. In Ref.~\cite{Aaij:2020hpf} they discuss a suppression of the charmonium-like state $\chi_{c1}(3872)$ production compared with that of the generic charmonium $\psi(2S)$ (see Fig.~\ref{fig:hadrons1}) in the proton-proton collisions at the energy 8 TeV. In addition, in both cases the production probability decreases with the multiplicity. Such a behaviour is unique for the prompt production of charmonia in $pp$ collisions since similar production probabilities in weak decays of $B$-mesons demonstrate a different pattern. In Ref.~\cite{Aaij:2020qga} the line shape of the $\chi_{c1}(3872)$ in the final state $\pi\pi J/\psi$ is studied in detail and the value of the Breit-Wigner width is obtained for the first time (as opposed to the upper limits set up and several times corrected previously at the $B$-factories). Besides that the pole of the amplitude responsible for this state was extracted. In Ref.~\cite{Aaij:2020xjx} branching ratios of the production of the $\psi_2(3823)$ and $\chi_{c1}(3872)$ charmonia in weak decays of the $B$-mesons were measured and the width of the $\chi_{c1}(3872)$ state was determined more precisely. Finally, in Ref.~\cite{Aaij:2020fnh} the probability of a double-$J/\psi$ production in the proton-proton collisions is measured at the energies from 6.2 to 7.4 GeV and nontirvial structures are observed in the line shape which are attributed to the existence of fully charmed tetraquark states. This observation has immediately initiated a live discussion in the scientific community and resulted in the appearence of several dozen theoretical papers devoted to various attempts to explain the data  \cite{liu:2020eha,Wang:2020ols,Jin:2020jfc,Yang:2020rih,Lu:2020cns,Chen:2020xwe,Wang:2020gmd,Sonnenschein:2020nwn,Albuquerque:2020hio,Giron:2020wpx,Maiani:2020pur,Richard:2020hdw,Wang:2020wrp,Chao:2020dml,Maciula:2020wri,Karliner:2020dta,Wang:2020dlo,Du:2020bqj}. 

Some studies of quarkonia were also performed in the experiments ATLAS \cite{Aad:2011ih} and CMS \cite{Chatrchyan:2013dma}.

\section{Conclusions}\label{chapter:conclusions}

For ages --- the entire time of its development --- science surprised the researches. Probably not a single important discovery or observation was done routinely and predictably but instead happened accidentally and not where it was expected. This must be so in the future too. Meanwhile this does not imply that advanced scientific investigations should not be well prepared and organised. The Belle II experiment is a good example of such a systematic approach. An important feature of this international project is the involvement of scientists from many leading institutes and universities all over the world which de\-mon\-strat\-es a strong cooperation between them. It might well be that this experiment will provide an important clue to our future understanding of what seems mysterious now and let us proceed beyond the SM towards the Theory of Everything. However, even if not, one can be absolutely certain that the enormous efforts spent on the preparation of this experiment will not be in vain and it will definitely advance us in understanding the fundamental foundations of our world. All opportunities related to the Belle II experiment  can not be described in a single review, so we tried to mention in our point of view the most important and promising investigations to be performed in it. Given the absence of other operating $B$-factories as direct competitors of Belle II\footnote{As was discussed in chapter \ref{chapter:history} another factory of beauty, SuperB \cite{Baszczyk:2013xua,Drutskoy:2012gt}, was supposed to work in parallel with Belle II, however because of the economical difficulties this project was eventually discontinued.}, all results obtained in it will have a very strong impact on the high energy physics. Time will tell how justified is our present judgement while now we would like to wish Belle II a long and productive lifetime and many interesting and surprising results. 

\bigskip

The authors are grateful to R. Mizuk for useful discussions. The reported study was funded by RFBR, project number 19-12-50167.


\begin{thebibliography}{100}


\bibitem{Chatrchyan:2012ufa}
Chatrchyan S, \bbletal{} \cc \emph{Phys. Lett.} \textbf{B716} 30 (2012)

\bibitem{Aad:2012tfa}
Aad G, \bbletal{} \ac \emph{Phys. Lett.} \textbf{B716} 1 (2012)

\bibitem{Danilov:1998yt}
Danilov M~V \emph{Phys. Usp.} \textbf{41} 559 (1998); \emph{UFN} \textbf{168} 631 (1998)

\bibitem{Bondar:2007zz}
Bondar A~E, Poluektov A~O, Pakhlov P~N \emph{Phys. Usp.} \textbf{50} 669 (2007); \emph{UFN} \textbf{697} (2007)

\bibitem{Kou:2018nap}
Altmannshofer W, \bbletal{} \bc \emph{PTEP} \textbf{2019} 123C01 (2019) [Erratum: \emph{PTEP} \textbf{2020} 029201 (2020)]

\bibitem{Boos:2014vpa}
Boos E~E \emph{Phys. Usp.} \textbf{57} 912 (2014); \emph{UFN} \textbf{184} 985 (2014)

\bibitem{Kazakov:2014ufa}
Kazakov D I \emph{Phys. Usp.} \textbf{57} 930 (2014); \emph{UFN} \textbf{184} 1004 (2014)

\bibitem{Kazakov:2019fil}
Kazakov D~I \emph{Phys. Usp} \textbf{62} 364 (2019); \emph{UFN} \textbf{189} 387 (2019)

\bibitem{Zyla:2020zbs}
Zyla P~A, \bbletal{} \pc \emph{Prog. Theor. Exp. Phys.} \textbf{2020} 083C01 (2020)

\bibitem{Jarlskog:1985ht}
Jarlskog C \emph{Phys. Rev. Lett.} \textbf{55} 1039 (1985)

\bibitem{Wolfenstein:1983yz}
Wolfenstein L \emph{Phys. Rev. Lett.} \textbf{51} 1945 (1983)

\bibitem{Vainshtein:1975sv}
Vainshtein A~I, Zakharov V~I, Shifman M~A \emph{JETP Lett.} \textbf{22} 55 (1975);
\emph{Pis'ma v ZhETF} \textbf{22} 123 (1975)

\bibitem{Shifman:1995hc}
Shifman M~A \bblin{} \emph{{ITEP Lectures in Particle Physics and Field Theory.
  Edited by M. Shifman. Singapore, World Scientific, 1999. Vol. 1, pp. v-xi}}
  (1995)

\bibitem{Buchalla:1995vs}
Buchalla G, Buras A~J, Lautenbacher M~E \emph{Rev. Mod. Phys.} \textbf{68} 1125
  (1996)

\bibitem{Chetyrkin:1996vx}
Chetyrkin K~G, Misiak M, Munz M \emph{Phys. Lett.} \textbf{B400} 206 (1997);
\emph{Phys. Lett.} \textbf{B425} 414 (1998) Erratum

\bibitem{Czakon:2006ss}
Czakon M, Haisch U, Misiak M \emph{JHEP} \textbf{03} 008 (2007)

\bibitem{Brambilla:2019esw}
Brambilla N, Eidelman S, Hanhart C, Nefediev A, Shen C~P, Thomas C~E, Vairo A,
Yuan C~Z \emph{Phys. Rept.} \textbf{873} 1 (2020)

\bibitem{Adinolfi:2002uk}
Adinolfi M, \bbletal{} \kc \emph{Nucl. Instrum. Meth.} \textbf{A488} 51 (2002)

\bibitem{Adinolfi:2002zx}
Adinolfi M, \bbletal{} \kc \emph{Nucl. Instrum. Meth.} \textbf{A482} 364 (2002)

\bibitem{AmelinoCamelia:2010me}
Amelino-Camelia G, \bbletal{} \kkc \emph{Eur. Phys. J.} \textbf{C68} 619 (2010)

\bibitem{Khazin:2008zz}
Khazin B \emph{Nucl. Phys. Proc. Suppl.} \textbf{181-182} 376 (2008)

\bibitem{Achasov:2009zza}
Achasov M~N, \bbletal{} \emph{Nucl. Instrum. Meth.} \textbf{A598} 31 (2009)

\bibitem{Ablikim:2009aa}
Ablikim M, \bbletal{} \bsc \emph{Nucl. Instrum. Meth.} \textbf{A614} 345 (2010)

\bibitem{Bondar:2013cja}
Bondar A~E, \bbletal{} \emph{Phys. Atom. Nucl.} \textbf{76} 1072 (2013); 
\emph{Yad. Fiz.} \textbf{76} 1132 (2013)

\bibitem{Albrecht:1988vy}
Albrecht H, \bbletal{} \arc \emph{Nucl. Instrum. Meth.} \textbf{A275} 1 (1989)

\bibitem{Prentice:1987ap}
Albrecht H, \bbletal{} \arc \emph{Phys. Lett.} \textbf{B192} 245 (1987)

\bibitem{Aubert:2001tu}
Aubert B, \bbletal{} \bac \emph{Nucl. Instrum. Meth.} \textbf{A479} 1 (2002)

\bibitem{Abashian:2000cg}
Abashian A, \bbletal{} \bc \emph{Nucl. Instrum. Meth.} \textbf{A479} 117 (2002)

\bibitem{Bona:2007qt}
Bona M, \bbletal{} \sbc, arXiv:0709.0451

\bibitem{Abe:2010gxa}
Abe T, \bbletal{} \bbc, arXiv:1011.0352

\bibitem{Bevan:2014iga}
Bevan A~J, \bbletal{} (BaBaR and Belle Collab.) \emph{Eur. Phys. J.} \textbf{C74} 3026 (2014)

\bibitem{cepc} http://cepc.ihep.ac.cn

\bibitem{fcc} http://fcc.web.cern.ch

\bibitem{Behnke:2013xla}
Behnke T, Brau J~E, Foster B, Fuster J, Harrison M, Paterson J~M, Peskin M,
  Stanitzki M, Walker N, Yamamoto H, arXiv:1306.6327

\bibitem{Ohnishi:2013fma}
Ohnishi Y, \bbletal{} \emph{PTEP} \textbf{2013} 03A011 (2013)

\bibitem{Akatsu:1999hi}
Akatsu M, \bbletal{} \emph{Nucl. Instrum. Meth.} \textbf{A440} 124 (2000)

\bibitem{Staric:2008zz}
Staric M, Inami K, Krizan P, Iijima T \emph{Nucl. Instrum. Meth.} \textbf{A595}
  252 (2008)

\bibitem{Iijima:2005qu}
Iijima T, \bbletal{} \emph{Nucl. Instrum. Meth.} \textbf{A548} 383 (2005)

\bibitem{Krizan:2006pc}
Krizan P, Korpar S, Iijima T \emph{Nucl. Instrum. Meth.} \textbf{A565} 457
  (2006)

\bibitem{Nishida:2008zz}
Nishida S, Adachi I, Iijima T, Ikeda H, Korpar S, Krizan P, Miyazawa Y,
  Nishizawa I, Sumiyoshi T \emph{Nucl. Instrum. Meth.} \textbf{A595} 150 (2008)

\bibitem{Nishida:2014gra}
Nishida S, \bbletal{} \emph{Nucl. Instrum. Meth.} \textbf{A766} 28 (2014)

\bibitem{Aushev:2014spa}
Aushev T, \bbletal{} \emph{Nucl. Instrum. Meth.} \textbf{A789} 134 (2015)

\bibitem{Balagura:2005gh}
Balagura V, Danilov M, Dolgoshein B, Klemin S, Mizuk R, Pakhlov P, Popova E,
  Rusinov V, Tarkovsky E, Tikhomirov I \emph{Nucl. Instrum. Meth.}
  \textbf{A564} 590 (2006)

\bibitem{London:1989ph}
London D, Peccei R~D \emph{Phys. Lett.} \textbf{B223} 257 (1989)

\bibitem{Gronau:1989ia}
Gronau M \emph{Phys. Rev. Lett.} \textbf{63} 1451 (1989)

\bibitem{Adachi:2012et}
Adachi I, \bbletal{} \bc \emph{Phys. Rev. Lett.} \textbf{108} 171802 (2012)

\bibitem{Adachi:2018itz}
Adachi I, \bbletal{} (BaBaR and Belle Collab.) \emph{Phys. Rev. Lett.} \textbf{121} 261801 (2018)

\bibitem{Gronau:1990ka}
Gronau M, London D \emph{Phys. Rev. Lett.} \textbf{65} 3381 (1990)

\bibitem{Gronau:2016idx}
Gronau M, Rosner J~L \emph{Phys. Lett.} \textbf{B763} 228 (2016)

\bibitem{Khachatryan:2015nza}
Khachatryan V, \bbletal{} \cc \emph{Phys. Lett.} \textbf{B757} 97 (2016)

\bibitem{Aaboud:2016bro}
Aaboud M, \bbletal{} \ac \emph{JHEP} \textbf{06} 081 (2016)

\bibitem{Charles:2004jd}
Charles J, Hocker A, Lacker H, Laplace S, Le~Diberder F~R, Malcles J, Ocariz J,
  Pivk M, Roos L \emph{Eur. Phys. J.} \textbf{C41} 1 (2005)

\bibitem{Ammar:1993sh}
Ammar R, \bbletal{} \clc \emph{Phys. Rev. Lett.} \textbf{71} 674 (1993)

\bibitem{Bertolini:1987pk}
Bertolini S, Borzumati F, Masiero A \emph{Nucl. Phys.} \textbf{B294} 321 (1987)

\bibitem{Baer:1996kv}
Baer H, Brhlik M \emph{Phys. Rev.} \textbf{D55} 3201 (1997)

\bibitem{Hewett:1996ct}
Hewett J~L, Wells J~D \emph{Phys. Rev.} \textbf{D55} 5549 (1997)

\bibitem{Carena:2000uj}
Carena M, Garcia D, Nierste U, Wagner C~E~M \emph{Phys. Lett.} \textbf{B499}
  141 (2001)

\bibitem{Fujikawa:1993zu}
Fujikawa K, Yamada A \emph{Phys. Rev.} \textbf{D49} 5890 (1994)

\bibitem{Babu:1993hx}
Babu K~S, Fujikawa K, Yamada A \emph{Phys. Lett.} \textbf{B333} 196 (1994)

\bibitem{Cho:1993zb}
Cho P~L, Misiak M \emph{Phys. Rev.} \textbf{D49} 5894 (1994)

\bibitem{Coan:1999kh}
Coan T~E, \bbletal{} \clc \emph{Phys. Rev. Lett.} \textbf{84} 5283 (2000)

\bibitem{Nakao:2004th}
Nakao M, \bbletal{} \bc \emph{Phys. Rev.} \textbf{D69} 112001 (2004)

\bibitem{Aubert:2009ak}
Aubert B, \bbletal{} \bac \emph{Phys. Rev. Lett.} \textbf{103} 211802 (2009)

\bibitem{Aaij:2012ita}
Aaij R, \bbletal{} \lc \emph{Nucl. Phys.} \textbf{B867} 1 (2013)

\bibitem{Horiguchi:2017ntw}
Horiguchi T, \bbletal{} \bc \emph{Phys. Rev. Lett.} \textbf{119} 191802 (2017)

\bibitem{Keum:2004is}
Keum Y~Y, Matsumori M, Sanda A~I \emph{Phys. Rev.} \textbf{D72} 014013 (2005)

\bibitem{Lyon:2013gba}
Lyon J, Zwicky R \emph{Phys. Rev.} \textbf{D88} 094004 (2013)

\bibitem{Beneke:2004dp}
Beneke M, Feldmann T, Seidel D \emph{Eur. Phys. J.} \textbf{C41} 173 (2005)

\bibitem{Ball:2006eu}
Ball P, Jones G~W, Zwicky R \emph{Phys. Rev.} \textbf{D75} 054004 (2007)

\bibitem{Kagan:2001zk}
Kagan A~L, Neubert M \emph{Phys. Lett.} \textbf{B539} 227 (2002)

\bibitem{Ahmady:2013cva}
Ahmady M, Sandapen R \emph{Phys. Rev.} \textbf{D88} 014042 (2013)

\bibitem{Paul:2016urs}
Paul A, Straub D~M \emph{JHEP} \textbf{04} 027 (2017)

\bibitem{Abe:2001dh}
Abe K, \bbletal{} \bc \emph{Phys. Rev. Lett.} \textbf{88} 021801 (2002)

\bibitem{Ishikawa:2003cp}
Ishikawa A, \bbletal{} \bc \emph{Phys. Rev. Lett.} \textbf{91} 261601 (2003)

\bibitem{Jager:2014rwa}
J{\"a}ger S, Martin~Camalich J \emph{Phys. Rev.} \textbf{D93} 014028 (2016)

\bibitem{Jager:2012uw}
J{\"a}ger S, Martin~Camalich J \emph{JHEP} \textbf{05} 043 (2013)

\bibitem{Becirevic:2011bp}
Becirevic D, Schneider E \emph{Nucl. Phys.} \textbf{B854} 321 (2012)

\bibitem{Grossman:2000rk}
Grossman Y, Pirjol D \emph{JHEP} \textbf{06} 029 (2000)

\bibitem{Aaij:2013qta}
Aaij R, \bbletal{} \lc \emph{Phys. Rev. Lett.} \textbf{111} 191801 (2013)

\bibitem{Aaij:2015oid}
Aaij R, \bbletal{} \lc \emph{JHEP} \textbf{02} 104 (2016)

\bibitem{Wehle:2016yoi}
Wehle S, \bbletal{} \bc \emph{Phys. Rev. Lett.} \textbf{118} 111801 (2017)

\bibitem{Altmannshofer:2017fio}
Altmannshofer W, Niehoff C, Stangl P, Straub D~M \emph{Eur. Phys. J.}
  \textbf{C77} 377 (2017)

\bibitem{Buras:2014fpa}
Buras A~J, Girrbach-Noe J, Niehoff C, Straub D~M \emph{JHEP} \textbf{02} 184
  (2015)

\bibitem{Altmannshofer:2009ma}
Altmannshofer W, Buras A~J, Straub D~M, Wick M \emph{JHEP} \textbf{04} 022
  (2009)

\bibitem{Kamenik:2009kc}
Kamenik J~F, Smith C \emph{Phys. Lett.} \textbf{B680} 471 (2009)

\bibitem{Kamenik:2011vy}
Kamenik J~F, Smith C \emph{JHEP} \textbf{03} 090 (2012)

\bibitem{Lutz:2013ftz}
Lutz O, \bbletal{} \bc \emph{Phys. Rev.} \textbf{D87} 111103 (2013)

\bibitem{Lees:2013kla}
Lees J~P, \bbletal{} \bac \emph{Phys. Rev.} \textbf{D87} 112005 (2013)

\bibitem{Grygier:2017tzo}
Grygier J, \bbletal{} \bc \emph{Phys. Rev.} \textbf{D96} 091101 (2017); 
\emph{Phys. Rev.} \textbf{D97} 099902 (2018) Addendum

\bibitem{delAmoSanchez:2010bk}
del Amo~Sanchez P, \bbletal{} \bac \emph{Phys. Rev.} \textbf{D82} 112002 (2010)

\bibitem{Lees:2012wv}
Lees J~P, \bbletal{} \bac \emph{Phys. Rev.} \textbf{D86} 051105 (2012)

\bibitem{Bhattacharya:2018msv}
  B.~Bhattacharya, C.~M.~Grant and A.~A.~Petrov, \emph{Phys. Rev.} \textbf{D99} 093010 (2019)

\bibitem{Hsu:2012uh}
Hsu C~L, \bbletal{} \bc \emph{Phys. Rev.} \textbf{D86} 032002 (2012)

\bibitem{Bennett:2006fi}
Bennett G~W, \bbletal{} \emph{Phys. Rev.} \textbf{D73} 072003 (2006)

\bibitem{Davier:2019can}
Davier M, Hoecker A, Malaescu B, Zhang Z \emph{Eur. Phys. J.} \textbf{C80} 241
  (2020) \emph{Eur. Phys. J.} \textbf{C80} 410 (2020) Erratum

\bibitem{Logashenko:2015xab}
Logashenko I, \bbletal{} \muc \emph{J. Phys. Chem. Ref. Data} \textbf{44} 031211
  (2015)

\bibitem{Abe:2019thb}
Abe M, \bbletal{} \emph{PTEP} \textbf{2019} 053C02 (2019)

\bibitem{Aoyama:2020ynm}
Aoyama T, \bbletal{} \emph{Phys. Rept.} \textbf{887} 1 (2020)

\bibitem{Druzhinin:2011qd}
Druzhinin V~P, Eidelman S~I, Serednyakov S~I, Solodov E~P \emph{Rev. Mod.
  Phys.} \textbf{83} 1545 (2011)

\bibitem{Shifman:1978by}
Shifman M~A, Vainshtein A~I, Zakharov V~I \emph{Nucl. Phys.} \textbf{B147} 448
  (1979)

\bibitem{Eidelman:1978xy}
Eidelman S~I, Kurdadze L~M, Vainshtein A~I \emph{Phys. Lett.} \textbf{82B} 278
  (1979)

\bibitem{Sirunyan:2017khh}
Sirunyan A~M, \bbletal{} \cc \emph{Phys. Lett.} \textbf{B779} 283 (2018)

\bibitem{Aaboud:2018pen}
Aaboud M, \bbletal{} \ac \emph{Phys. Rev.} \textbf{D99} 072001 (2019)

\bibitem{Ablikim:2014uzh}
Ablikim M, \bbletal{} \bsc \emph{Phys. Rev.} \textbf{D90} 012001 (2014)

\bibitem{Albrecht:1992td}
Albrecht H, \bbletal{} \emph{Phys. Lett.} \textbf{B292} 221 (1992)

\bibitem{Abe:2006vf}
Abe K, \bbletal{} \arc \emph{Phys. Rev. Lett.} \textbf{99} 011801 (2007)

\bibitem{Belous:2013dba}
Belous K, \bbletal{} \bc \emph{Phys. Rev. Lett.} \textbf{112} 031801 (2014)

\bibitem{Anastassov:1996tc}
Anastassov A, \bbletal{} \clc \emph{Phys. Rev.} \textbf{D55} 2559 (1997); 
\emph{Phys. Rev.} \textbf{D58} 119904 (1998) Erratum

\bibitem{Aubert:2009qj}
Aubert B, \bbletal{} \bac \emph{Phys. Rev. Lett.} \textbf{105} 051602 (2010)

\bibitem{Lees:2015gea}
Lees J~P, \bbletal{} \bac \emph{Phys. Rev.} \textbf{D91} 051103 (2015)

\bibitem{Fael:2015gua}
Fael M, Mercolli L, Passera M \emph{JHEP} \textbf{07} 153 (2015)

\bibitem{Fetscher:1986uj}
Fetscher W, Gerber H~J, Johnson K~F \emph{Phys. Lett.} \textbf{B173} 102 (1986)

\bibitem{Tamai:2003he}
Tamai K \emph{Nucl. Phys.} \textbf{B668} 385 (2003)

\bibitem{Epifanov:2017kly}
Epifanov D~A \emph{Nucl. Part. Phys. Proc.} \textbf{287-288} 7 (2017)

\bibitem{Inami:2002ah}
Inami K, \bbletal{} \bc \emph{Phys. Lett.} \textbf{B551} 16 (2003)

\bibitem{Bernreuther:1993nd}
Bernreuther W, Nachtmann O, Overmann P \emph{Phys. Rev.} \textbf{D48} 78 (1993)

\bibitem{Atwood:1991ka}
Atwood D, Soni A \emph{Phys. Rev.} \textbf{D45} 2405 (1992)

\bibitem{Eidelman:2007sb}
Eidelman S, Passera M \emph{Mod. Phys. Lett.} \textbf{A22} 159 (2007)

\bibitem{Abdallah:2003xd}
Abdallah J, \bbletal{} \emph{Eur. Phys. J.} \textbf{C35} 159 (2004)

\bibitem{Eidelman:2016aih}
Eidelman S, Epifanov D, Fael M, Mercolli L, Passera M \emph{JHEP} \textbf{03}
  140 (2016)

\bibitem{Laursen:1983sm}
Laursen M~L, Samuel M~A, Sen A \emph{Phys. Rev.} \textbf{D29} 2652 (1984);
\emph{Phys. Rev.} \textbf{D56} 3155 (1997) Erratum

\bibitem{Amhis:2016xyh}
Amhis Y, \bbletal{} \hfc \emph{Eur. Phys. J.} \textbf{C77} 895 (2017)

\bibitem{BABAR:2011aa}
Lees J~P, \bbletal{} \bac \emph{Phys. Rev.} \textbf{D85} 031102 (2012);
\emph{Phys. Rev.} \textbf{D85} 099904 (2012) Erratum

\bibitem{Bigi:2005ts}
Bigi I~I, Sanda A~I \emph{Phys. Lett.} \textbf{B625} 47 (2005)

\bibitem{Grossman:2011zk}
Grossman Y, Nir Y \emph{JHEP} \textbf{04} 002 (2012)

\bibitem{Bischofberger:2011pw}
Bischofberger M, \bbletal{} \bc \emph{Phys. Rev. Lett.} \textbf{107} 131801 (2011)

\bibitem{Ackerstaff:1998yj}
Ackerstaff K, \bbletal{} \oc \emph{Eur. Phys. J.} \textbf{C7} 571 (1999)

\bibitem{Davier:2013sfa}
Davier M, H{\"o}cker A, Malaescu B, Yuan C~Z, Zhang Z \emph{Eur. Phys. J.}
  \textbf{C74} 2803 (2014)

\bibitem{Pich:2016bdg}
Pich A, Rodriguez-S{\'a}nchez A \emph{Phys. Rev.} \textbf{D94} 034027 (2016)

\bibitem{Tsai:1971vv}
Tsai Y~S \emph{Phys. Rev.} \textbf{D4} 2821 (1971);
\emph{Phys. Rev.} \textbf{D13} 771 (1976)

\bibitem{Thacker:1971hy}
Thacker H~B, Sakurai J~J \emph{Phys. Lett.} \textbf{36B} 103 (1971)

\bibitem{Marciano:1988vm}
Marciano W~J, Sirlin A \emph{Phys. Rev. Lett.} \textbf{61} 1815 (1988)

\bibitem{Eidelman:1990pb}
Eidelman S~I, Ivanchenko V~N \emph{Phys. Lett.} \textbf{B257} 437 (1991)

\bibitem{Davier:2002dy}
Davier M, Eidelman S, Hocker A, Zhang Z \emph{Eur. Phys. J.} \textbf{C27} 497
  (2003)

\bibitem{Davier:2003pw}
Davier M, Eidelman S, Hocker A, Zhang Z \emph{Eur. Phys. J.} \textbf{C31} 503
  (2003)

\bibitem{Jegerlehner:2011ti}
Jegerlehner F, Szafron R \emph{Eur. Phys. J.} \textbf{C71} 1632 (2011)

\bibitem{Asner:1999kj}
Asner D~M, \bbletal{} \clc \emph{Phys. Rev.} \textbf{D61} 012002 (2000)

\bibitem{Antonelli:2013usa}
Antonelli M, Cirigliano V, Lusiani A, Passemar E \emph{JHEP} \textbf{10} 070
  (2013)

\bibitem{Barate:1999hj}
Barate R, \bbletal{} \alc \emph{Eur. Phys. J.} \textbf{C11} 599 (1999)

\bibitem{Abbiendi:2004xa}
Abbiendi G, \bbletal{} \oc \emph{Eur. Phys. J.} \textbf{C35} 437 (2004)

\bibitem{Aston:1988wf}
Aston D, \bbletal{} \emph{AIP Conf. Proc.} \textbf{176} 750 (1988)

\bibitem{Aaij:2017kbo}
Aaij R, \bbletal{} \lc \emph{Eur. Phys. J.} \textbf{C78} 443 (2018)

\bibitem{Guler:2010if}
Guler H, \bbletal{} \bc \emph{Phys. Rev.} \textbf{D83} 032005 (2011)

\bibitem{Asner:2000nx}
Asner D~M, \bbletal{} \clc \emph{Phys. Rev.} \textbf{D62} 072006 (2000)

\bibitem{Epifanov:2007rf}
Epifanov D, \bbletal{} \bc \emph{Phys. Lett.} \textbf{B654} 65 (2007)

\bibitem{Grube:2019hoa}
Grube B \bblin{} \emph{{18th International Conference on Hadron Spectroscopy
  and Structure (HADRON 2019) Guilin, Guangxi, China, August 16-21, 2019}}
  (2019)

\bibitem{Weinberg:1958ut}
Weinberg S \emph{Phys. Rev.} \textbf{112} 1375 (1958)

\bibitem{Leroy:1977pq}
Leroy C, Pestieau J \emph{Phys. Lett.} \textbf{72B} 398 (1978)

\bibitem{Inami:2008ar}
Inami K, \bbletal{} \bc \emph{Phys. Lett.} \textbf{B672} 209 (2009)

\bibitem{Descotes-Genon:2014tla}
Descotes-Genon S, Moussallam B \emph{Eur. Phys. J.} \textbf{C74} 2946 (2014)

\bibitem{Chung:1968zz}
Chung S~U, Dahl O~I, Kirz J, Miller D~H \emph{Phys. Rev.} \textbf{165} 1491
  (1968)

\bibitem{Aubert:2009an}
Aubert B, \bbletal{} \bac \emph{Phys. Rev. Lett.} \textbf{103} 041802 (2009)

\bibitem{Paver:2012tq}
Paver N, Riazuddin \emph{Phys. Rev.} \textbf{D86} 037302 (2012)

\bibitem{Pakhlova:2010zza}
Pakhlova G~V, Pakhlov P~N, Eidelman S~I \emph{Phys. Usp.} \textbf{53} 219;
\emph{UFN} \textbf{180} 225 (2010)

\bibitem{Xu:2018uye}
Xu Q~N, \bbletal{} \bc \emph{Phys. Rev.} \textbf{D98} 072001 (2018)

\bibitem{Fulsom:2018hpf}
Fulsom B~G, \bbletal{} \bc \emph{Phys. Rev. Lett.} \textbf{121} 232001 (2018)

\bibitem{Mizuk:2012pb}
Mizuk R, \bbletal{} \bc \emph{Phys. Rev. Lett.} \textbf{109} 232002 (2012)

\bibitem{Tamponi:2015xzb}
Tamponi U, \bbletal{} \bc \emph{Phys. Rev. Lett.} \textbf{115} 142001 (2015)

\bibitem{Chilikin:2017evr}
Chilikin K, \bbletal{} \bc \emph{Phys. Rev.} \textbf{D95} 112003 (2017)

\bibitem{Abdesselam:2019gth}
Mizuk R, \bbletal{} \bc \emph{JHEP} \textbf{10} 220 (2019)

\bibitem{Katrenko:2019vdd}
Katrenko P, \bbletal{} \bc \emph{Phys. Rev. Lett.} \textbf{124} 122001 (2020)

\bibitem{Ferretti:2013faa}
Ferretti J, Galat{\`a} G, Santopinto E \emph{Phys. Rev.} \textbf{C88} 015207
  (2013)

\bibitem{Bondar:2016hva}
Bondar A~E, Mizuk R~V, Voloshin M~B \emph{Mod. Phys. Lett.} \textbf{A32}
  1750025 (2017)

\bibitem{Zhukova:2017pen}
Zhukova V, \bbletal{} \bc \emph{Phys. Rev.} \textbf{D97} 012002 (2018)

\bibitem{Uglov:2016orr}
Uglov T~V, Kalashnikova Yu~S, Nefediev A~V, Pakhlova G~V, Pakhlov P~N \emph{JETP Lett.} \textbf{105} 1 (2017);
\emph{Pis'ma v ZhETF} \textbf{105} 3 (2017)

\bibitem{Eichten:1978tg}
Eichten E, Gottfried K, Kinoshita T, Lane K~D, Yan T~M \emph{Phys. Rev.} \textbf{D17} 3090 (1978); \emph{Phys. Rev.} \textbf{D21} 313 (1980) Erratum

\bibitem{Choi:2003ue}
Choi S~K, \bbletal{} \bc \emph{Phys. Rev. Lett.} \textbf{91} 262001 (2003)

\bibitem{Kalashnikova:2018vkv}
Kalashnikova Yu~S, Nefediev A~V \emph{Phys. Usp.} \textbf{62} 568 (2019);
\emph{UFN} \textbf{189} 603 (2019)

\bibitem{Aaij:2020xjx}
Aaij R, \bbletal{} \lc \emph{JHEP} \textbf{2008} 123 (2020) 

\bibitem{Belle:2011aa}
Bondar A, \bbletal{} \bc \emph{Phys. Rev. Lett.} \textbf{108} 122001 (2012)

\bibitem{Collaboration:2011gja}
Adachi I \bblin{} \bc \emph{{Flavor physics and CP violation. Proceedings, 9th
  International Conference, FPCP 2011, Maale HaChamisha, Israel, May 23-27,
  2011}} (2011)

\bibitem{Adachi:2012cx}
Adachi I, \bbletal{} \bc, arXiv:1209.6450

\bibitem{Garmash:2015rfd}
Garmash A, \bbletal{} \bc \emph{Phys. Rev. Lett.} \textbf{116} 212001 (2016)

\bibitem{Guo:2017jvc}
Guo F~K, Hanhart C, Mei{\ss}ner U-G, Wang Q, Zhao Q, Zou B~S \emph{Rev. Mod.
  Phys.} \textbf{90} 015004 (2018)

\bibitem{Esposito:2014rxa}
Esposito A, Guerrieri A~L, Piccinini F, Pilloni A, Polosa A~D \emph{Int. J.
  Mod. Phys.} \textbf{A30} 1530002 (2015)

\bibitem{Bondar:2011ev}
Bondar A~E, Garmash A, Milstein A~I, Mizuk R, Voloshin M~B \emph{Phys. Rev.}
  \textbf{D84} 054010 (2011)

\bibitem{Voloshin:2011qa}
Voloshin M~B \emph{Phys. Rev.} \textbf{D84} 031502 (2011)

\bibitem{Mehen:2011yh}
Mehen T, Powell J~W \emph{Phys. Rev.} \textbf{D84} 114013 (2011)

\bibitem{Baru:2017gwo}
Baru V, Epelbaum E, Filin A~A, Hanhart C, Nefediev A~V \emph{JHEP} \textbf{06}
  158 (2017)

\bibitem{Baru:2019xnh}
Baru V, Epelbaum E, Filin A~A, Hanhart C, Nefediev A~V, Wang Q \emph{Phys.
  Rev.} \textbf{D99} 094013 (2019)

\bibitem{Aaij:2019vzc}
Aaij R, \bbletal{} \lc \emph{Phys. Rev. Lett.} \textbf{122} 222001 (2019)

\bibitem{Aaij:2019evc}
Aaij R, \bbletal{} \lc \emph{JHEP} \textbf{07} 035 (2019)

\bibitem{Aaij:2020hpf}
Aaij R \bbletal{} \lc, arXiv:2009.06619  

\bibitem{Aaij:2020qga}
Aaij R, \bbletal{} \lc \emph{Phys. Rev.} \textbf{D102} 092005 (2020) 

\bibitem{Aaij:2020fnh}
Aaij R, \bbletal{} \lc Aaij R, \bbletal{} \emph{Sci.Bull.} \textbf{65} 1983 (2020)

\bibitem{liu:2020eha}
liu M~S, Liu F~X, Zhong X~H, Zhao Q, arXiv:2006.11952  

\bibitem{Wang:2020ols}
Wang Z~G \emph{Chin. Phys.} \textbf{C44} 113106 (2020) 

\bibitem{Jin:2020jfc}
Jin X, Xue Y, Huang H, Ping J \emph{Eur. Phys. J.} \textbf{C80} 1083 (2020) 

\bibitem{Yang:2020rih}
Yang G, Ping J, He L, Wang Q, arXiv:2006.13756  

\bibitem{Lu:2020cns}
L{\"u} Q~F, Chen D~Y, Dong Y~B \emph{Eur. Phys. J.} \textbf{C80} 871 (2020)  

\bibitem{Chen:2020xwe}
Chen H~X, Chen W, Liu X, Zhu S~L \emph{Sci. Bull.} \textbf{65} 1994 (2020)

\bibitem{Wang:2020gmd}
Wang X~Y, Lin Q~Y, Xu H, Xie Y~P, Huang Y, Chen X \emph{Phys. Rev.} \textbf{D102} 116014 (2020)

\bibitem{Sonnenschein:2020nwn}
Sonnenschein J, Weissman D \emph{Eur. Phys. J.} \textbf{C81} 25 (2021)

\bibitem{Albuquerque:2020hio}
Albuquerque R~M, Narison S, Rabemananjara A, Rabetiarivony D, Randriamanatrika G \emph{Phys. Rev.} \textbf{D102} 094001 (2020) 

\bibitem{Giron:2020wpx}
Giron J~F, Lebed R~F \emph{Phys. Rev.} \textbf{D102} 074003 (2020)

\bibitem{Maiani:2020pur}
Maiani L, arXiv:2008.01637  

\bibitem{Richard:2020hdw}
Richard J~M \emph{Sci. Bull.} \textbf{65} 1954 (2020)

\bibitem{Wang:2020wrp}
Wang J~Z, Chen D~Y, Liu X, Matsuki T, arXiv:2009.02100  

\bibitem{Chao:2020dml}
Chao K~T, Zhu S~L \emph{Sci.Bull.} \textbf{65} 1952 (2020) 

\bibitem{Maciula:2020wri}
Maciula R, Sch{\" a}fer W, Szczurek A \emph{Phys. Lett.} \textbf{B812} 136010 (2021)

\bibitem{Karliner:2020dta}
Karliner M, Rosner J~L \emph{Phys. Rev.} \textbf{D102} 114039 (2020) 

\bibitem{Wang:2020dlo}
Wang Z~G \emph{Int. J. Mod. Phys.} \textbf{A36} 2150014 (2021)

\bibitem{Du:2020bqj}
Du M~L, Baru V, Guo F~K, Hanhart C, Mei{\ss }ner U~G, Nefediev A, Strakovsky I \emph{Eur. Phys. J.} \textbf{C80} 1053 (2020)

\bibitem{Aad:2011ih}
Aad G, \bbletal{} \ac \emph{Phys. Rev. Lett.} \textbf{108} 152001 (2012)

\bibitem{Chatrchyan:2013dma}
Chatrchyan S, \bbletal{} \cc \emph{Phys. Lett.} \textbf{B734} 261 (2014)

\bibitem{Baszczyk:2013xua}
Baszczyk M, \bbletal{} \sbc, arXiv:1306.5655

\bibitem{Drutskoy:2012gt}
Drutskoy A~G, Guo F~K, Llanes-Estrada F~J, Nefediev A~V, Torres-Rincon J~M
  \emph{Eur. Phys. J.} \textbf{A49} 7 (2013)

\end{thebibliography}

\end{document}